\documentclass{emulateapj}
\usepackage[]{natbib, graphicx, float, enumerate}
\usepackage[FIGTOPCAP]{subfigure}
\usepackage[section]{placeins}
\usepackage{wrapfig}
\usepackage[fleqn]{amsmath}
\begin{document}

\begin{abstract}
The broad-band variability of many accreting systems displays characteristic structure; log-normal flux distributions, RMS-flux relations, and long inter-band lags. These characteristics are usually interpreted as inward propagating fluctuations in an accretion disk driven by stochasticity of the angular momentum transport mechanism. We present the first analysis of propagating fluctuations in a long-duration, high-resolution, global three-dimensional magnetohydrodynamic (MHD) simulation of a geometrically-thin ($h/r\approx0.1$) accretion disk around a black hole.  While the dynamical-timescale turbulent fluctuations in the Maxwell stresses are too rapid to drive radially-coherent fluctuations in the accretion rate, we find that the low-frequency quasi-periodic dynamo action introduces low-frequency fluctuations in the Maxwell stresses which then drive the propagating fluctuations.  Examining both the mass accretion rate and emission proxies, we recover log-normality, linear RMS-flux relations, and radial coherence that would produce inter-band lags.  Hence, we successful relate and connect the phenomenology of propagating fluctuations to modern MHD accretion disk theory.
\end{abstract}


\title{Testing the Propagating Fluctuations Model with a Long, Global Accretion Disk Simulation}

\author{J~Drew~Hogg\altaffilmark{1,2} and Christopher~S.~Reynolds\altaffilmark{1,2}}

\altaffiltext{1}{Department of Astronomy, University of Maryland, College Park, MD20742}
\altaffiltext{2}{Joint Space Science Institute (JSI), University of Maryland, College Park, MD20742}

\maketitle

\section{Introduction}

While accretion of gas onto compact objects has been studied for over fifty years, much uncertain remains about how angular momentum is transported.  There is increasing evidence that the interplay between the local angular momentum transport from magnetohydrodynamic (MHD) turbulence and the more global accretion flow shapes the behavior of the mass accretion \citep[e.g.][]{2010ApJ...712.1241S}.  The primary goal of this paper is to connect MHD accretion disk theory with one of the most widely discussed phenomenological models for disk variability, the propagating fluctuations model.  We explore how a stochastically varying effective viscosity leads to the growth of so-called ``propagating fluctuations" in our long, global accretion disk simulation and relate the behavior of our simulation to observable properties from accreting black holes.

The idealized $\alpha$-disk from the seminal work of \cite{1973A&A....24..337S} has long served as the canonical disk model.  Assuming turbulence was present in the disk, they argued that an effective kinematic viscosity originating from turbulent eddies spanning a radial range provided the internal viscous stress required to transport angular momentum and allow for accretion.  In their model, the turbulent viscosity takes the form $\nu_{t}=v_{t} l$, where $v_{t}$ is the turbulent velocity and $l$ is a characteristic length scale of the turbulent eddies.  This viscosity can be parameterized with an ``efficiency" factor in terms of the local sound speed, $c_{s}$, and the local scale height of the disk, h, $\nu_{t}=\alpha c_{s} h$.  With this prescription, \cite{1973A&A....24..337S} restrict the $\alpha$ parameter to a range of $0<\alpha \leq 1$.  If the turbulent velocity is greater than the sound speed, the gas will shock, convert excess kinetic energy to thermal energy, and return to the subsonic domain, ensuring $v_{t} / c_{s} \leq 1$.  Likewise, they postulated $l / h \leq 1$ because the differential shear in the disk will quickly break-up large turbulent eddies leading to localized turbulence.  An alternative but equivalent view, motivated by dimensional analysis, is to say that the relevant viscous stresses in the disk are proportional to the pressure, $T_{R \phi} = \alpha^\prime P$, where $T_{R \phi}$, where $T_{R \phi}$ is the off-diagonal component of the stress energy tensor and $P$ is pressure.  $\alpha$ and $\alpha^\prime$ are related to each other by a factor of order unity that depends upon the equation of state.

Even in the early work of \citet{1973A&A....24..337S}, the influence of magnetic stress was intuitively assumed to be an essential part of the turbulent injection mechanism, though a viable driving mechanism was absent until the magneto-rotational instability (MRI) was put forth as a candidate to drive the disk turbulence.  The instability of a differentially rotating, magnetized plasma was originally discovered by \citet{Velikhov59} and \citet{1960PNAS...46..253C}, but its astrophysical significance was not realized until \citet{1991ApJ...376..214B} studied it in the context of an ionized accretion disk.  They showed that radial perturbations of gas elements are linearly unstable if $k v_{A} < \Omega$, where $k$ is the wavenumber, $v_{A}$ is the Alfv\'en speed, and $\Omega$ is the orbital angular frequency.  As a fluid element is perturbed, the displacement acts to enhance a weak seed field until it goes nonlinear, which subsequent numerical MHD simulations have verified drives and sustains turbulence \citep{1991ApJ...376..223H}.  To date, a large and rich body of simulation work has been dedicated to studying how MRI driven turbulence behaves and transports angular momentum including studies using unstratified shearing boxes \citep[e.g.][]{1991ApJ...376..223H, 1995ApJ...440..742H}, stratified shearing boxes \citep[e.g.][]{1995ApJ...446..741B,1996ApJ...463..656S}, unstratified global models \citep[e.g.][]{1998ApJ...501L.189A, 2001ApJ...554..534H, 2001ApJ...548..868A, 2003MNRAS.341.1041A}, stratified global disk models \citep[e.g.][]{2000ApJ...528..462H, 2001ApJ...548..348H, 2009ApJ...692..869R, 2010ApJ...712.1241S}, relativistic global models \citep[e.g.][]{2003ApJ...599.1238D, 2003ApJ...589..444G, 2006MNRAS.368.1561M}, and global models with radiative transfer \citep[e.g.][]{2014ApJ...796..106J,2014MNRAS.441.3177M, 2015MNRAS.447...49S}.

The viscous behavior of the disk ultimately arises because of net internal stress due to correlated fluctuations in the magnetic field and correlated fluctuations in gas velocity \citep{1999ApJ...521..650B}.  These correlated fluctuations lead to the Maxwell stress ($M_{R \phi} = -B_{R} B_{\phi}/4\pi$) and the Reynolds stress ($R_{R \phi} = \rho v_{R} \delta v_{\phi}$), respectively.   Following the $\alpha$-prescription, the effective $\alpha$ from the net stress is thus the ratio of the volume averaged stress to the volume averaged pressure,
\begin{equation} \label{eqn-alpha}
\alpha = \frac{\langle M_{R \phi} + R_{R \phi} \rangle}{\langle P \rangle}.  
\end{equation}
This serves to parameterize the effectiveness of angular momentum loss by the stresses acting on the gas in the disk.  In this paper we use angled brackets to denote a volume average for a given quantity, $U$, where
\begin{equation}
\langle U(t) \rangle = \int U(t, r, \theta, \phi) r^2 \textrm{sin} \theta \textrm{d}r \textrm{d}\theta \textrm{d}\phi .
\end{equation}

Turbulent fluctuations in the disk will naturally lead to stochastic variability in the effective $\alpha$ and provide an intuitive explanation for the existence of broadband photometric variability observed in both galactic black hole binaries (GBHBs) and active galactic nuclei (AGNs).  The photometric variability is observed across several decades in frequency, typically in the form of aperiodic ``flicker" type noise.  Employing a stochastic $\alpha$, \citet{1997MNRAS.292..679L} developed a model for this flicker noise by linearizing the standard geometrically thin, optically thick accretion disk evolution equations.  The model was limited to small perturbations in $\alpha$, but it reproduced the power spectrum of the fluctuations and laid the foundation for the subsequent development of the propagating fluctuations model.

The propagating fluctuations model has since taken a more generalized form and served as a phenomenological description for the structure and organization of the broadband variability for GBHBs and AGNs.  The evolution of a geometrically thin, optically thick accretion disk (in units where $G = M = c = 1$) is governed by a nonlinear diffusion equation \citep{1981ARA&A..19..137P}
\begin{equation} \label{eqn-can_disk}
\frac{\partial \Sigma}{\partial t} = \frac{3}{R}\frac{\partial}{\partial R}\Big[R^{\frac{1}{2}} \frac{\partial}{\partial R}(\nu \Sigma R^{\frac{1}{2}})\Big]
\end{equation} 
where R is the distance from the central engine, $\nu(\Sigma; R, t)$ is the local effective viscosity in the disk, and $\Sigma(R,t)$ is the local surface density.  In this equation, the diffusive redistribution of angular momentum depends on the product of the effective viscosity and the local surface density.  If the effective viscosity fluctuates, the instantaneous mass accretion rate ($\dot{M}$) at each point in the disk is thus the product of the many prior stochastic events.  A consequence of the multiplicative combination is that fluctuations in $\dot{M}$ will tend to be preserved as they accrete as higher density regions will typically have higher $\dot{M}$ and lower density regions will typically have lower $\dot{M}$.  Additionally, since the viscous time increases with radii, higher frequency fluctuations will be imposed on the accretion flow as material moves inwards.  This will lead to the development of a hierarchical structure in the fluctuations with the smaller, high frequency fluctuations imposed upon the larger, low frequency fluctuations.

The appeal of the propagating fluctuations model is that it explains three properties of black hole variability that have been difficult to reproduce with other variability models.  First, GBHB and AGN light curves are log-normally distributed \citep{2005MNRAS.359..345U, 2004ApJ...612L..21G} which is accounted for through the multiplicative combination of many independent fluctutations. Second, and related to the log-normal distribution, the root-mean square (RMS) of the flux variability in the light curves of both GBHBs and AGNs depends linearly on the flux level \citep{2002PASJ...54L..69N, 2004ApJ...612L..21G}, the so-called RMS-flux relationship.  This indicates the mass accretion rate variation has a constant fractional amplitude and that the mechanism driving variability is, therefore, independent of the accretion rate \citep{2001MNRAS.323L..26U}.  Finally, coherent fluctuations in the emission at different wavebands from GBHBs have frequency dependent time lags with variability in higher energy bands lagging variability in lower energy bands \citep{1999ApJ...510..874N, 1999ApJ...517..355N}.  This ``hard lag" is believed to be a signature of fluctuations in the accretion flow moving to smaller radii in the disk where the emission temperature is higher. 

Despite the success the propagating fluctuations model has had in explaining the phenomenology of GBHB and AGN variability, the model has yet to be examined in the context of modern MHD theory.  To date, the model has been examined in the context of geometrically-thin (vertically-integrated) $\alpha$-disks in which some stochasticity is imposed upon $\alpha$.   How well this viscous behavior translates into an MHD disk is unknown.  A particularly challenge is the timescales on which $\alpha$ fluctuates and drives variability within the accretion flow.  \citet{1997MNRAS.292..679L} prescribes $\alpha$ fluctuations to be slow, of order the viscous time.  However, if fluctuations in $\alpha$ are solely due to turbulent fluctuations, we might expect them to be on the dynamical timescale.  \citet{2014ApJ...791..126C} show, using a semi-analytic model, that propagating fluctuations only exist if fluctuations in $\alpha$ occur at sufficiently low frequencies.  If the frequency of $\alpha$ fluctuations is too high, they find that $\dot{M}$ fluctuations are damped out and the nonlinear features of the variability do not develop.  Therefore, fluctuations in the effective-$\alpha$ on timescales longer than the dynamical timescale are needed to reproduce the phenomenology seen in the variability.  

In this Paper, we perform the first detailed analysis of propagating fluctuations in the context of a long-duration, MHD accretion disk simulation.  We show that the well-known quasi-periodic dynamo induces slow fluctuations in the effective $\alpha$ that then drive propagating fluctuations.  In Section \ref{sec-model_conv} we introduce the simulation and perform a convergence study.  Section \ref{sec-overview} presents a brief discussion of the nature of the turbulence.  In Section \ref{sec-dynamo}, we investigate the relationship of the quasi-periodic disk dynamo and angular momentum transport, finding that it imposes an intermediate-timescale modulation of the effective $\alpha$.  We then examine the variability of both the mass accretion rate (Section \ref{sec-prop_flucs}) and a proxy for bolometric radiative luminosity (Section \ref{sec-proxy_emission}) finding that both quantities display the non-linear variability characteristics of real sources and that the coherence of these quantities across the disk is exactly as expected in the propagating fluctuations picture.  In Section \ref{sec-discussion}, we place our results into a broader context, and then conclude in Section \ref{sec-summary}.

\section{Numerical Model and Convergence}
\label{sec-model_conv}

In this Section, we describe the construction and convergence properties of our numerical model.   In brief, we evolve a three-dimensional ideal MHD simulation in order to study the local and global dynamics of a geometrically-thin (constant opening angle $h/r=0.1$) accretion disk.  Given our focus on the time-variability of the accretion flow, the crux of our problem is the description of the MHD turbulent dynamics and how it influences the long-timescale behavior of the disk.  Thus, we concentrate our computational resources into the high-spatial resolution and the long duration nature of the simulation. We demonstrate below that the numerical grid is of high enough resolution that the macroscopic behavior of the disk turbulence is converged and insensitive to increases in resolution.   As a penalty, we simplify the physics to the bare minimum. We employ non-relativistic ideal MHD, and a pseudo-Newtonian gravitational potential to emulate the dynamical effects of a general relativistic potential, including the presence of an inner-most circular orbit (ISCO).  We use a simple cooling function to keep the disk thin, but otherwise neglect all radiation physics.  Renderings of density and magnetic turbulent structure of our fiducial disk simulation are shown in Figure \ref{fig-diskimg}.

\subsection{Simulation Setup}

Our simulation employs the finite-difference MHD code \emph{Zeus-MP} \emph{v2} \citep{1992ApJS...80..753S, 1992ApJS...80..791S, 2006ApJS..165..188H}.  \emph{Zeus-MP} solves the differential equations of ideal compressible MHD, 
\begin{equation}
\frac{D\rho}{Dt}=-\rho{\bf\nabla\cdot{v}},
\end{equation}
\begin{equation}
\rho\frac{D{\bf{v}}}{Dt}=-{\bf\nabla}P+\frac{1}{4\pi}({\bf\nabla\times{B}}){\bf\times{B}}-\rho{\bf\nabla}\Phi,
\end{equation}
\begin{equation}
\rho\frac{D}{Dt}\left(\frac{e}{\rho}\right)=-{P}{\bf\nabla\cdot{v}}-\Lambda,
\end{equation}
\begin{equation}
\frac{\partial{\bf{B}}}{\partial{t}}={\bf\nabla\times}({\bf{v}\times{B}}),
\end{equation}
where 
\begin{equation}
\frac{D}{Dt}\equiv\frac{\partial}{\partial{t}}+{\bf{v\cdot\nabla}}.
\end{equation}
to second order accuracy in space.  The method of constrained transport is used to maintain zero divergence in the magnetic field to machine precision.  The explicit integration time step is set by the usual Courant conditions, and is first-order accurate in time.  

\begin{figure}[t]
  \subfigure[Gas Density]{\includegraphics[width=0.5\textwidth]{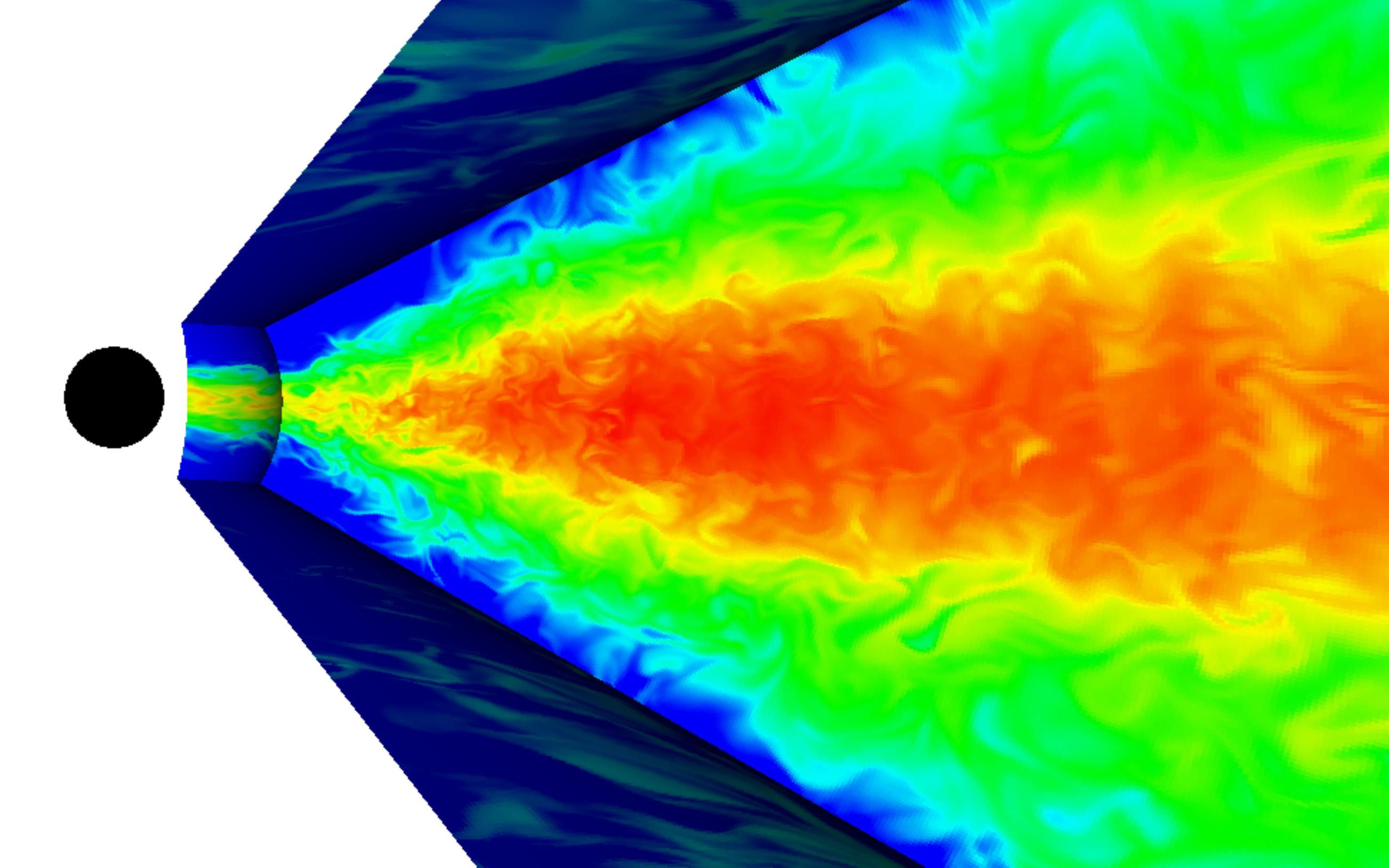}}
   \subfigure[$\bf{|B^2|}$]{\includegraphics[width=0.5\textwidth]{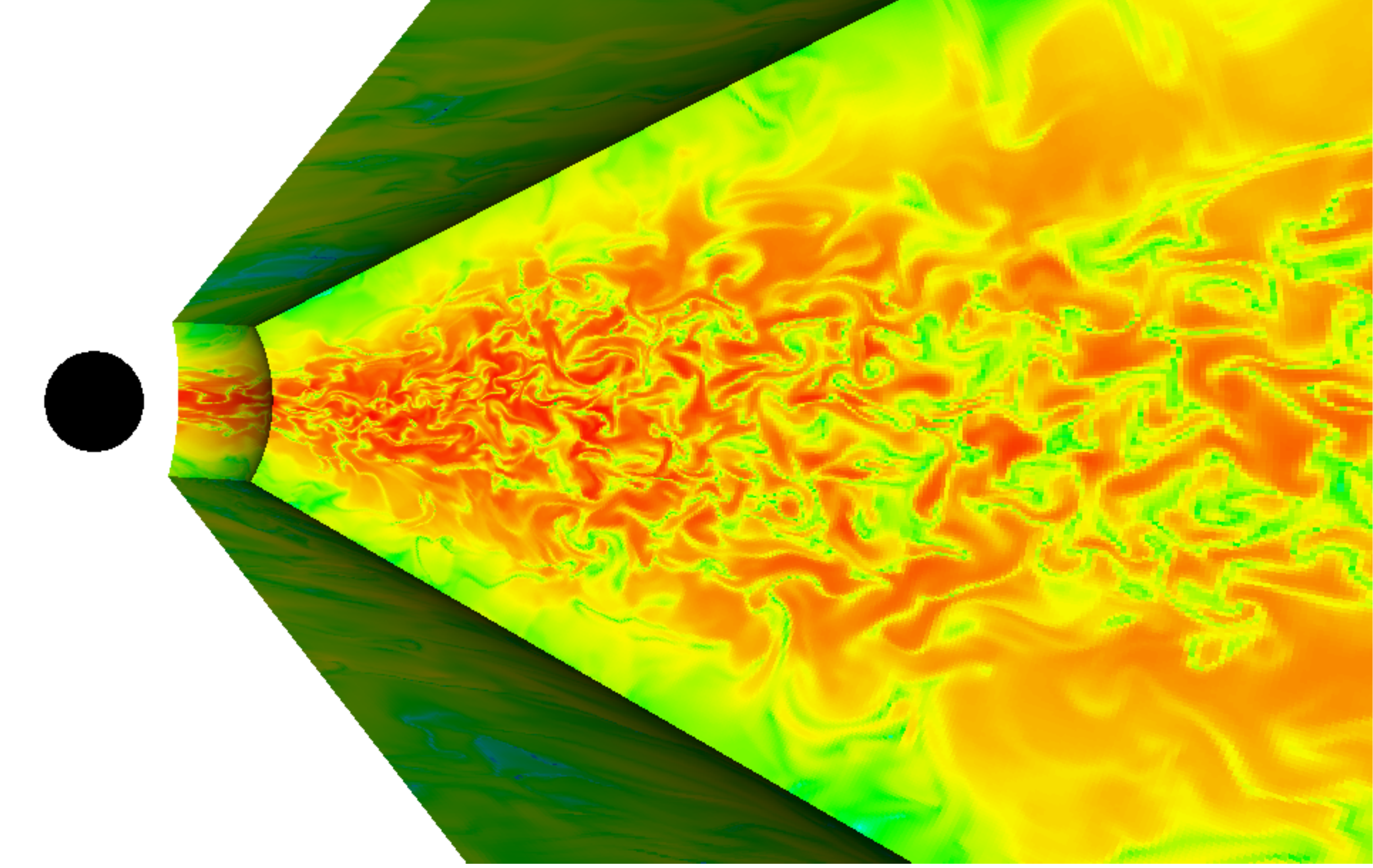}}
\caption{Snapshots of $\rho$ (a) and $\bf{|B^2|}$ (b) in the fiducial simulation used in our analysis at t=0, the point we take to be the beginning of science analysis after the initial transient behavior has died off and the turbulence had saturated.
\label{fig-diskimg}}
\end{figure}

We begin by defining our fiducial simulation.  The simulation domain is covered by spherical coordinates ($R,\theta,\phi$) and spans $R \in [4 {\rm r_g},145 {\rm r_g}], \theta \in [ \pi/2 -0.5,\pi/2+0.5], \phi \in [0,\pi/3)$, shown in Figure \ref{fig-diskimg}.  The zone aspect ratio is approximately $\Delta R: R\Delta \theta: R\Delta \phi = 2:1:2$ and each scale height is resolved with $28\: \theta$-zones.  Outflowing boundary conditions were used for the $\pm R$ and $\pm \theta$ boundaries and periodic boundary conditions were used in the $\phi$ direction.  We modified the standard ZEUS-MP boundaries to conserve $\nabla \cdot \bf{B} = 0$ across the $R$- and $\theta$-boundaries. 

The simulation domain was broken into a well-resolved inner region ($r < 45 r_{g}$), used for our analysis, and an outer region with poorer resolution to act as a gas reservoir.  The gas reservoir mitigates the effect of secular decay from the draining of material from the disk over the course of the simulation.  The radial spacing in the inner region logarithmically increases in R from 4-45 r$_{g}$.  In all, the simulation has $N_{R} \times N_{\theta} \times N_{\phi} = 512 \times 288 \times 128 = 1.89 \times 10^{7}$ zones in this region.  The $\Delta R$ spacing in the outer region logarithmically increases from $45-145 r_{g}$ and a total of $N_{R} \times N_{\theta} \times N_{\phi} = 128 \times 288 \times 128 = 4.72 \times 10^{6}$ zones in this region.

We approximate the gravitational field around a nonrotating black hole with a psuedo-Newtonian gravitational potential \citep{1980A&A....88...23P} of the form:
\begin{equation}
\label{eq:pnpot}
\Phi=-\frac{GM}{R-2r_{\rm g}},\qquad r_{\rm g}\equiv\frac{GM}{c^2}.
\end{equation}
With this potential, several important aspects of a general relativistic gravity field are captured including an ISCO (at $6r_{g}$) and the qualitative change of radial shear in the disk.    

A $\gamma = 5/3$ adiabatic equation of state is used for the gas.  An initially axisymmetric thin disk was initialized with constant midplane density and radially decreasing pressure corresponding to an isothermal vertical profile:
\begin{equation}
\rho(R,\theta)=\rho_{\rm 0}\exp\left(-\frac{\cos^2\theta}{2(h/r)^2\sin^2\theta}\right),
\end{equation}
and 
\begin{equation}
p(R,\theta)=\frac{GMR(h/r)^2\sin^2\theta}{(R-2r_{\rm g})^2}~\rho(R,\theta)
\end{equation}

The disk is initialized to a scale height of $h/r = 0.1$ and was maintained by an \emph{ad hoc}, optically thin cooling function similar to \citet{2009ApJ...692..411N}.  The cooling function plays the role of the real physical processes and removes energy at a rate equivalent to the thermal time of the disk ($\tau_{\rm cool}=10 \tau_{\rm orb}\sim \tau_{\rm orb}/\alpha$), as was done in \citet{2011ApJ...736..107O}.  If the gas energy in a cell is above the target energy, $e_{\rm targ} \propto \rho v_{\phi}^2 (h/r)^2$, it is cooled over the cooling time.  The cooling function used is $\Lambda = f(e-e_{\rm targ})/ \tau_{\rm cool}$, where $f = 0.5 [(e-e_{\rm targ})/|e-e_{\rm targ}|+1]$.  The switch function, $f$, acts as a threshold function and is zero when $e_{\rm targ} > e$.  A protection routine is implemented to ensure the density and pressure values do not become artificially small and/or negative.

The gas was initialized with a purely azimuthal velocity field and a weak magnetic field.  The local velocity is set so that the effective centripetal force balances the gravitational force.  A series of weak, large-scale poloidal magnetic field loops are set for the initial magnetic field configuration (e.g. \citealt{2009ApJ...692..869R}) to perturb the disk and allow for the growth of the MRI.  The loops were set from the vector potential ${\bf A}=(A_{r}, A_{\theta}, A_{\phi})$ to ensure the initial field was divergence free and had the form: \begin{equation} \label{eqn-potential}
A_{\phi}=A_{0}p^{\frac{1}{2}}f(r,\theta)\sin \bigg( \frac{2\pi r}{5h}\bigg), A_{r}=A_{z}=0,
\end{equation} where $A_{0}$ is a normalization constant, $f(r,\theta)$ is an envelope function that is 1 in the body of the disk and smoothly goes to 0 at $r=r_{ISCO}, r_{out}$ and 3 scale heights above the disk.  The alternating field polarity is set by the final multiplicative term such that loops have a radial wavelength of $5h$.  The $A_{0}$ constant was set to initialize the disk to an average ratio of gas- to magnetic-pressure of $\beta = 500$.

The following analysis treats the simulation after the MRI has saturated and transient behavior from initialization has died away.  $\bf{B}$ and $e$ reach a quasi-steady state in the well-resolved region of the disk and the MRI saturates, as measured by the convergence metrics discussed in Section \ref{sec-conv}, after approximately 200 ISCO orbits ($12,320 \:GM/c^{3}$).  We take this point to be $t=0$.  The simulation was then run for 1410 ISCO orbits ($86,856 \:GM/c^{3}$).  In the interest of data management, the values of $\bf{B}$, $\bf{v}$, $e$, and $\rho$ were output every two ISCO orbits ($123.2 \:GM/c^{3}$).

\subsection{Convergence}
\label{sec-conv}

The issue of convergence is forefront when considering the reliability of results from any \emph{ab initio} accretion disk simulation because the MHD turbulence that mediates angular momentum transport has structure on the smallest accessible scales.  With this type of simulation, the macroscopic behavior can have a resolution dependence if the turbulent scales dominating the shear stresses are not adequately resolved \citep{2009ApJ...694.1010G, 2012ApJ...749..189S, 2013ApJ...772..102H}.  As a preface to our study of variability, here we assess the degree to which we are adequately resolving the turbulent dynamics using two measures.   Firstly, we examine the resolvability of the fastest growing linear MRI mode through the commonly employed quality factors introduced by \citet{2010ApJ...711..959N}.  Secondly, to characterize the non-linear saturation of the turbulence, we use the magnetic tilt angle.  Using these diagnostics, we compare our fiducial (long) simulation with higher resolution comparison simulations. 

The first set of diagnostics used to measure resolvability were the vertical and azimuthal quality factors \citep{2010ApJ...711..959N}, $Q_{\theta} = \lambda_{MRI} / R\Delta \theta$ and $Q_{\phi} = \lambda_{C} / R\Delta \phi$, respectively.  $Q_{\theta}$ and $Q_{\phi}$ measure resolvability through the number of grid cells per fastest-growing MRI wavelength, $\lambda_{MRI}$, and critical toroidal field length, $\lambda_{C}$ where
\begin{equation}
\lambda_{MRI}=\frac{2 \pi | B_{\theta} |}{\sqrt{\rho} \: \Omega(r)} \quad,  \quad \lambda_{C}=\frac{2 \pi | B_{\phi} |}{\sqrt{\rho} \: \Omega(r)}.
\end{equation}  

Using the quality factors, we characterized the global resolvability of the disk.  The local values of $Q_{\theta}$ and $Q_{\phi}$ were calculated in every voxel of the well-resolved region ($4-45 \: r_{g}$) for every data dump after the transient behavior in the disk had died away ($t=0$ onwards).  The volume-weighted average was calculated for each quality factor by averaging over in the region within one scale height above and below the disk mid-plane to get the instantaneous value and then over the entire duration of the simulation to get a temporal average.  The averaging was restricted to the body of the disk because we want the most conservative estimate of the resolvability of the simulation.  In the highly magnetized, low-density coronal regions the quality factors are very high and, if included, the quality factors would appear artificially high, overestimating how well we resolve the MRI.  We find $\langle Q_{\theta} \rangle = 7.0$ and $\langle Q_{\phi} \rangle = 15.2$, a bit below the nominal resolvability criteria of $\langle Q_{\theta} \rangle = 10$ \citep{2011ApJ...738...84H}.

\begin{figure}[b]
\subfigure[$Q_{\theta}$]{\includegraphics[width=0.5\textwidth]{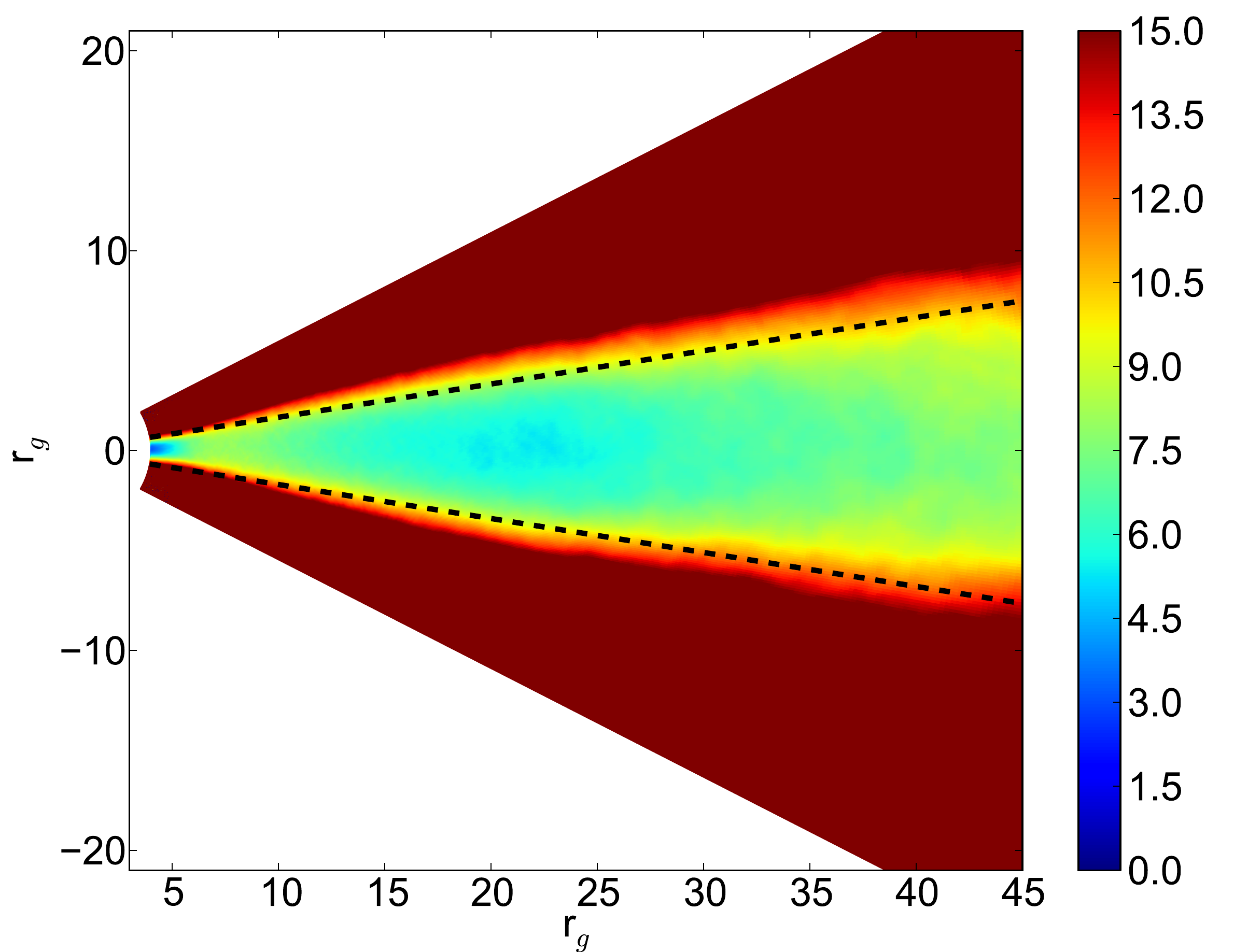}}
\subfigure[$Q_{\phi}$]{\includegraphics[width=0.5\textwidth]{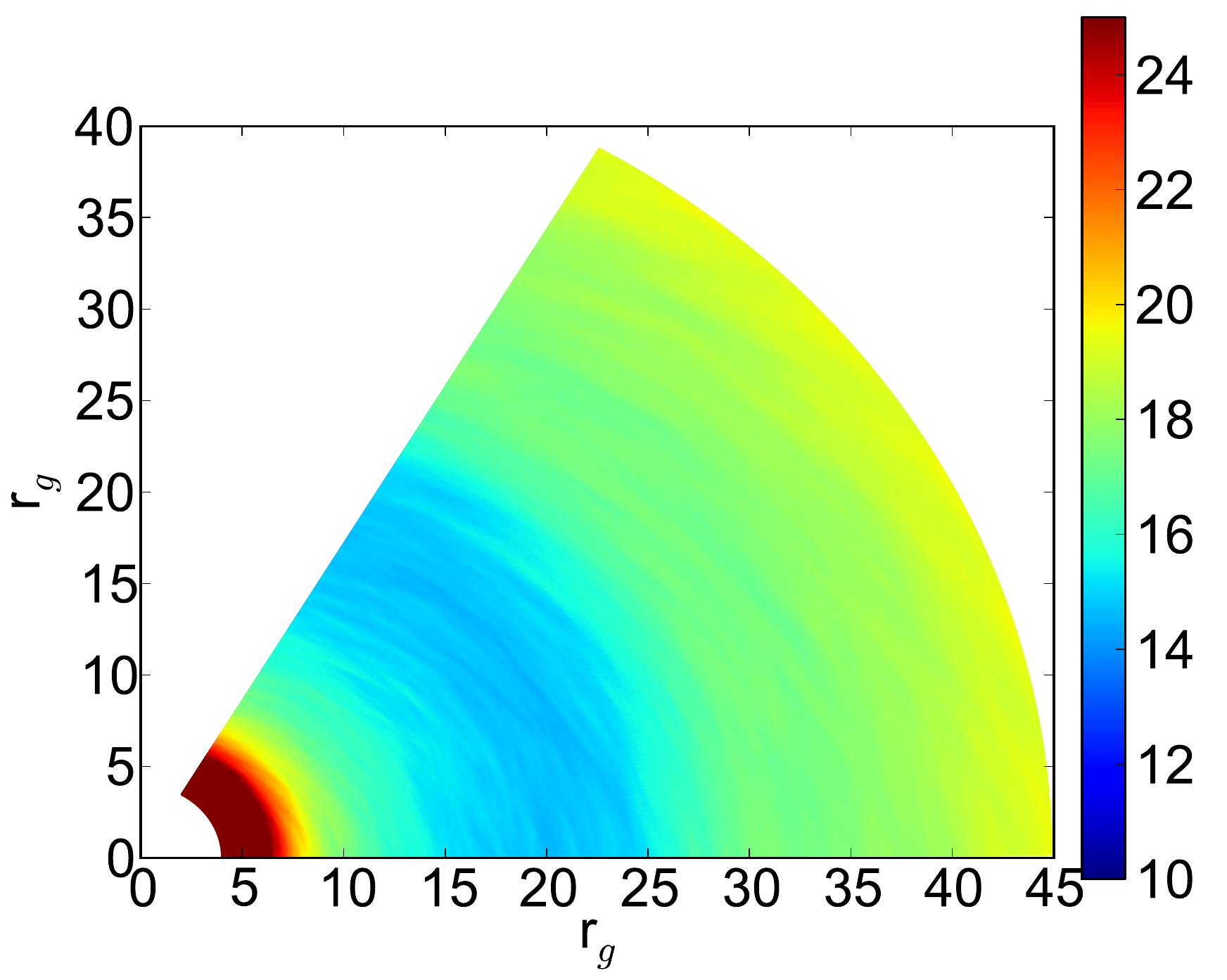}}
\caption{Plots of temporally and azimuthally averaged $Q_{\theta}$ (a) and temporally and vertically averaged $Q_{\phi}$ (b).  One scale height (1h) above and below the disk is shown as dashed black lines in (a).
\label{fig-QzQp}}
\end{figure}

In addition to the global averages, we are interested in how the resolvability depends on the position within the grid.  Shown in Figure \ref{fig-QzQp} is the time and azimuthally averaged profile of $Q_{\theta}$, as well as the time and polar average profile of $Q_{\phi}$.  The average of $Q_{\theta}$ was calculated as a function of $r$ and $\theta$ by averaging over entire $\phi$ domain.  As we did in our global average, the vertical averaging over the polar angle in our calculation of $Q_{\phi}$ is only done in the region within one scale height above and below the disk to exclude the coronal region.  The spatial variation can be seen in both quantities, but no apparent ``dead zones" of gross under-resolution are present.

\begin{deluxetable*}{cccccccccccc}
\tablewidth{\textwidth}
\tablecaption{Simulation Parameters
\label{tab-ellips_fit}}
\tablehead{& \colhead{Azimuthal} \vspace{-0.05cm} & && && \\
\colhead{Simulation} & & \colhead{ISCO Orbits} & \colhead{N$_{R}$} & \colhead{N$_{\theta}$} & \colhead{N$_{\phi}$} & \colhead{$H /\Delta z$}& \colhead{$\langle Q_{z} \rangle$}& \colhead{$\langle Q_{\theta} \rangle$} & \colhead{$\langle \alpha \rangle$}& \colhead{$\langle \beta \rangle$}&\colhead{$\Theta_{B}$}\\
&\colhead{Range}& && && }
\startdata
Fiducial & $\pi / 3$ & $1610$ & $640$ & $288$ & $128$ & $28.8$ & $8.1$ & $20.5$ & $0.12$ & $23.4$ &$10.5^\circ$\\
HiResWedge & $\pi / 3$ & $204$ & $1280$ & $576$ & $256$ & $57.6$ & $27.2$ & $43.0$ & $0.15$ & $6.5$ &$12.1^\circ$ \\
HiRes2$\pi$ & $2\pi$ & $110$ & $1280$ & $576$ & $256$ & $57.6$ & $18.4$ & $11.8$ & $0.09$ & $7.9$ & $11.6^\circ$\\
\enddata
\end{deluxetable*}

These quality factors probe the ability of the simulation to resolve linear MRI modes but do not address the structure of the (non-linear) saturated turbulence.   Thus, we additionally measure the average in-plane magnetic tilt angle,
\begin{equation}
\Theta_B=-\textrm{arctan}\left(\Big\langle\frac{B_r}{B_\phi}\Big\rangle\right),
\end{equation}
the value of which is closely connected to the processes by which the MHD turbulence saturates \citep{2010ApJ...716.1012P,2011ApJ...738...84H, 2012ApJ...749..189S, 2013ApJ...772..102H}.  We note that, given the dominance of $B_\phi$ over the other field components, the tilt angle is approximately given by 
 \begin{equation}
\Theta_{B}=\textrm{arcsin}(\alpha_{M} \beta)/2,
 \end{equation}
where $\alpha_{M}$ is the Maxwell stress normalized by the magnetic pressure,
\begin{equation}
\alpha_{M} = \frac{\langle -2 B_{R}B_{\phi} \rangle}{\langle P \rangle},
\end{equation}
and $\beta$ is the ratio of gas pressure to magnetic pressure,
\begin{equation}
\beta = \frac{P}{\langle |B^2| \rangle}.
\end{equation}
Thus, the magnetic tilt angle is directly tied to the effectiveness of angular momentum transport at a given field strength and can be used to probe whether a given simulation has correctly captured the non-linear saturation of the MHD turbulence.

\begin{figure}[t]
\includegraphics[width=0.5\textwidth]{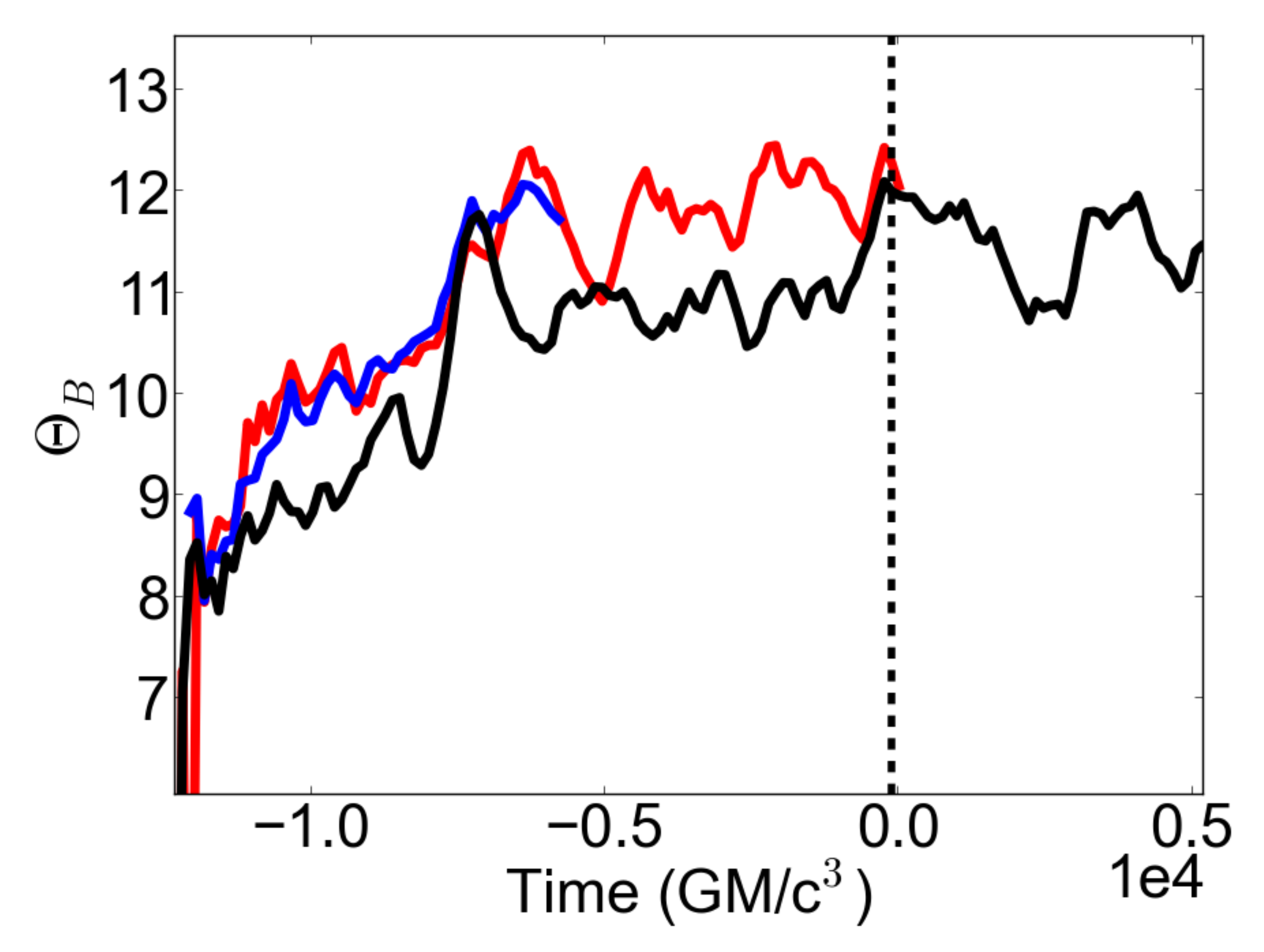}
\caption{Evolution of magnetic tilt angle, $\Theta_{B}$, for long, fiducial disk simulation (black line), high resolution wedge simulation (red line), and high resolution $2\pi$ simulation (blue line).  The t=0 time we set use in our analysis is marked by the black dashed line.  The evolution of the long disk simulation used for the science run in our paper has been truncated to only show the initial saturation of the turbulence and a short time period short after, but continues for $86,856 \:GM/c^{3}$.  $\Theta_{B}$ grows similarly in all three simulations and saturates at the same value, $\Theta_{B}  \approx 11-12^{\circ}$.  The black dashed line indicates t=0 used in our analysis.
\label{fig-mag_tilt}}
\end{figure}

We calculate $\Theta_{B}$ within the mid-plane regions of the disk, restricting to half a scale height above and below the disk midplane in the well-resolved region.  The temporally averaged magnetic tilt (neglecting the first 200 ISCO orbits) is $\overline{\Theta_{B}}\approx 11.3^\circ$ (Fig.~\ref{fig-mag_tilt}), comparable to estimates from both analytic theory  \citep{2010ApJ...716.1012P} and previous high-resolution local \citep{2011ApJ...738...84H} and global \citep{2012ApJ...749..189S, 2013ApJ...772..102H} simulations.  

We further assess the fidelity of our fiducial simulation by performing two additional high-resolution comparison runs.  The first comparison simulation had the exact same geometry as the fiducial simulation, but twice the resolution in each of the three dimensions (hence voxels of $1/8^{th}$ the volume).  The second comparison simulation also doubles the number of zones in each of the three dimensions, but then extends the domain to include the full 2$\pi$ in azimuth.   Both of these comparison simulations possess $N_{R} \times N_{\theta} \times N_{\phi} = 1280 \times 576 \times 256 = 1.89 \times 10^{8}$ total zones with a vertical resolution of $56 \: \theta$-zones per disk scale-height.  Of course, the $\Delta R : R\Delta \phi$ ratio was larger in the $2\pi$ higher resolution simulation due to the extension of the $\phi$ domain.  The evolution of $\Theta_{B}$ for the simulation used in our analysis and the two test simulations is shown in Figure \ref{fig-mag_tilt}.  

In the higher resolution simulations the evolution of $\Theta_{B}$ is slightly faster than our fiducial run, but settles at similar levels as the MRI-driven turbulence saturates.  One difference between the development of the turbulence in the comparison simulations and our fiducial run is that our fiducial run has a transient spike in magnetic tilt angle $5,790\:GM/c^{3}$ (94 ISCO orbits) into the initialization.  From visual inspection, we see the growth and break-up of a large channel flow in the disk body in the fiducial run.  This does not develop in the higher resolution comparison simulations because the non-axisymmetric parasitic instabilities are more effective at breaking up nascent channel flows.  With this minor difference, the similarity in the evolution and saturation of the turbulence between the three simulations indicates that our fiducial run is adequately resolved.

In all, we can be confident in the convergence of our model.  By standard measures the simulation is adequately resolved and has reached a level of saturated turbulence.  The quality factors slightly are lower than what has been deemed well-resolved by prior convergence studies \citep[i.e.][]{2011ApJ...738...84H, 2012ApJ...749..189S, 2013ApJ...772..102H}, but the similarities between $\Theta_{B}$ in our long science run and those of our two shorter, high-resolution comparison simulations demonstrate the turbulence has reached a saturated level.

\section{Turbulence and the Effective $\alpha$ Description of the Model Disk}
\label{sec-overview}

\subsection{Fluctuating Effective $\alpha$}
\label{sec-measuring_alpha}

We will now look at the behavior of the effective $\alpha$ in the disk.  Of particular interest is the variable behavior of the effective $\alpha$ as this is required to drive fluctuations in $\dot{M}$.  Using Equation \ref{eqn-alpha}, we calculate the effective $\alpha$ by restricting the domain over which we average to the disk midplane, $\theta = [\pi / 2 \pm h]$, where h is one disk scale height.  For each data output from our simulation ($\Delta t = 123.2 GM/c^3$) we calculate the average stress and pressure within one scale height across the entire azimuthal domain for each radial bin.  The ratio of the stress to pressure was then taken to provide the effective $\alpha$ as a function of radius and time.  

\begin{figure}[t]
\includegraphics[width=0.5\textwidth]{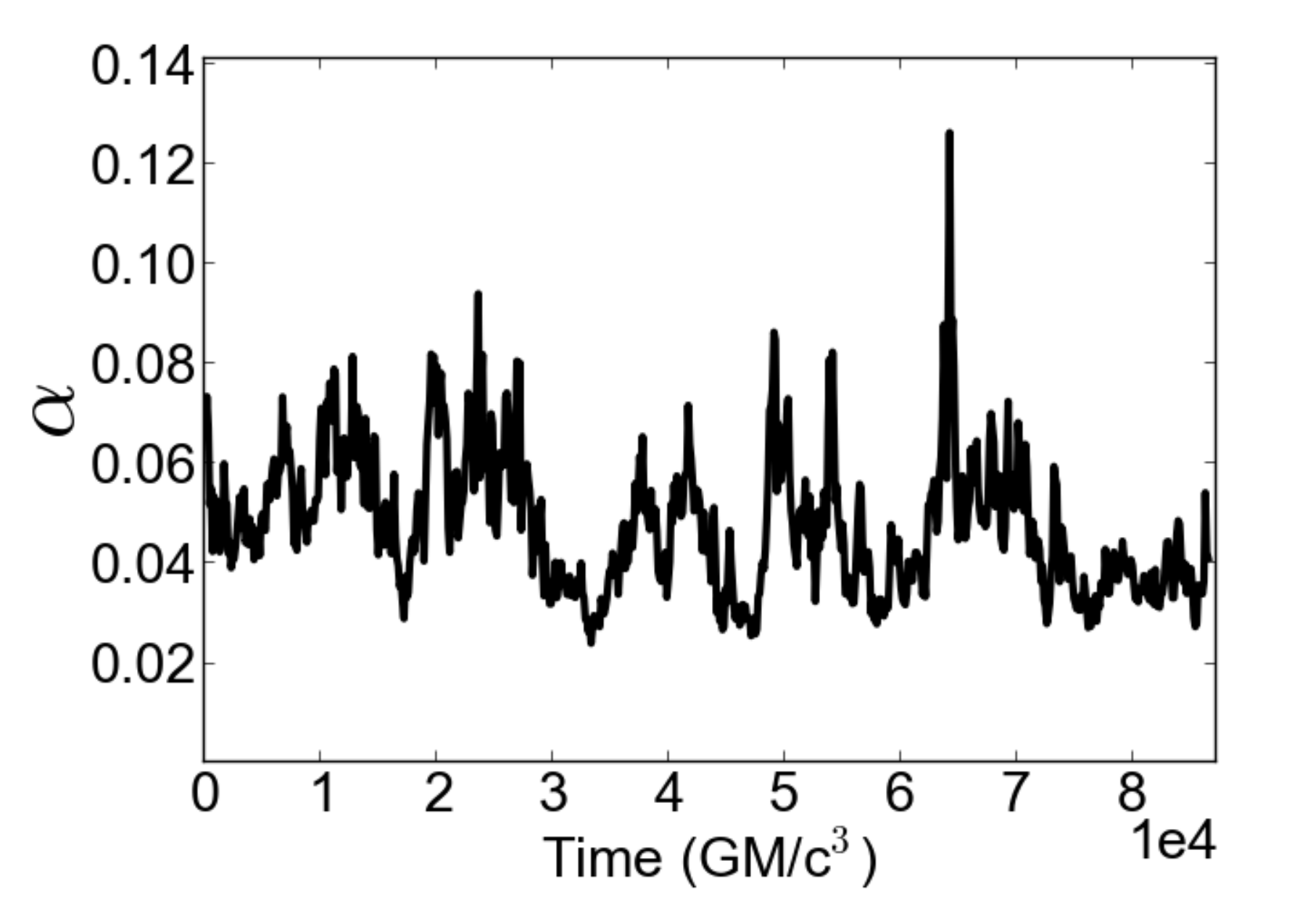}
\caption{ Time trace of the effective $\alpha$ parameter in the disk after averaging over radius.  The $\alpha$ parameter is highly variable and fluctuates between $\alpha \approx 0.04$ and $\alpha \approx 0.08$ in a quasi-periodic manner.
\label{fig-alpha_var}}
\end{figure}

Figure \ref{fig-alpha_var} shows the time variability of the effective $\alpha$ parameter in the disk after averaging over radius.  Large variability occurs on an intermediate timescale, longer than a dynamical time and shorter than a viscous time.  The value of the effective $\alpha$ in the disk typically fluctuates between $\alpha \approx 0.03$ and $\alpha \approx 0.08$ in a quasi-periodic manner.  On the lower end of the fluctuating range, $\alpha \approx 0.03$ seemingly acts as a floor where the minima consistently returns.  The upper range on the effective $\alpha$ is less well-defined with maxima found at different levels as low as $\alpha=0.07$ and as high as $\alpha=0.13$.  

\begin{figure}[t]
\includegraphics[width=0.5\textwidth]{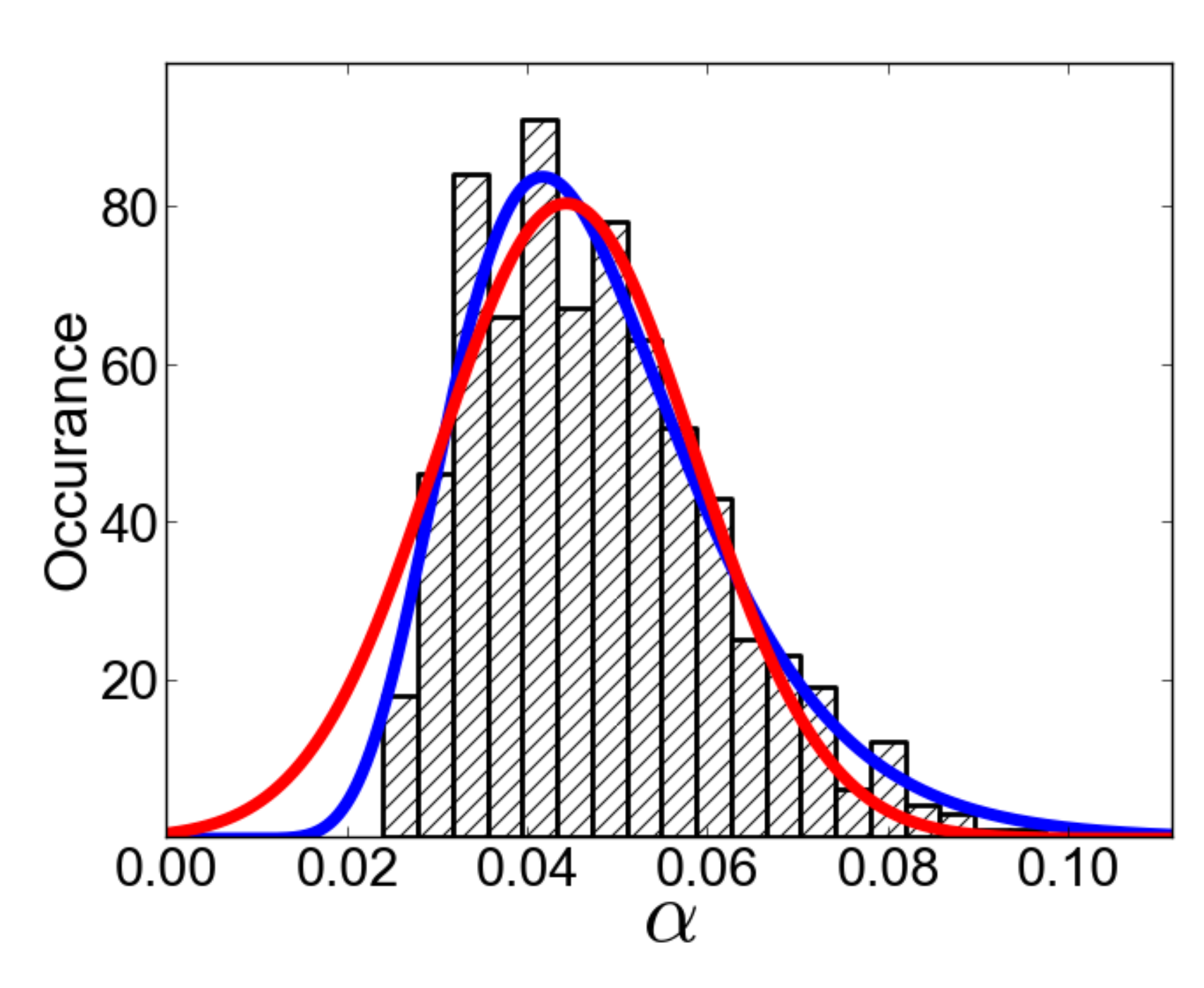}
\caption{Histogram of $\alpha$ fit with a normal distribution (red line) and log-normal distribution (blue-line).  The normal distribution provides a better fit to the distribution.  The distribution is best fit by a log-normal distribution.
\label{fig-alpha_hist}}
\end{figure}  

Figure \ref{fig-alpha_hist} shows the histogram of effective $\alpha$ with normal (Gaussian) and log-normal fits to the probability density function (PDF).  Statistically, the PDF of the effective $\alpha$ is better fit by a log-normal distribution.  The best fit with a normal distribution has $\chi^{2} / D.o.F = 44.7 / 24 = 1.9$ and the best fit with a log-normal distribution has $\chi^{2} / D.o.F = 29.6 / 24 = 1.2$.

\subsection{Characterizing Velocity and Length Scales of the Turbulence}
\label{sec-ss_disk}

Here, we have a brief digression in order to characterize the velocity and length scales of the turbulence.  Given the range of time- and length-scales involved in a global disk simulation, we choose to perform this exercise over a limited range of radii; we select the region $R \in [15 {\rm r_g},18 {\rm r_g}], \theta \in [ \pi/2 -0.35,\pi/2+0.35], \phi \in [0,\pi/3)$.  In physical coordinates, this domain spans $\Delta R = 3 r_{g}, R\Delta\theta = 10.5 r_{g}, R\Delta\phi = 5 \pi r_{g}$.  The size of the subdomain was chosen such that any localized turbulent structure would be well contained within the box, but also so that any medium-scale structure would be captured by our analysis if it is present. 

\subsubsection{Measuring Turbulent Velocity}
\label{sec-turb_vel}

Within this region, we first calculated the ratio of the turbulent speed to the sound speed within the disk midplane as a function of time.  The turbulent velocity is defined as 
\begin{equation} 
|v|=\sqrt{v_{r}^2+v_{\theta}^2+(v_{\phi}-\langle v_{\phi}\rangle)^2} 
\end{equation} and the sound is defined in its usual form 
\begin{equation} 
c_{s}=\sqrt{\frac{\gamma P}{\rho}}.
\end{equation}  
For each data dump, we calculated the volume weighted spatial average of $|v| / c_{s}$ to serve as a diagnostic for the average velocity of the turbulence within the disk.

\begin{figure}[!t]
  \subfigure[t][Gas Velocity]{\includegraphics[width=0.5\textwidth]{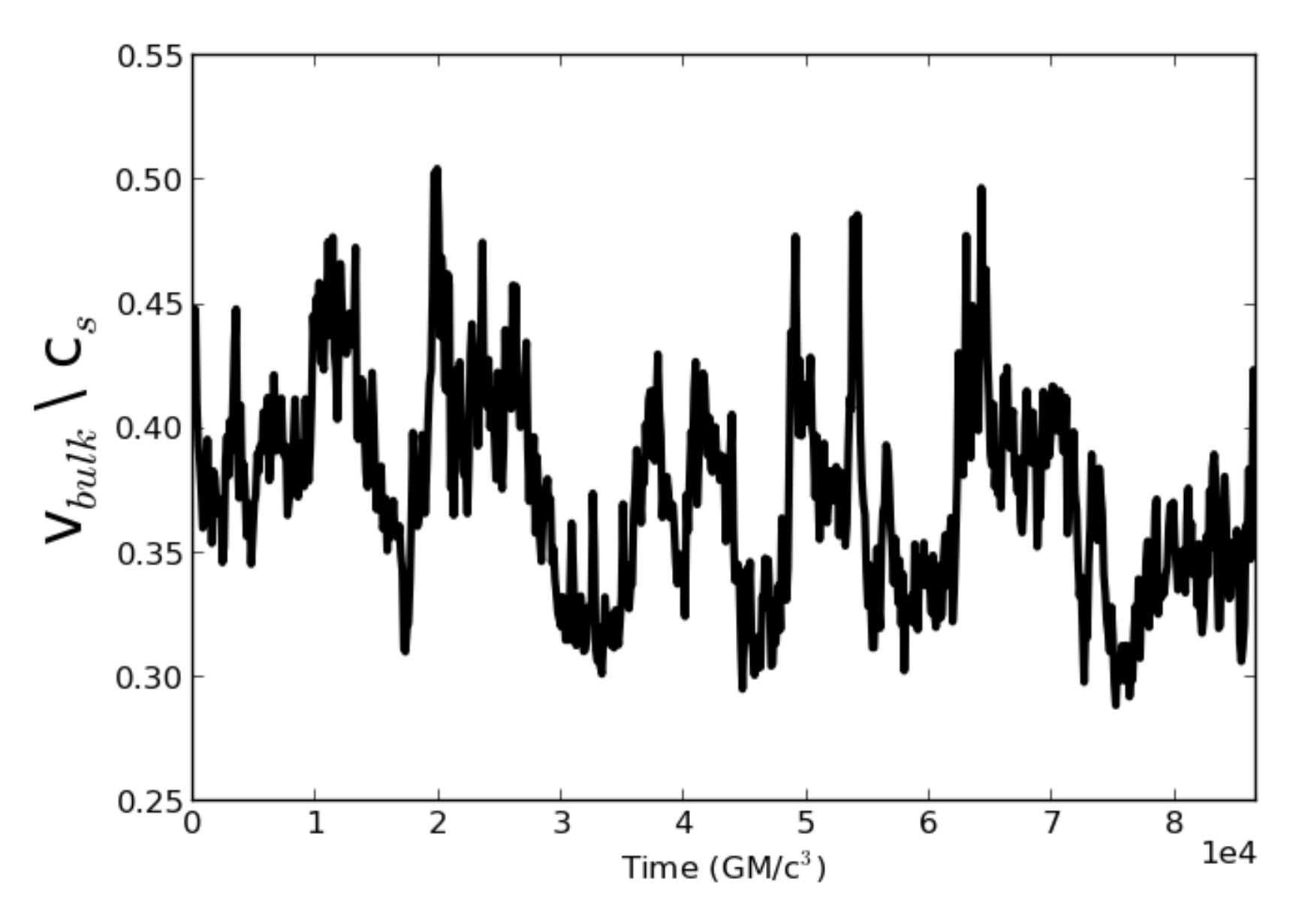}}
  \subfigure[t][Radial Correlation Length]{\includegraphics[width=0.5\textwidth]{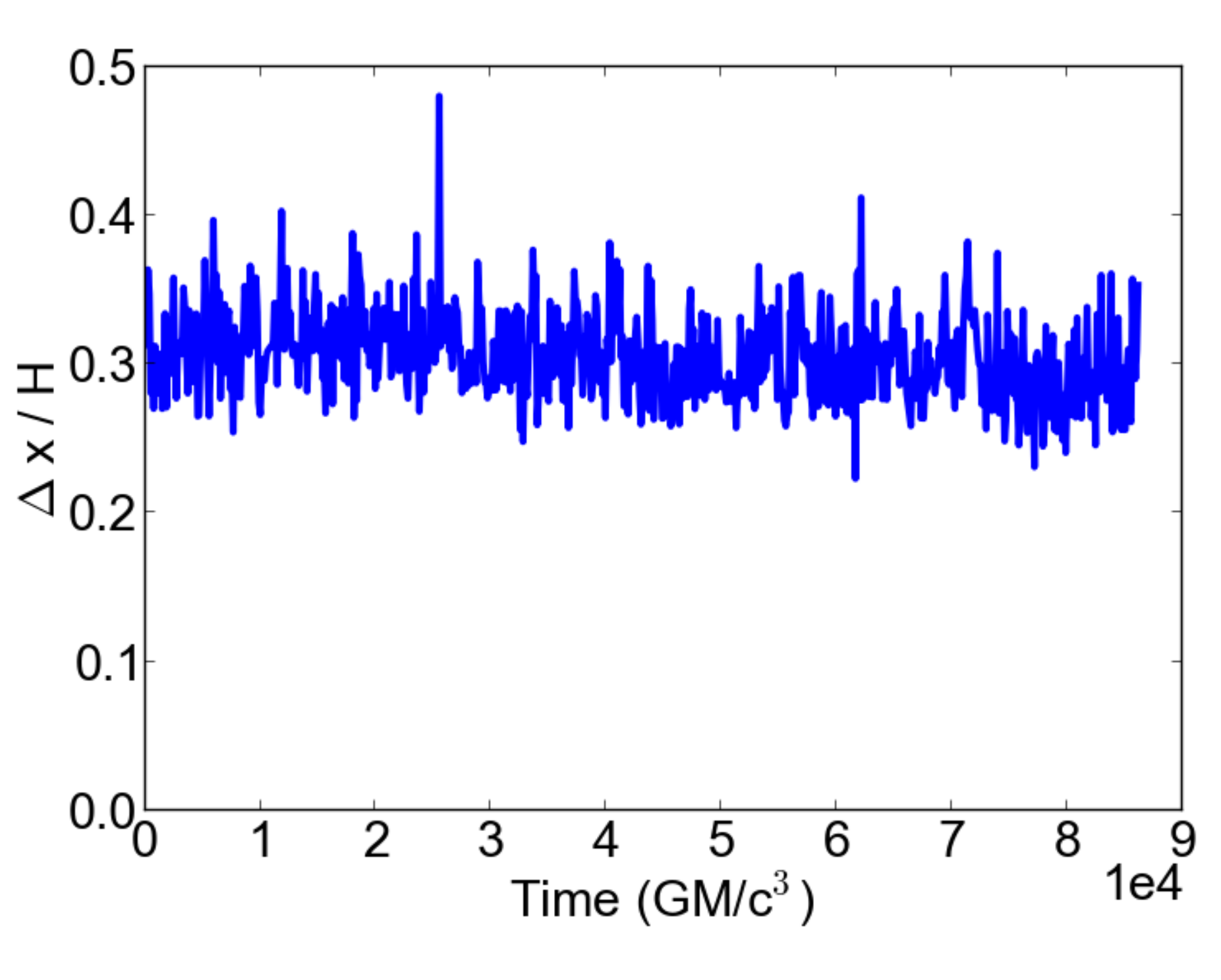}}
\caption{Instantaneous values of $v_{turb} / c_{s}$.  The fluctuations in $v_{turb} / c_{s}$ are correlated with the changes in $\alpha$ seen in Figure \ref{fig-alpha_var} indicating that as the stress increases in the disk the gas is accelerated.  Instantaneous values of $\Delta x / h$.  $\Delta x / h$ fluctuates stochastically around $l=0.32$ and is not correlated with $\alpha$.  The turbulent spatial scales are independent of the disk stress at all times.
\label{fig-turb_variability}}
\end{figure}

Figure \ref{fig-turb_variability} shows the time variability of $\langle |v| / c_{s} \rangle$.  This ratio is highly variable and fluctuates within a range of $|v| / c_{s} \approx 0.3$ to $|v| / c_{s} \approx 0.5$.  Comparing with the time variability of $\alpha$ in Figure \ref{fig-alpha_var}, the correlation between $\alpha$ and $|v| / c_{s}$ can be readily observed.  The average value is $\langle |v| / c_{s} \rangle = 0.37$.  

\subsubsection{Measuring Turbulent Scales}
\label{sec-turb_scale}

Measuring the spatial scales of the turbulence is significantly more involved than measuring the turbulent velocity, but, nevertheless, provides a method for characterizing the turbulence in simple terms.  The spatial scales of the density, magnetic field components and Maxwell stress of the turbulence were measured in our $r=15-18 r_{g}$ subdomain using the autocorrelation the method of \citet{2009ApJ...694.1010G} \& \citet{2011MNRAS.416..361B} with some modifications.  For each data dump after the initialization of the simulation, the two-point autocorrelation of a given quantity of interest, $F(r, \theta, \phi, t)$, was calculated within this subdomain in the following way.  First, large-scale structure of $F(r, \theta, \phi, t)$ was removed. For a given time step, t, the average over $\Delta T = 1232$ GM/c$^3$ (11 data dumps) for each cell in the subdomain was calculated.  The time average was centered on the timestep of interest such that the average for a single cell included its values 5 data dumps before and after.  This was then subtracted off of the present cell value:
\begin{equation}
\begin{split}
F_{sub}(r, \theta, \phi, t) = F(r, \theta, \phi, t)- \\
& \mkern-108mu \frac{1}{\Delta T}\int_{t-0.5\Delta T}^{t+0.5\Delta T}F_{norm}(r, \theta, \phi, t) dt
\end{split}
\end{equation} leaving only the high-frequency perturbation.  Since the average was determined for an individual cell, this also removes vertical gradients.

\begin{deluxetable*}{ccccc}
\tablewidth{\textwidth}
\tablecaption{Ellipsoid Fit Parameters
\label{tab-ellips_fit}}
\tablehead{&& \colhead{Fit Parameters}\vspace{-0.2cm}  & \colhead{Fit Parameters} & \colhead{Fit Parameters} \\
& \colhead{$\theta$} &&&\\
&&\colhead{$0.5$} \vspace{-0.cm} & \colhead{e$^{-1}$} & \colhead{e$^{-2}$}}
\startdata
&& a - 0.32 & a - 0.41 & a - 0.66 \\
$\rho$ & 16.7$^{\circ}$ & b - 1.17 & b - 1.67 & b - 3.49 \\
&& c - 0.24 & c - 0.42 & c - 1.08 \\ \hline
&& a - 0.29 & a - 0.38 & a - 0.60 \\
B$_R$ & 30.2$^{\circ}$ & b - 0.75 & b - 0.97 & b - 1.79 \\
&& c - 0.07 & c - 0.10 & c - 0.17 \\ \hline
&& a - 0.20 & a - 0.24 & a - 0.36 \\
B$_\theta$ & 14.8$^{\circ}$ & b - 0.56 & b - 0.74 & b - 1.20 \\
&& c - 0.14 & c - 0.21 & c - 0.38 \\ \hline
&& a - 0.27 & a - 0.35 & a - 0.58 \\
B$_\phi$ & 22.2$^{\circ}$ & b - 0.97 & b -1.31 & b - 2.52 \\
&& c - 0.10 & c - 0.14 & c - 0.24 \\ \hline
&& a - 0.20 & a - 0.28 & a - 0.47 \\
B$^2$ & 22.2$^{\circ}$ & b - 0.78 & b - 1.03 & b - 2.03 \\
&& c - 0.07 & c - 0.10 & c - 0.21 \\ \hline
&& a - 0.23 & a - 0.30 & a - 0.53 \\
B$_R$B$_{\phi}$ & 26.4$^{\circ}$ & b - 0.70 & b - 0.92 & b - 1.64 \\
&& c - 0.07 & c - 0.10 & c - 0.17 \\ \hline
\enddata
\end{deluxetable*}
Next, the physical coordinates were normalized by the local scale height to express the turbulent scales in terms of $h$.  This converts the wedge geometry into a rectangular geometry and allows for easier comparison with prior work.  In this normalized geometry, $r$ maps to $x$, $\phi$ maps to $y$, and $\theta$ maps to $z$.

After the normalization of the coordinates, the three-dimensional Fourier transform was then taken of $F_{sub}$:
\begin{equation}
\begin{split}
\mathcal{F}(k_x, k_z, k_{y}, t)= \\ 
 & \mkern-54mu \int \!\!\! \int \!\!\! \int F(x, z, y) e^{i (k_x x +k_{z} z +k_{y}y)} dx dz dy.  
\end{split}
\end{equation}From this, the autocorrelation, $C_{\mathcal{F}}(\Delta x, \Delta z, \Delta y)$, was calculated according to:
\begin{equation}
\begin{split}
C_{\mathcal{F}}(\Delta x, \Delta z, \Delta y) = \int \!\!\! \int \!\!\!\int |\mathcal{F}(k_x, k_z, k_{y}, t)|^2 \quad \times  \\
& \mkern-206mu e^{i (k_x x +k_{z} z +k_{y}y)} dk_{x} dk_{z} dk_{y}.
\end{split}
\end{equation}

\begin{figure}[!b]
  \centering
   \subfigure[$C_{\mathcal{F}}(\Delta x, \Delta z, \Delta y)$ in x-y plane]{\includegraphics[width=0.37\textwidth]{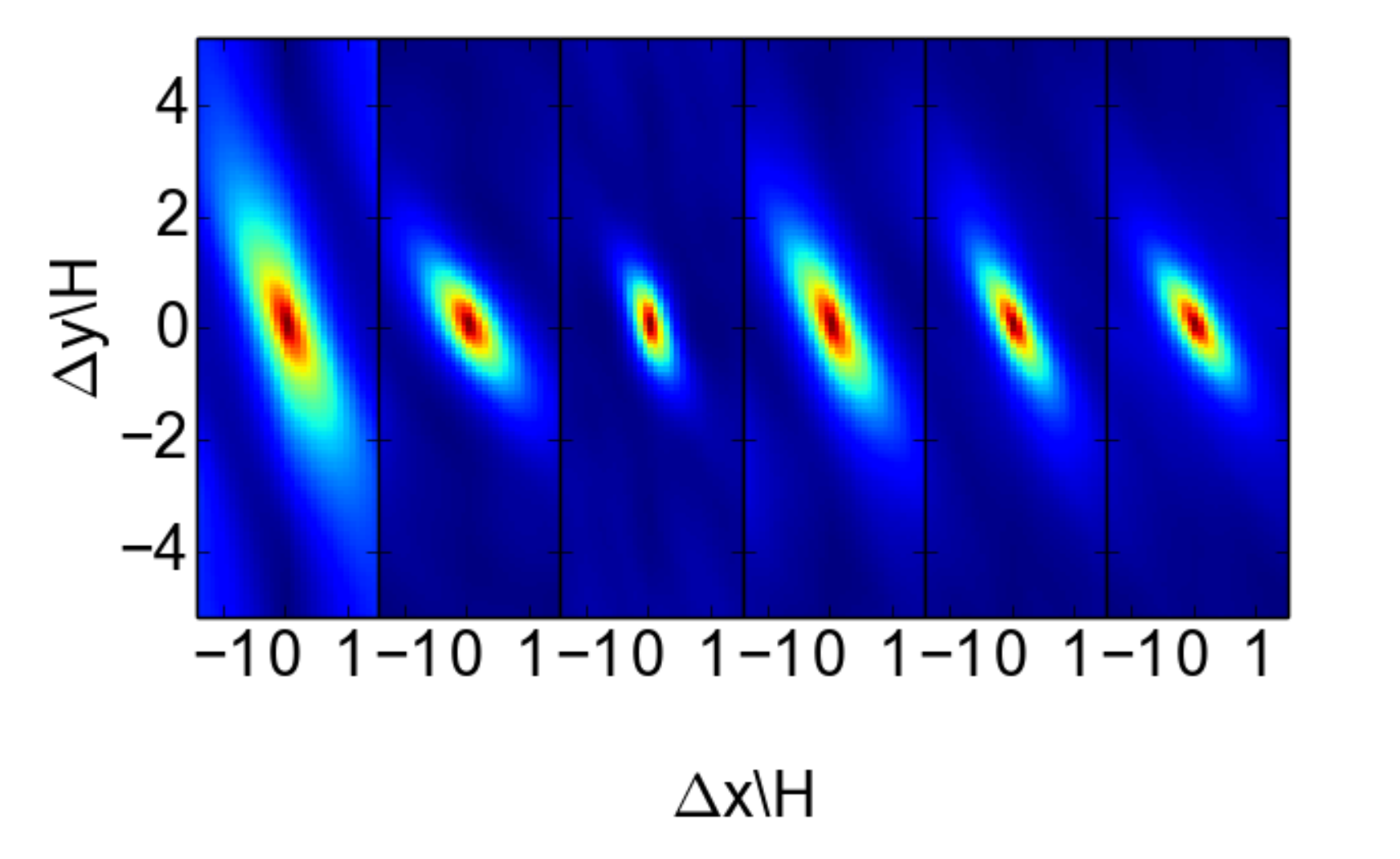}}
   \vspace{-0.25cm}
  \subfigure[$C_{\mathcal{F}}(\Delta x, \Delta z, \Delta y)$ in x-z plane]{\includegraphics[width=0.37\textwidth]{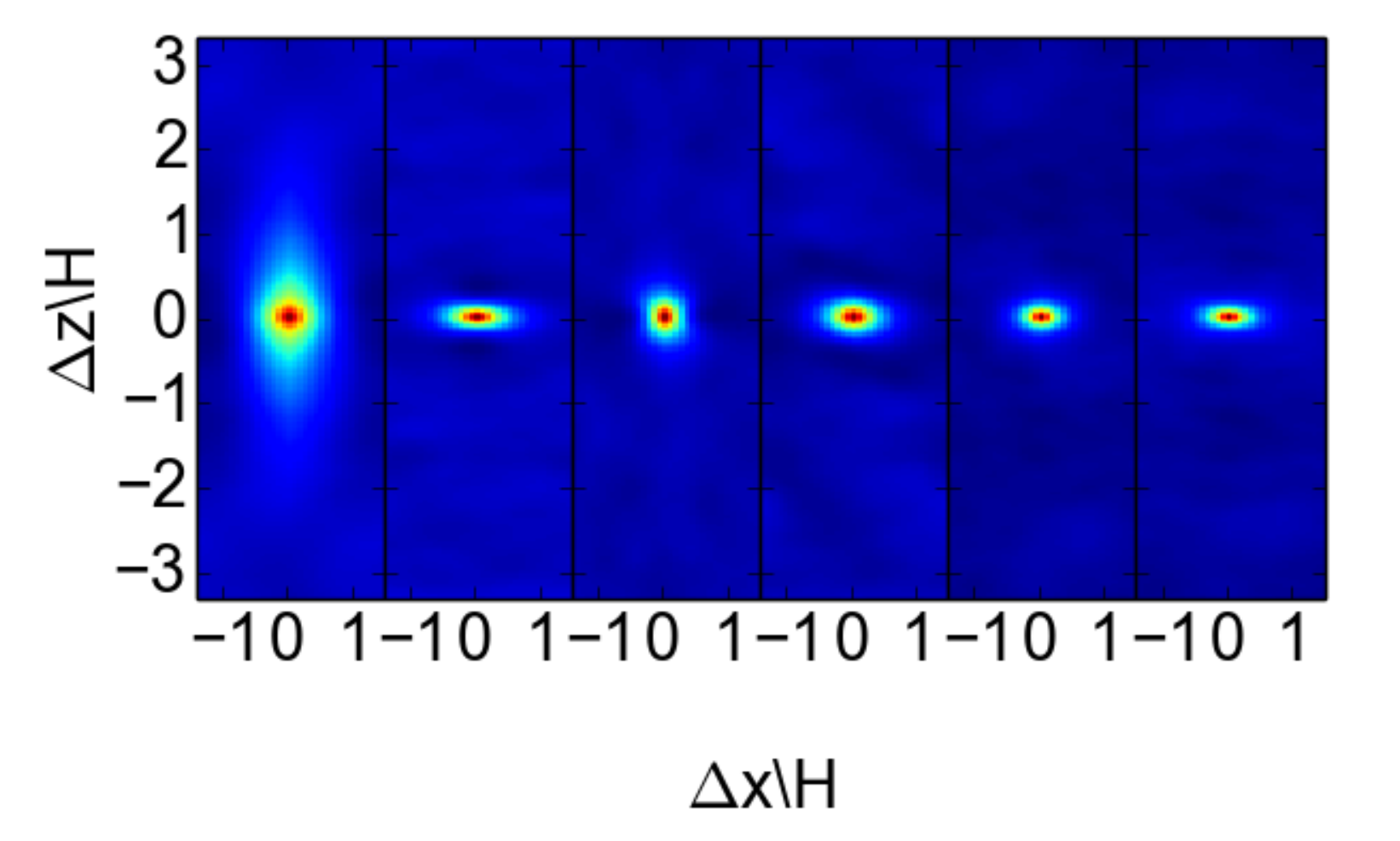}}
  \vspace{-0.25cm}
  \subfigure[$C_{\mathcal{F}}(\Delta x, \Delta z, \Delta y)$ in y-z plane]{\includegraphics[width=0.37\textwidth]{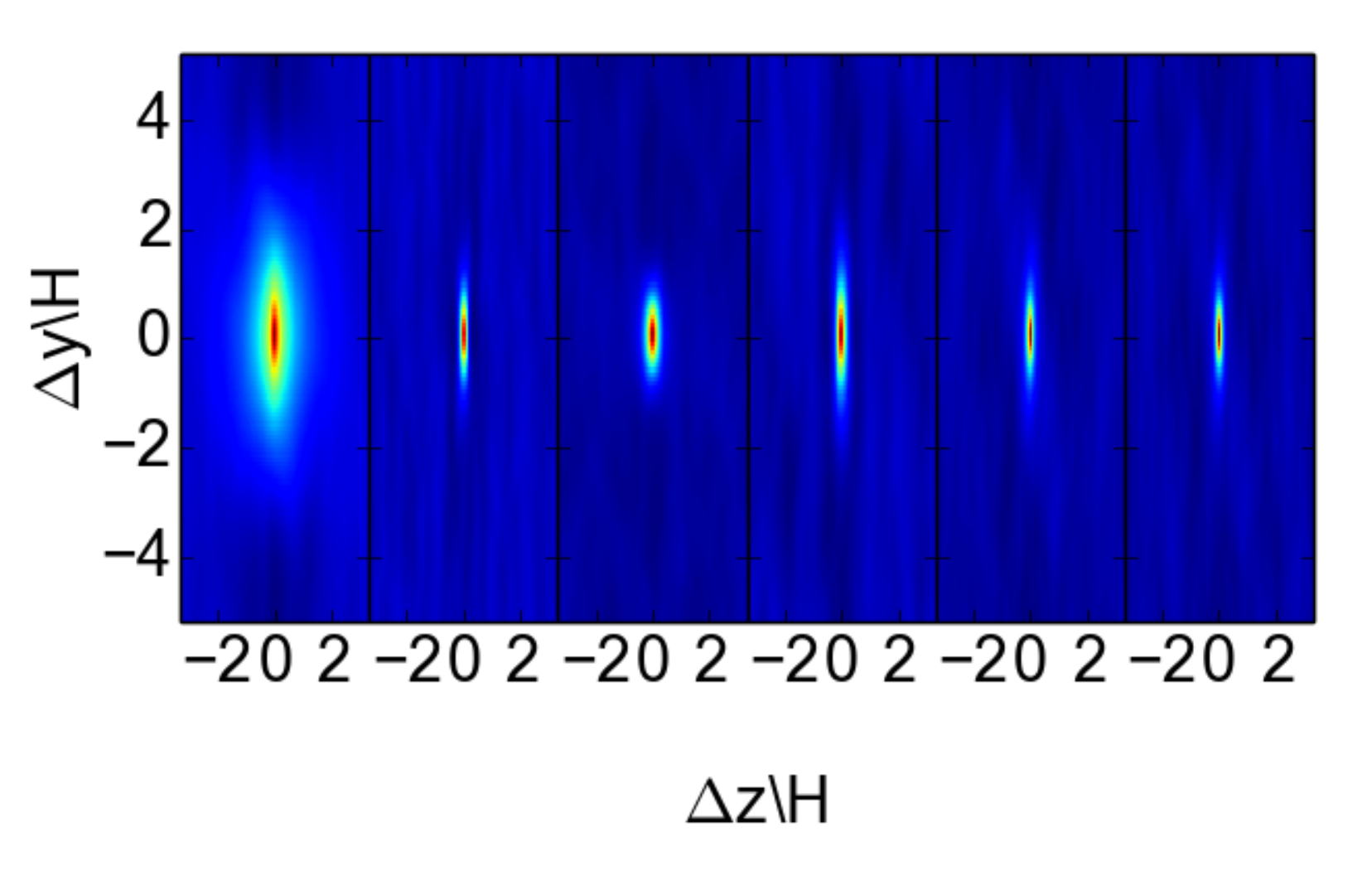}}
    \vspace{-0.25cm}
\caption{Slices of 3D autocorrelation function along x-y, x-z, and y-z planes for $\rho$, $B_{r}$, $B_{\theta}$, $B{\phi}$, $B^2$, and $B_{R}B_{\phi}$.
\label{fig-auto_corr}}
\end{figure}

For the first part of our analysis into the spatial scales, we fit the time average three-dimensional autocorrelations.  Two-dimensional slices through the origin along the x-y, x-z, and y-z planes are shown in Figure \ref{fig-auto_corr}.  To measure the shape of the three-dimensional autocorrelations, we defined three surfaces corresponding to where the value of the autocorrelation falls to $C_{\mathcal{F}}(\Delta x, \Delta z, \Delta y)$ = 0.5, $e^{-1}$, and $e^{-2}$.  The standard definition of the correlation length is the full-width, half-max of the autocorrelation which is where the autocorrelation falls to 0.5.  However, other works use different definitions of the correlation length, so we present those values to aid in comparing the turbulent scales between simulations.

A least-squares minimization was used to fit these surfaces with an ellipsoid of the form:
\begin{equation}
\begin{split}
1 = \\
& \mkern-54mu \frac{[x\cos(\theta)-y\sin(\theta]^2}{a^2} + \frac{[y\cos(\theta)-x\sin(\theta)]^2}{b^2} +\frac{z^2}{c^2}.
\end{split}
\end{equation} where $\theta$ is the tilt angle introduced by the disk shear and $a$, $b$, and $c$ are the axes corresponding to the $x$, $y$, and $z$ directions, respectively.  The tilt angle and best fits for the three different correlation lengths are given in Table \ref{tab-ellips_fit}.

The radial correlation length of the gas density, $\rho$, provides an estimate for the characteristic length scale, $l$, of the turbulence.  In the Shakura \& Sunyaev $\alpha$-prescription the turbulence is less than the disk scale height.  This requirement does not strictly hold for the y- and z-directions, but it does hold for the x-direction, which is the important direction for the angular momentum transport.  Figure \ref{fig-turb_variability} shows the time variability of $l$ calculated at individual data dumps.  Unlike $|v| / c_{s}$, the variability of $l$ is stochastic and lacks the structure seen in the time variability of $v_{turb} / c_{s}$.  Instead of varying in lock step with $\alpha$, the time trace of $l$ is flat and simply fluctuates around $l=0.32$.  This means that increases in stress act to drive the turbulence more quickly, rather than injecting energy at a larger scale and allowing it to cascade down to smaller scales and that the turbulent eddy scales are preserved regardless of the magnitude of the stress in the disk. 

\section{Disk Dynamo and Intermediate Timescale $\alpha$ Variability}
\label{sec-dynamo}

\begin{figure}[!b]
\includegraphics[width=0.5\textwidth]{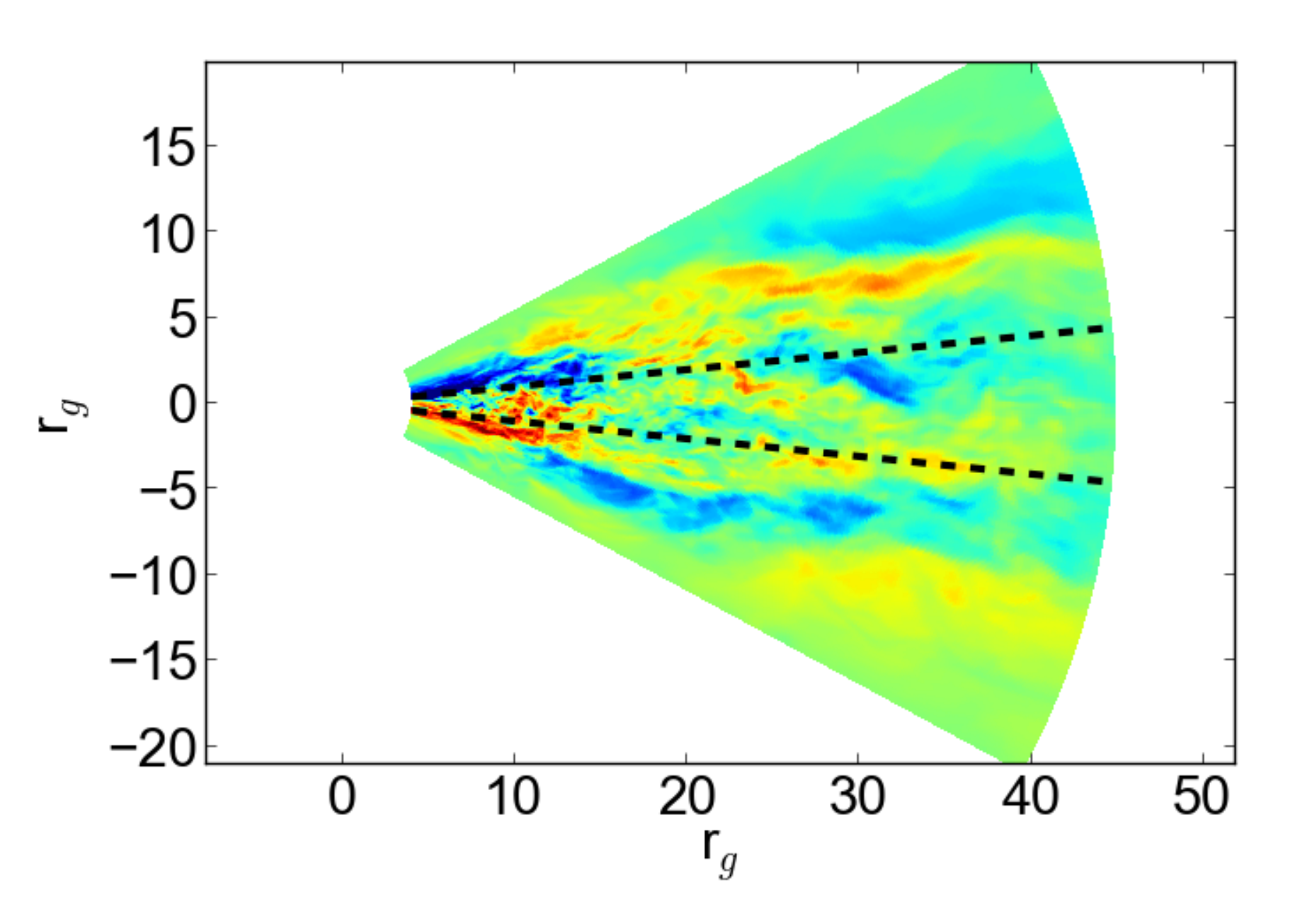}
\caption{Azimuthally averaged $B_{\phi}$ at $t=3.7\times10^{4}$ $GM/c^3$.  Within the disk midplane the fluctuations in $B_{\phi}$ are largely random.  Large, stratified regions of coherent magnetic field develop in the corona as buoyancy lifts highly magnetized gas vertically and regions of similar polarity grow through magnetic reconnection.  The coronal regions are symmetric across the midplane, but have opposite polarity.
\label{fig-bfield_strat}}
\end{figure}  

We will now study the disk dynamo and its influence on the variability of the effective $\alpha$ in the simulation.  The cyclical magnetic dynamo is a well-established phenomenon in accretion disks and is seen in both local \citep{1995ApJ...446..741B, 1996ApJ...463..656S, 1996ApJ...464..690H, 2004ApJ...605L..45T, 2009ApJ...697.1269J, 2009ApJ...691L..49S, 2010ApJ...713...52D, 2015ApJ...799...20B} and global \citep{2011ApJ...736..107O, 2011MNRAS.416..361B, 2012ApJ...744..144F, 2013ApJ...763...99P} MHD simulations.  It is characterized by rising bundles of toroidal magnetic field in which the toroidal field flips polarity between two such events.  The dynamo is quasi-periodic in nature with $\omega\sim 10^{-1} \: \omega_{dyn}$, where $\omega_{dyn}$ is the local dynamical frequency.

\begin{figure}
\centering
  \subfigure[PSD of $B_{\phi}$ in midplane]{\includegraphics[width=0.4\textwidth]{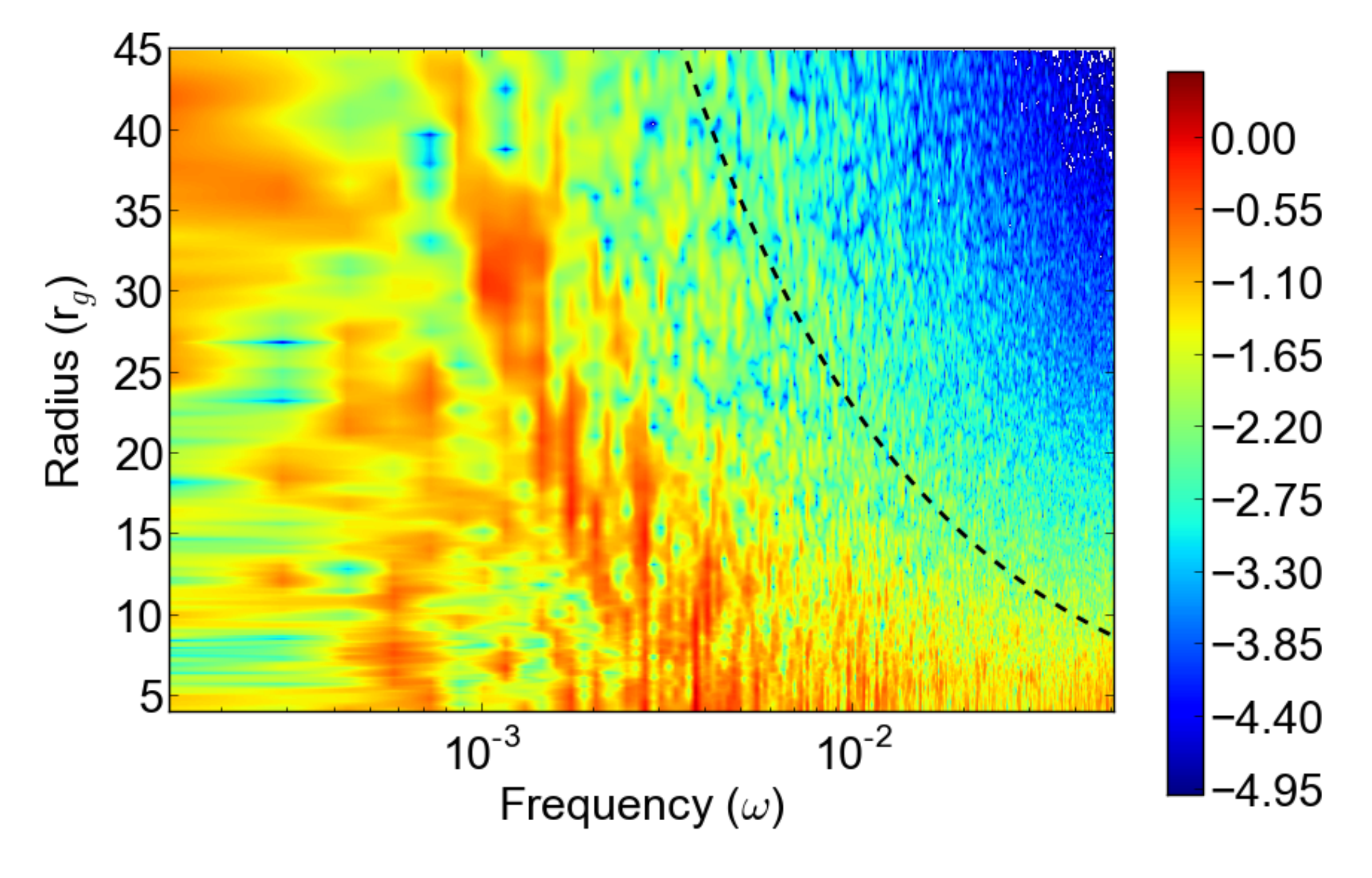}}
  \subfigure[PSD of $B_{\phi}$ at 2$h$]{\includegraphics[width=0.4\textwidth]{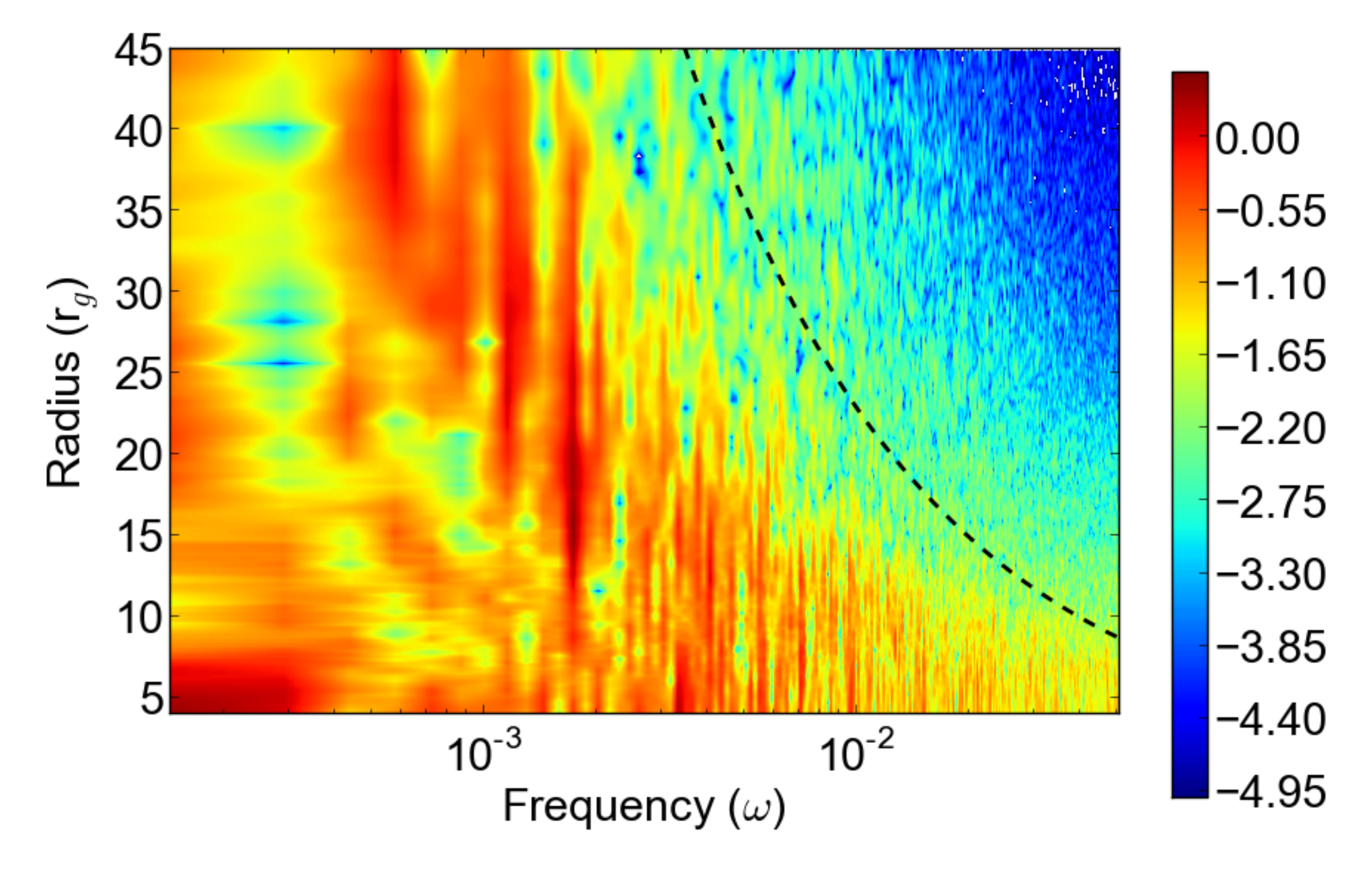}}
  \subfigure[PSD of $B_{\phi}$ at 3$h$]{\includegraphics[width=0.4\textwidth]{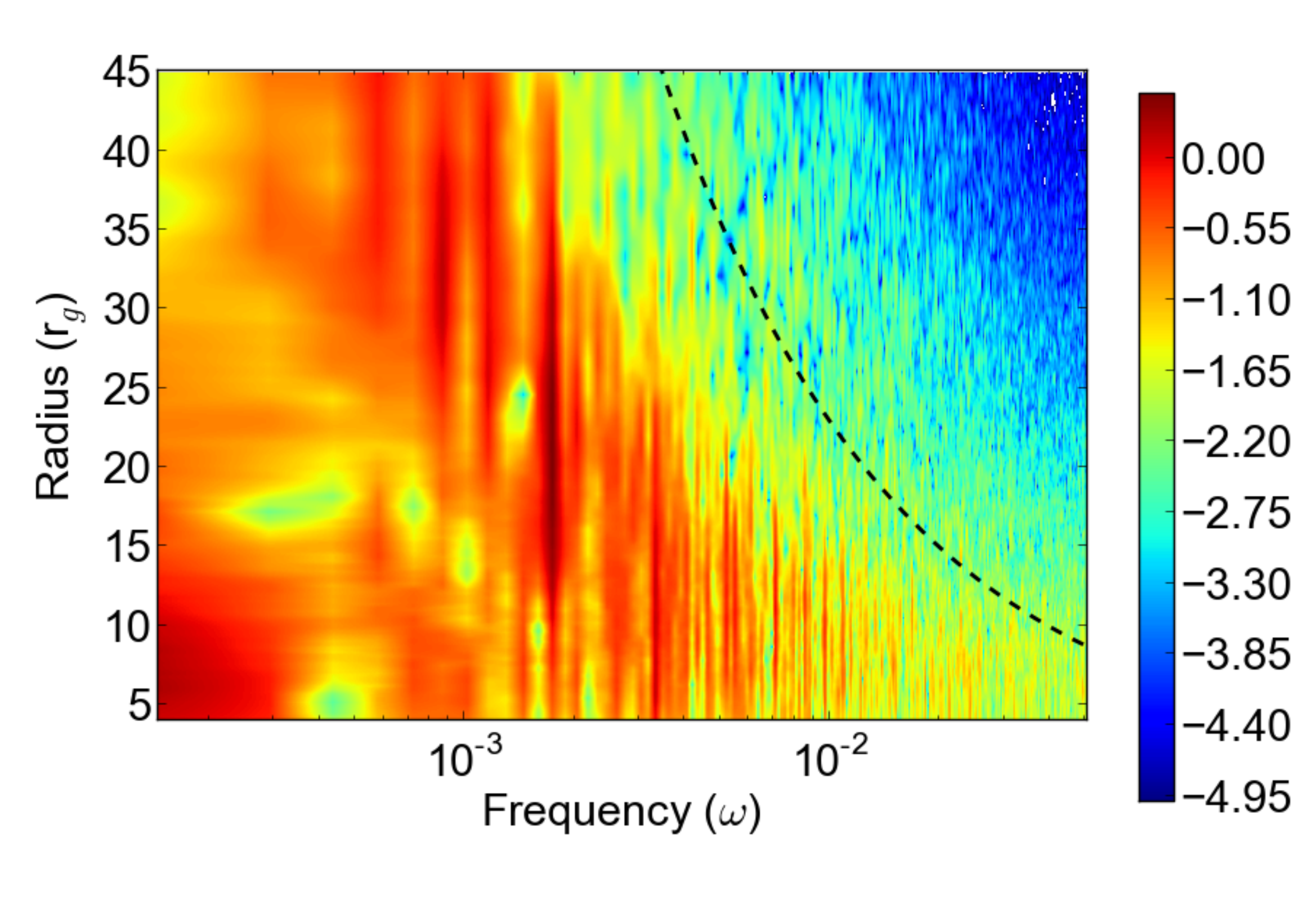}}
  \subfigure[PSD of $B_{\phi}$ at 4$h$]{\includegraphics[width=0.4\textwidth]{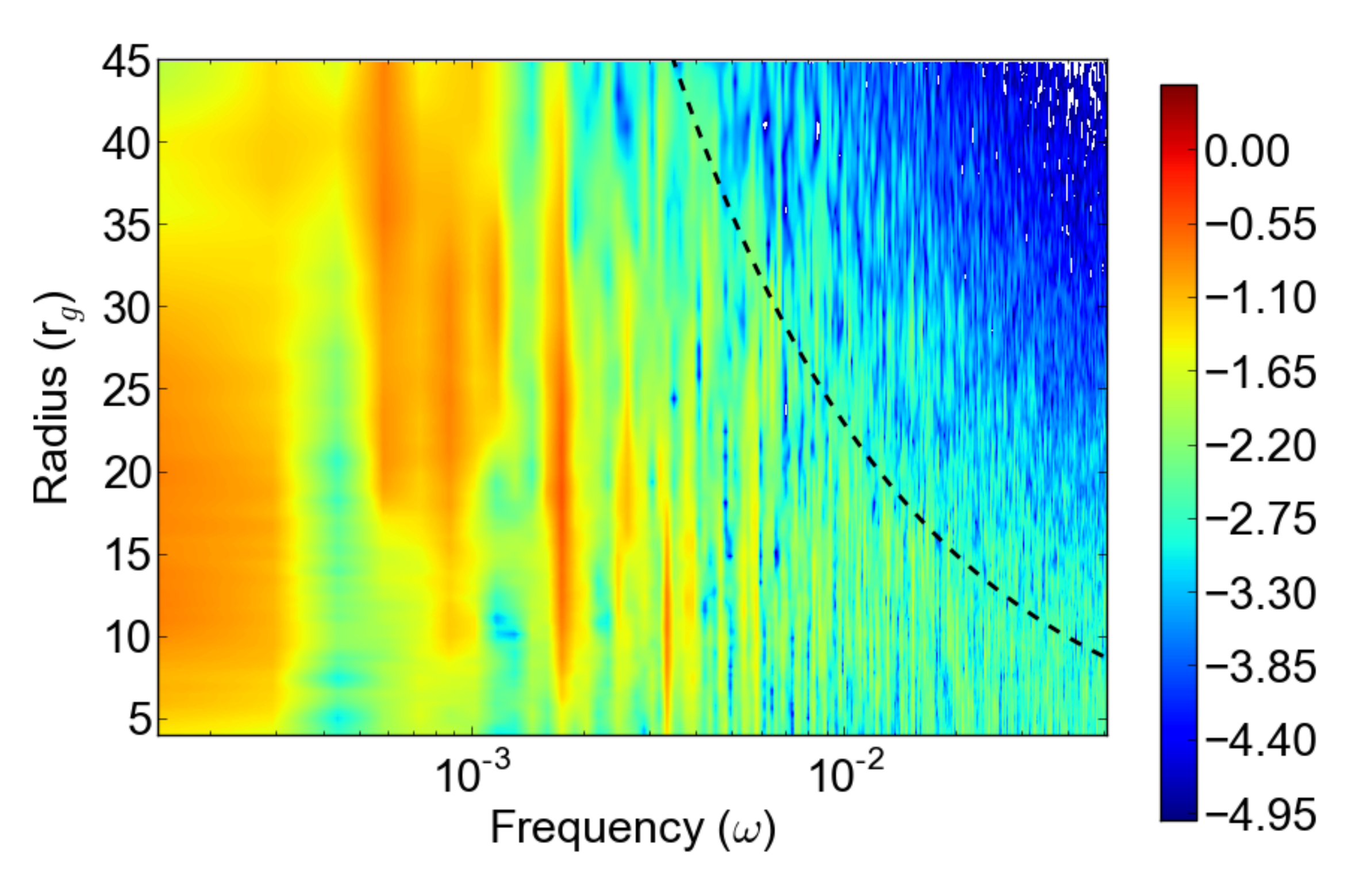}}
\caption{PSDs of $B_{\phi}$ at different elevations above the disk with the orbital frequency overlaid as the black, dashed line.  Within the midplane the standard $\alpha\Omega$ dynamo behavior is seen.  A band of power vertically spanning $\approx10$ r$_{g}$ follows the shape of the orbital frequency at 10$\times$ lower frequency.  The radial correlation of $B_{\phi}$ increase with increasing height above the disk due to the stratification that develops in the corona.
\label{fig-dynamo_FFTs}}
\end{figure}

The influence the dynamo has on the disk is not fully understood.  Since the Maxwell stress and, thus, turbulent injection from the MRI is proportional to $B_{R}B_{\phi}$, we might imagine that the stress would be modulated by the large-scale toroidal magnetic field generated by the dynamo.  Indeed, a correlation between stress and toroidal magnetic field has been found \citep{2010ApJ...713...52D, 2012ApJ...744..144F}, however, the coupling of Maxwell stress to the low-frequency, periodic fluctuations of the dynamo has not been well-investigated.   

Of additional interest is how the dynamo could provide a local, net poloidal magnetic field, thereby increasing the turbulent transport of angular momentum.  Local, shearing box simulations with a net poloidal magnetic field typically find higher $\alpha$-values than similar simulations with zero net magnetic field \citep{1995ApJ...440..742H,2012ApJ...749..189S}.  When considering the impact of the dynamo on the global disk, we can think of the global domain as a collection of many local regions.  The large-scale magnetic field generated by the dynamo could produce a net local poloidal field through these regions, even if the overall magnetic field in the global domain is zero.

\subsection{General Dynamo Behavior}

We begin with an overview of the spatio-temporal properties of the magnetic field in the disk.  After the intialization of the simulation, a stratified magnetic field geometry developed and was sustained.  Figure \ref{fig-bfield_strat} shows a snapshot of the azimuthally average of $B_{\phi}$ at an arbitrary time ($t=3.7\times10^{4}$ $GM/c^3$).  Correlated regions of field in the corona are seen spanning large portions of the disk.  Like other simulations, the large-scale magnetic field is generated in the disk and is then slowly lifted by magnetic buoyancy \citep{2010ApJ...713...52D,2011ApJ...736..107O}.

This growth can be seen in the PSDs of the azimuthally averaged $B_{\phi}$. Shown in Figure \ref{fig-dynamo_FFTs} are the PSDs of $B_{\phi}$ as a function of radius for the midplane and 2h, 3h, and 4h above the disk midplane.  The PSDs were calculated similarly to \citet{2009ApJ...692..869R} and are defined as $P(\nu)=|\widetilde{f}(\nu)|^{2}$ where $\widetilde{f}(\nu)$ is the Fourier transform of the time sequence of the variable of interest,
\begin{equation}
\widetilde{f}(\nu) = \int f(t) e^{-2 \pi i \nu t} dt.
\end{equation}  In our case, the time sequence is the time variability of the azimuthal average of $B_{\phi}$ in the disk mid-plane.  The Fourier transform was calculated for each point on the radial grid.  The large gas reservoir supplies additional material to the disk during the simulation which protected from secular changes due to the draining from accretion.  Therefore, there was no need ``pre-whiten" before taking Fourier transforms of the basic fluid variability, as \citep{2009ApJ...692..869R} discuss the need for in the case of significant secular evolution.

In the disk midplane the dynamo behaves similarly to the dynamo in \citet{2010ApJ...713...52D}.  Rather than having a single characteristic frequency, a band of enhanced power parallels the orbital frequency at frequencies $\approx0.1\Omega$.  For a given frequency, a band of power extends 10-15 r$_{g}$, highlighting the radial coherence of the dynamo action pointed out by \citet{2011ApJ...736..107O}.  For a given radius, the band of power spans approximately a factor of three.  

The PSDs of the azimuthal magnetic field above the disk show a larger radial range of enhanced power.  As the the height above the disk midplane increases, the vertical bands of power at a given frequency stretches to smaller radii.  At $4h$, power at the lowest frequencies is enhanced along the entire radial domain of the simulation.  All the while, the high frequency envelope seen in the PSD of the disk midplane is preserved.  

\subsection{Modulation of Stress By Dynamo}

\begin{figure}[t]
\includegraphics[width=0.5\textwidth]{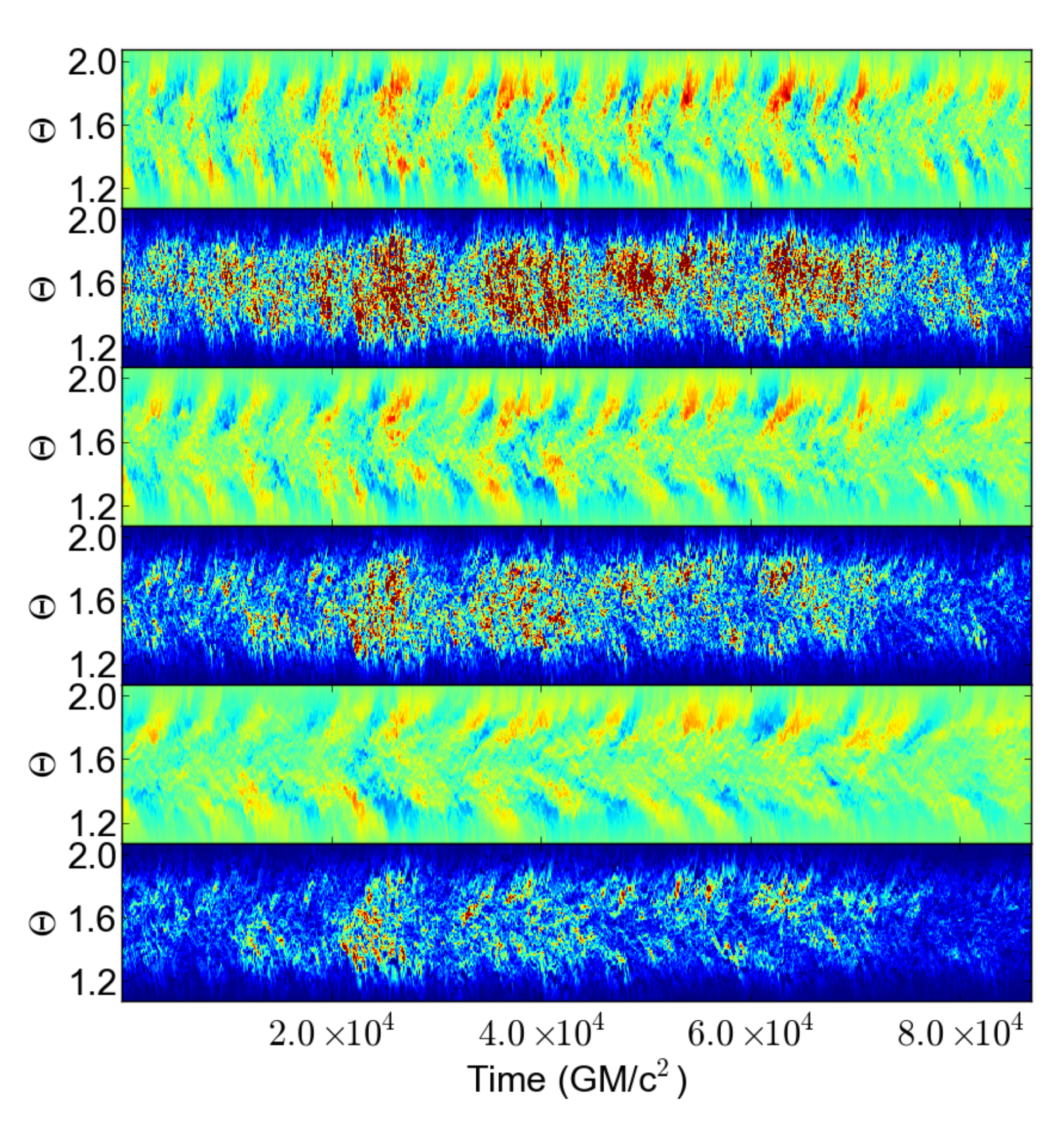}
\caption{Spacetime diagrams of azimuthally averaged $B_{\phi}$ and $B_{R}B_{\phi}$ for 15, 20, and 25 r$_{g}$.  The characteristic ``butterfly" pattern is seen in the spacetime diagrams of $B_{\phi}$.  When $B_{\phi}$ is stronger, $B_{R}B_{\phi}$ is stronger, although the variability is mostly dominated by high-frequency variability. 
\label{fig-dynamo_vs_brbp}}
\end{figure}

Figure \ref{fig-dynamo_vs_brbp} shows the spatio-temporal evolution of the dynamo (probed by the azimuthal average of $B_{\phi}$) compared with the Maxwell stress for three distinct radii (15 r$_{g}$, 20 r$_{g}$, \& 25 r$_{g}$).  We will refer to these three radii in order to provide a comparison of different properties related to the magnetic field throughout the disk.  As we saw in the previous section, the dynamo plays a central role in the disk's magnetic field evolution and, as we can see from the butterfly diagrams, this translates into modulation of the Maxwell stress --- when the magnetic field is stronger at the maxima and minima of the dynamo cycle, the stress is largest.  This behavior is quite similar to that seen in the stratified shearing boxes of \citet{2010ApJ...713...52D}.  Since the polarity of the global field flips during the dynamo cycle the stress is always positive.  It is also worth noting that while the dynamo defines the trend within a given cycle, the fluctuations in the stress are dominated by high-frequency variability.

\begin{figure}
\includegraphics[width=0.5\textwidth]{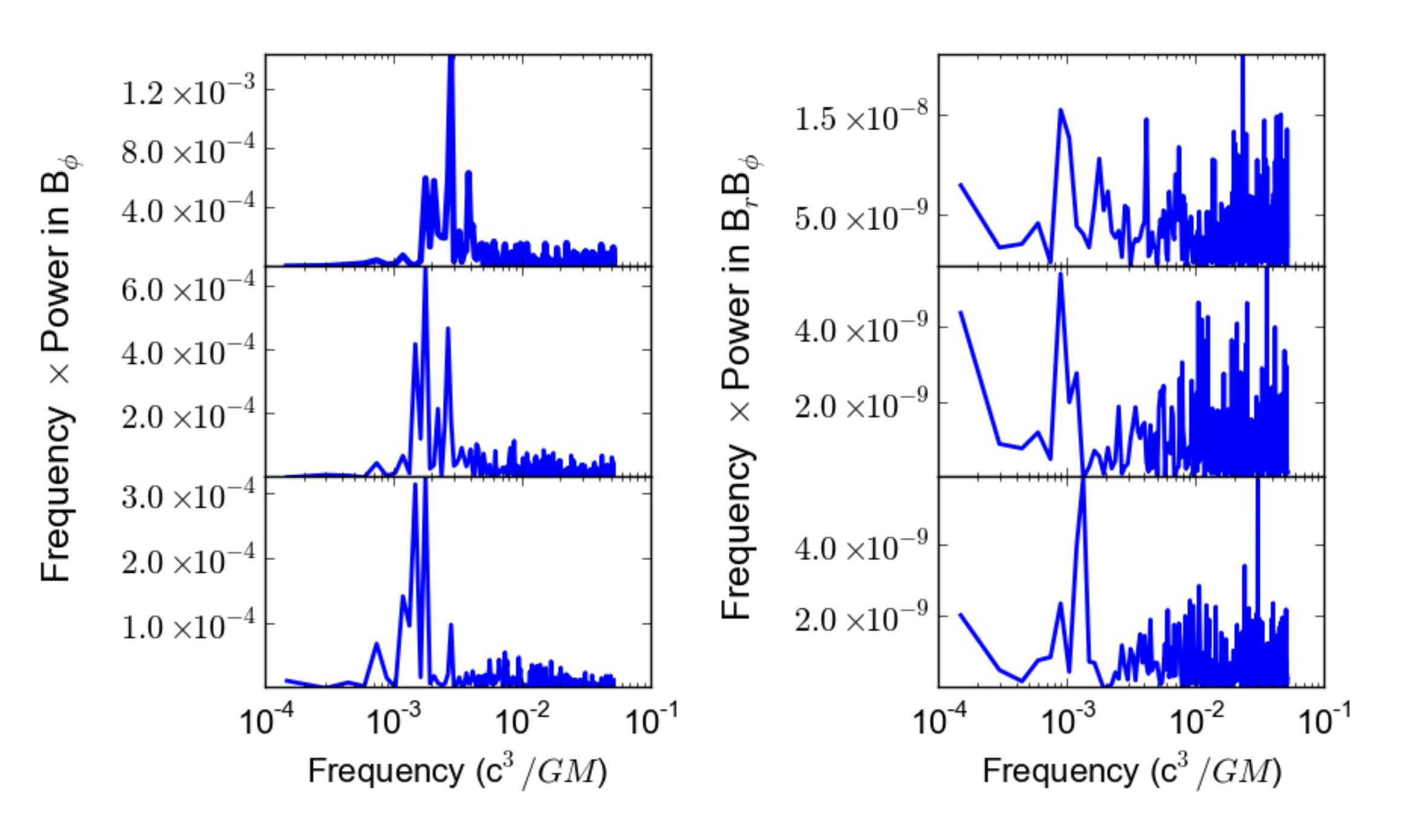}
\caption{Frequency-weighted PSDs ($\nu P$) of $B_{\phi}$ (left panel) and $B_{R}B_{\phi}$ (right panel) for 15 (top), 20 (middle), and 25 (bottom) r$_{g}$.  The strong, multi-peaked bands of power  correspond to the oscillations seen in Figure \ref{fig-dynamo_vs_brbp}.  A noticeable difference between the frequency-weighted PSDs of $B_{\phi}$ and $B_{R}B_{\phi}$ is the strong power at high-frequencies in $B_{R}B_{\phi}$.
\label{fig-alpha_2dFFT}}
\end{figure} 

Shown in Figure \ref{fig-alpha_2dFFT} are the frequency-weighted PSDs of $B_{\phi}$ and $B_{R}B_{\phi}$ for 15 , 20, and 25 r$_{g}$.  The modulation of the Maxwell stress by the dynamo is more clearly seen in the strong, multi-peaked bands of power in the PSDs of both $B_{\phi}$ and $B_{R}B_{\phi}$ corresponding to oscillations of the low-frequency dynamo.  However, the PSDs of $B_{R}B_{\phi}$ also have a high-frequency component to its variability from fluctuations due to MRI-driven turbulence.  This component is much more significant than the high-frequency variability in $B_{\phi}$.  Interestingly, the bands of low-frequency variability in the frequency-weighted PSDs of $B_{R}B_{\phi}$ are found around the local dynamo frequency, but the frequency of the variability is constant across the radii we probe.  The strong band of power in the PSDs of $B_{\phi}$ show the expected shift to lower frequency with increasing radii due to the decrease in orbital frequency.  This discrepancy seems to indicate that while the dynamo does introduce the low frequency variability in $\alpha$ needed to drive the propagating fluctuations, the frequency at which $\alpha$ is modulated is subject to a more global interaction of the magnetic field.

\section{Propagating Fluctuations in the Model Disk}
\label{sec-prop_flucs}

\subsection{Overview of Global Behavior}
In the analysis of propagating fluctuations we are interested in the behavior of the instantaneous mass accretion rate,
\begin{equation}
\dot{M}(R) = \int \rho v_{R} R \textrm{sin}(\theta) d\phi d\theta,
\end{equation} 
and how it is related to the local disk variables, specifically the surface density and the effective $\alpha$-parameter. Additionally, we are interested in how evolution of the accretion disk might appear if we could observe our simulated disk. We therefore employ a dissipation proxy to generate synthetic time-dependent luminosity.  In accretion disks, the outward transport of angular momentum liberates gravitational potential energy from the disk gas.  This is then converted to thermal energy through the turbulent cascade of MRI driven turbulence and removed from the system through radiative processes.  Assuming the disk is radiatively-efficient, the magnetic stress can be used as a tracer of energy dissipation since the energy injected by the stress is quickly radiated away.  Adopting Eqn 9 from \citet{1998ApJ...505..558H}, the local flux at the photosphere of the disk (at one scaleheight) due to dissipation is given by 
\begin{equation}
F=\sqrt{\frac{GM}{r^{3}}}\Bigg(\frac{A}{B}\Bigg) \int_{0}^{h} B_{r}B_{\phi} dz
\end{equation}\label{eqn-proxy} where A and B are relativistic correction factors given by 
\begin{equation}
A=1-\frac{2GM}{r c^2}
\end{equation} and
\begin{equation}
B=1-\frac{3GM}{r c^2}.
\end{equation} This scheme has been used to produce synthetic light curves in other studies of global accretion disks, e.g. \citet{2001ApJ...548..348H} and \citet{2003MNRAS.341.1041A}.  

\begin{figure*}[t]
\includegraphics[width=\textwidth]{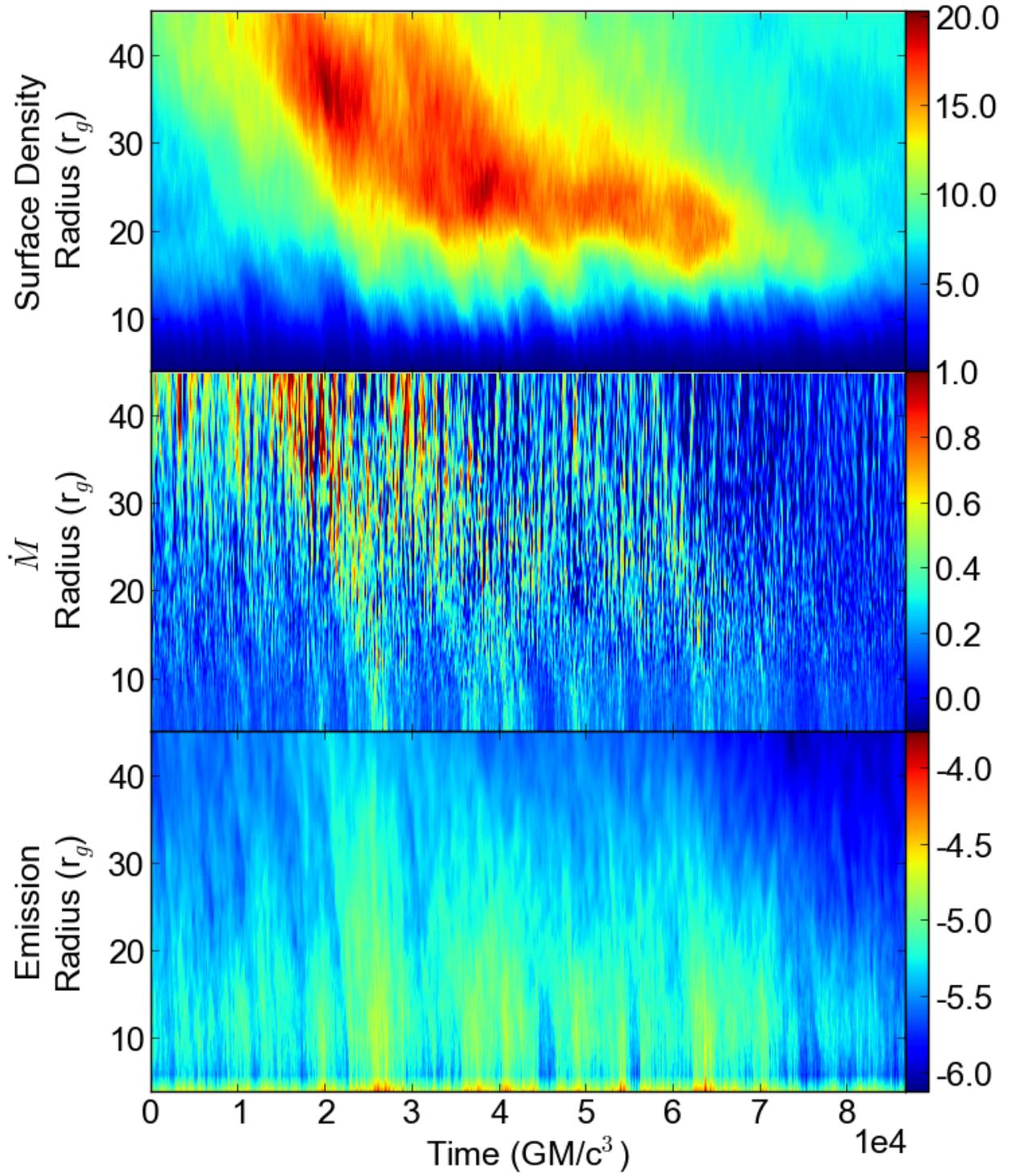}
\caption{Spacetime diagrams of $\Sigma$ (top panel), $\dot{M}$ (middle panel), and logarithm of the synthetic emission (bottom panel).  The preservation of fluctuations in the accretion flow that gives the appearance of ``propagating fluctuations" is readily observed in the spacetime diagram of $\Sigma$ and $\dot{M}$.  While a simple estimate of the radiation, our emission proxy does track the other two quantities well, demonstrating that the behavior of the accretion flow will be observable.
\label{fig-globalplot}}
\end{figure*}

Shown in Figure \ref{fig-globalplot} are the spacetime diagrams of $\Sigma$, $\dot{M}$ and the emission proxy.  As can be seen in the figure, periods and regions of enhanced accretion (higher $\dot{M}$) are coincident with higher $\Sigma$.  The correlation of $\dot{M}$ with surface density acts to (partially) preserve the pattern of fluctuations in surface density as disk material loses angular momentum and moves to smaller radii.  Additionally, when $\dot{M}$ is greater at a given radii, there is a corresponding increase in the emission proxy.  It is worth noting that the name ``propagating fluctuations" is a slight misnomer as there is no real propagation mechanism at play in the disk.  Rather, the mass accretion is a diffusive process that scales with $\Sigma$, providing the qualitative appearance of ``propagating" to smaller radii.

\begin{figure*}[t]
  \centering
  \subfigure[$\dot{M}$ at ISCO]{\includegraphics[width=0.5\textwidth]{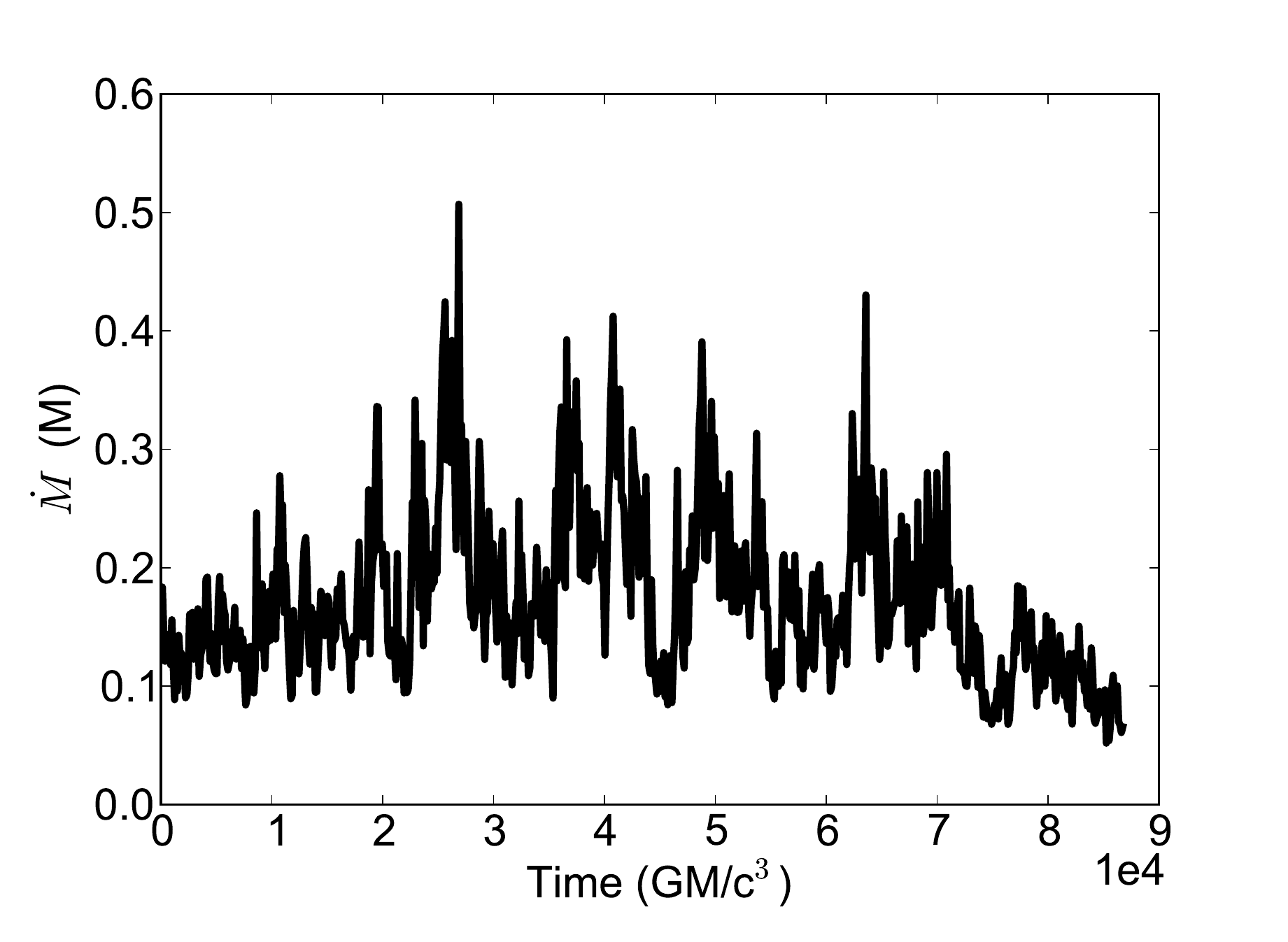}}
  \hspace{-0.5cm}
   \subfigure[PSD of $\dot{M}$ at ISCO]{\includegraphics[width=0.5\textwidth]{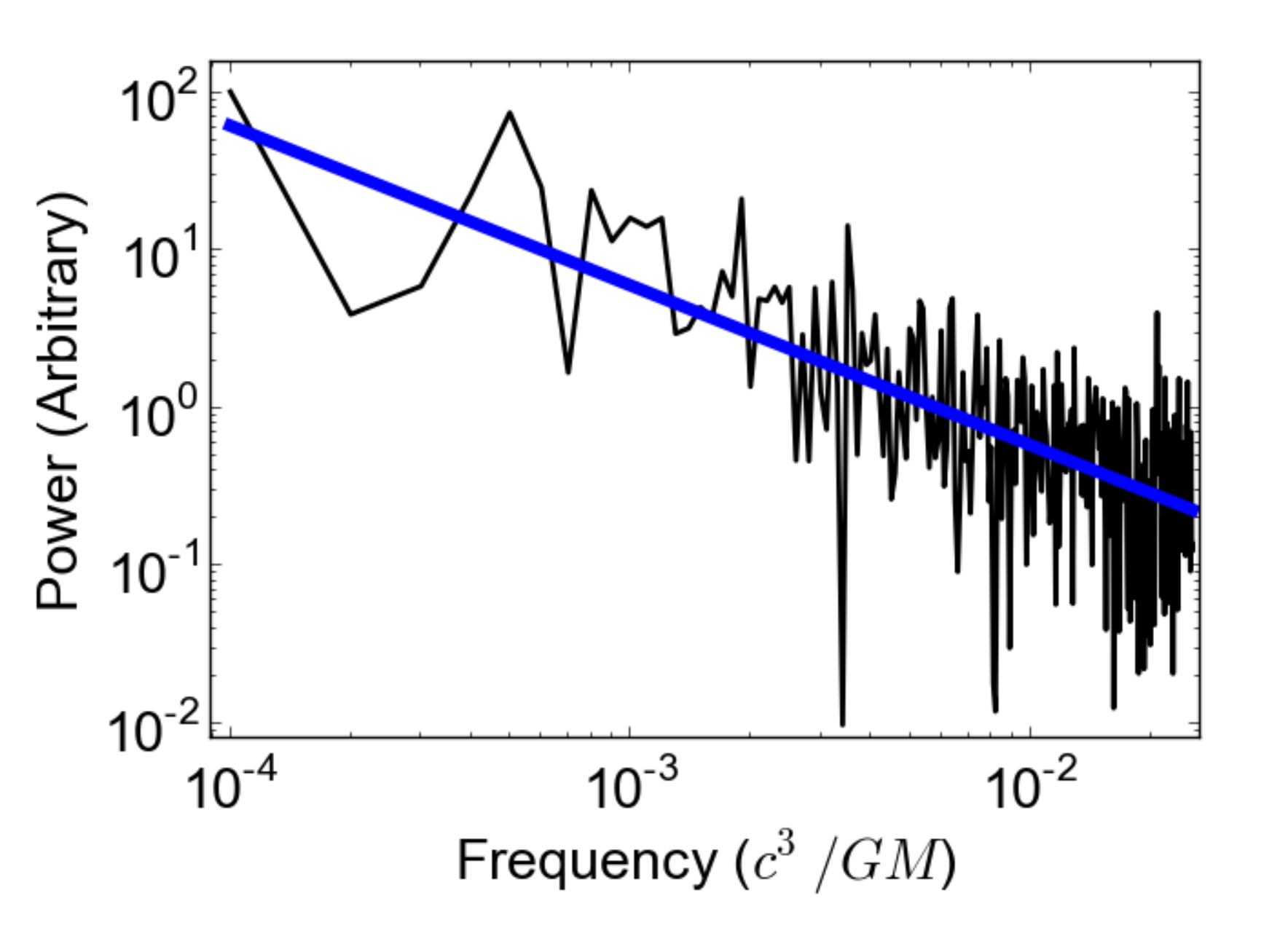}}
  \caption{$\dot{M}$ at the ISCO (left panel) and its PSD (right panel).  The PSD is best fit by a single power law with a slope of $\Gamma=-1$, shown by the blue line.
  	} \label{fig-mdottime}
\end{figure*}

Figure \ref{fig-mdottime} shows the time trace and PSD of $\dot{M}$ at the ISCO (6 r$_{g}$). As expected, there are stochastic accretion events when $\dot{M}$ dramatically increases (by a factor of 3-5), and then returns to a stationary baseline value, a value of $0.1$.  The PSD is featureless and is best fit with a powerlaw with index of $\Gamma=-1.01 \pm 0.07$, corresponding to flicker-type noise.

\subsection{Log-normal $\dot{M}$ Distribution}

\begin{figure}[t]
\centering
\includegraphics[width=0.5\textwidth]{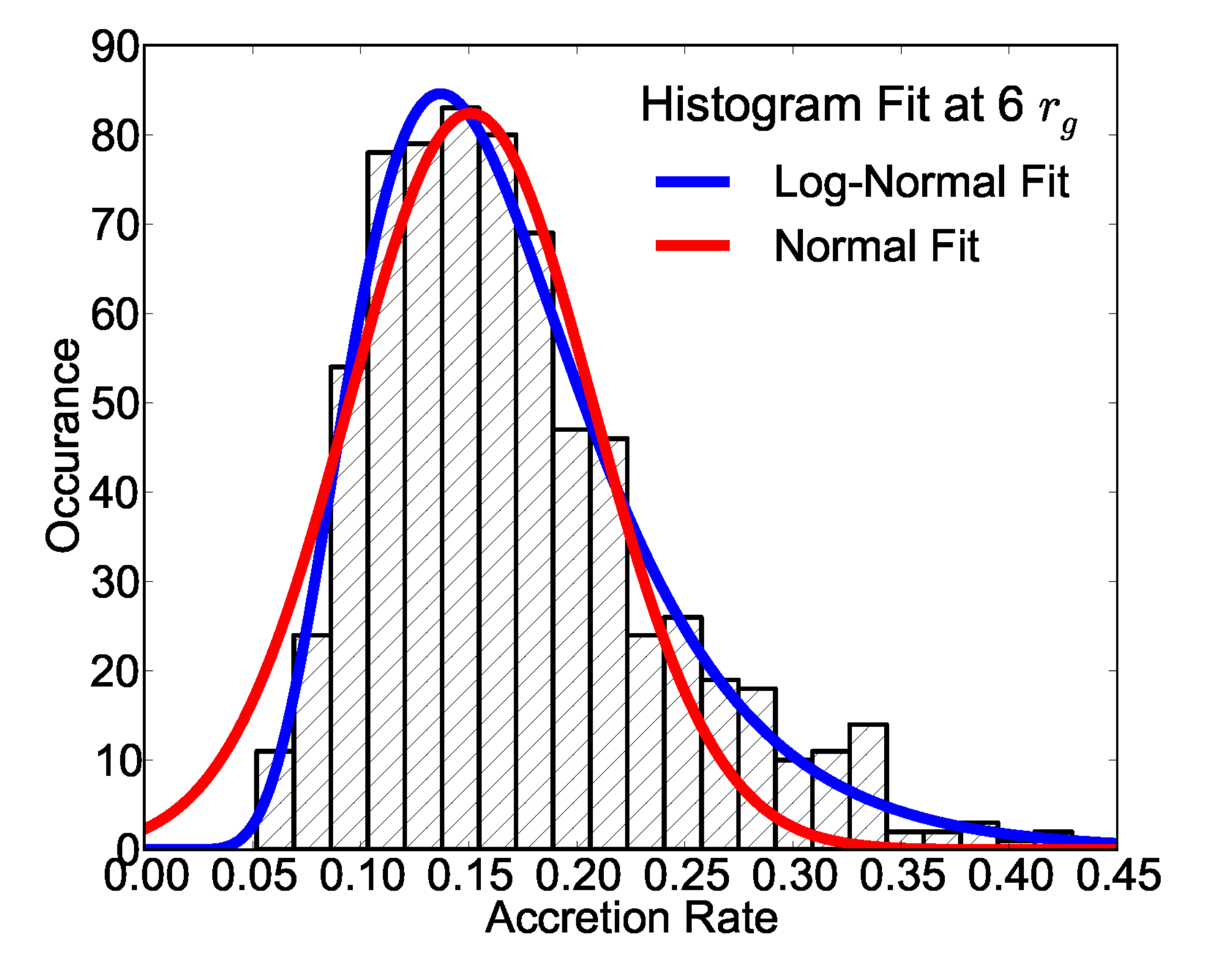}
\caption{Histogram of $\dot{M}$ at the ISCO fit with a normal (red line) and log-normal (blue line) distribution.  The log-normal distribution provides a superior fit. 
\label{fig-dist}}
\end{figure}

The histogram of the instantaneous mass accretion rate at the ISCO in our simulation is shown in Figure \ref{fig-dist}.  The shape of the distribution resembles the distribution of GBHB systems like Cygnus X-1 \citep{2005MNRAS.359..345U} and is, similarly, very well fit by a log-normal distribution of the form 
\begin{equation}
f(x; \mu, \sigma) = P_{0} \textrm{exp} \Big[- \frac{(\textrm{ln} x - \mu)^2}{2 \sigma^{2}} \Big],
\end{equation} 
where $P_{0}$ is the normalization, $\mu$ is the log of the mean, and $\sigma$ is the standard deviation.  The best-fitting model parameters for the probability density function (PDF) of the mass accretion rate at $6 r_{g}$ are $\mu=-1.84$ and $\sigma=-0.38$ and gives a $\chi^{2}/D.O.F$ of $17.9/24$.  For comparison, the best fit with a normal distribution has a $\chi^{2}/D.O.F$ of $79.8/24$.  Statistically, the log-normal distribution is overwhelmingly favored ($\Delta \chi^{2} = 62$) and, visually, we can see the normal distribution fails to reproduce the fast rise of the distribution at low $\dot{M}$ values and shallower tail of the distribution at higher $\dot{M}$ values.  The implication of this distribution is, hence, that fluctuations in $\dot{M}$ combine multiplicatively, rather than additively in our MHD disk. 

We can also quantify the asymmetry of the distribution by measuring skewness.  Skewness is the third momentum of the distribution and is given by, 
\begin{equation}
\gamma_{1}=\frac{\sum\limits_{i=1}^N (X_{i}-\mu)^3 / N}{\sigma^{3}}
\end{equation} 
where $\mu$ is the mean, $\sigma$ is the standard deviation, and $N$ is the number of data points.  The distribution of $\dot{M}$ at the ISCO has a value of $\gamma_{1}=1.08$, indicating a strongly positively skewed distribution. 

\subsection{Correlations Between $\alpha$, $\dot{M}$, $\&$ $\Sigma$}
\label{sec-ams_corr}

We will now look at the correlations in $\alpha$, $\dot{M}$, $\&$ $\Sigma$ to verify that $\dot{M}$ does indeed scale with both $\alpha$ and $\Sigma$.   It is typically assumed $\alpha$ is independent of $\Sigma$ thus, according the canonical disk equation, $\dot{M}$ is expected to be higher when $\alpha$ and $\Sigma$ are larger.  However, there is a degeneracy between these three parameters and we must remove the stochasticity of the third in order to tease out relationships between any two quantities.  The easiest way to find the underlying trend is to simply bin the data, which essentially averages out the stochasticity of the variable we are not interested in. Since there are radial gradients in the disk, we focused on the behavior at $r=15$ $r_{g}$.  Ten bins were used with 70 data points per bin.  The bars indicate the standard deviation of the bin.

\begin{figure*}[t]
  \centering
  \subfigure[$\Sigma$ vs. $\dot{M}$]{\includegraphics[width=0.45\textwidth]{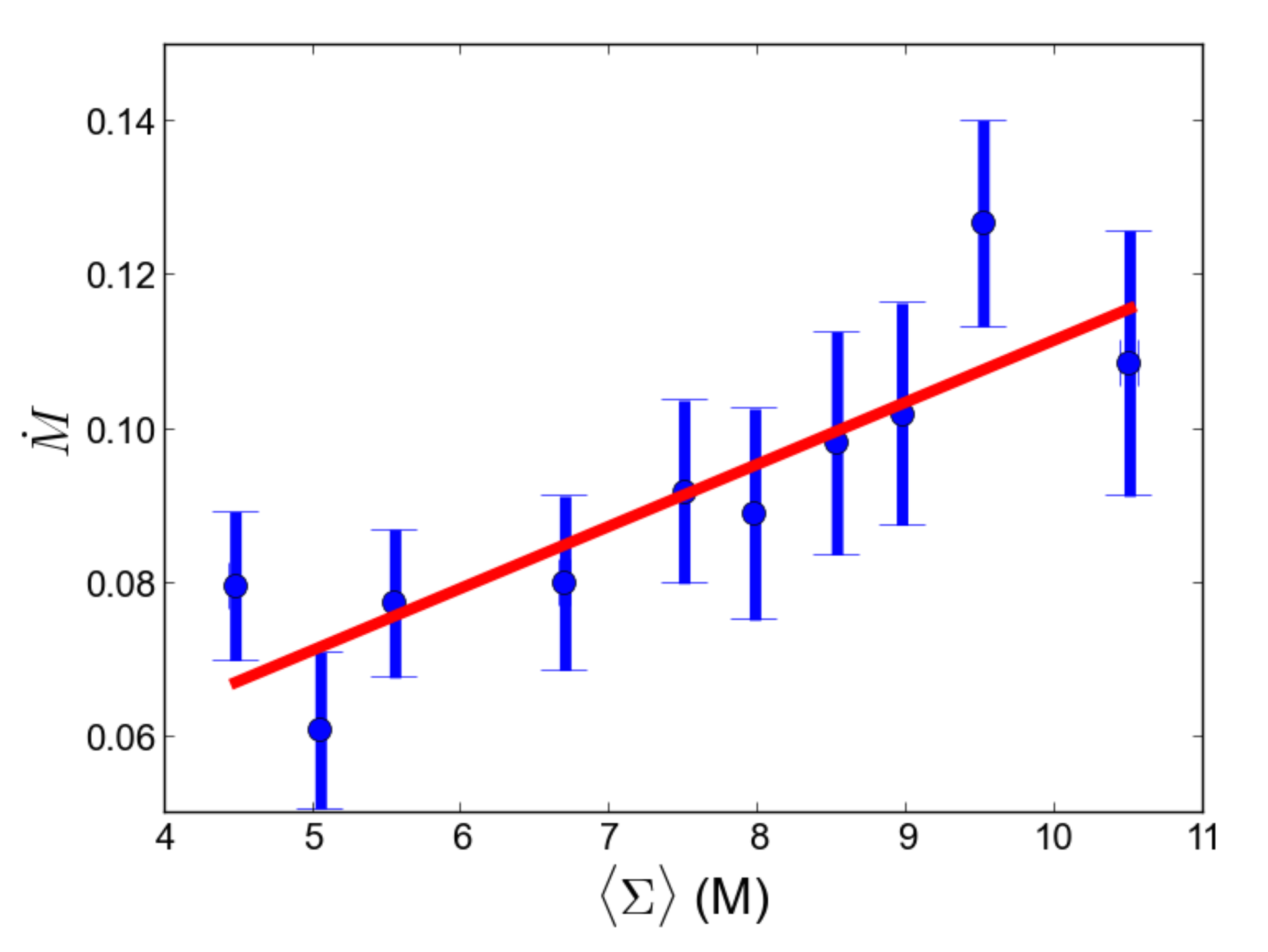}}
   \subfigure[$\Sigma$ vs $\alpha$]{\includegraphics[width=0.45\textwidth]{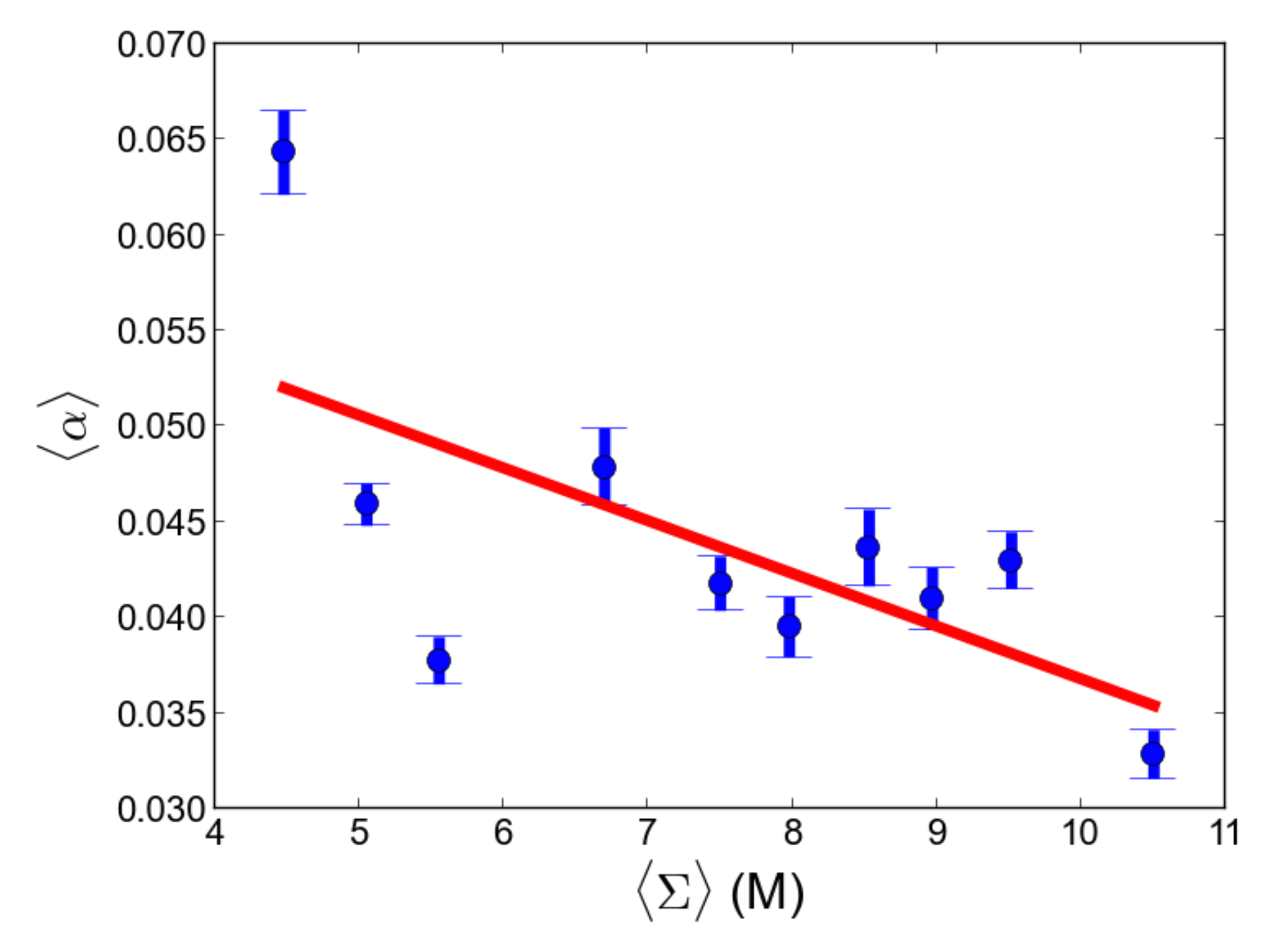}}
  \subfigure[$\alpha$ vs $\dot{M}$]{\includegraphics[width=0.45\textwidth]{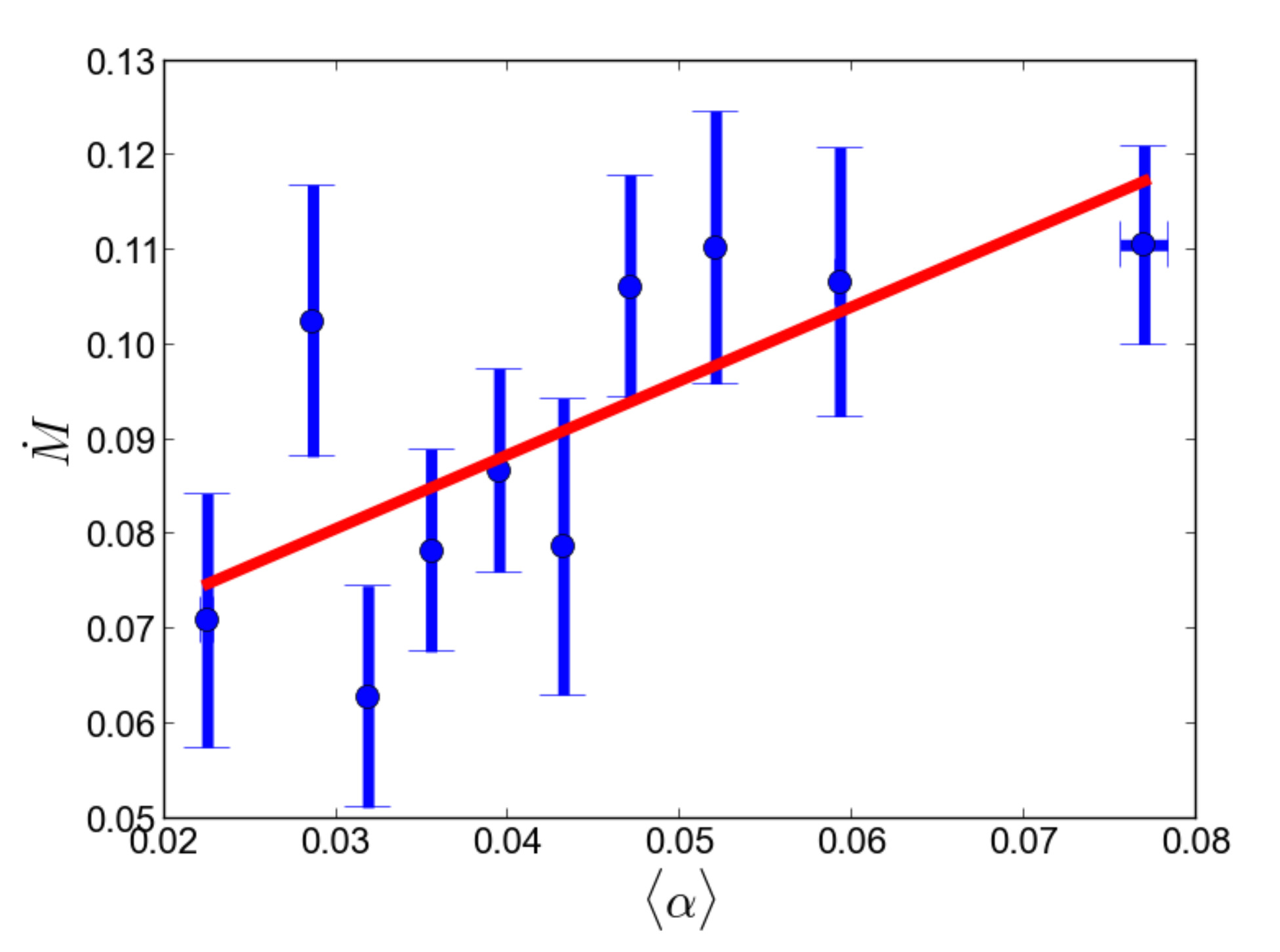}}
  \caption{Correlations between $\dot{M}-\Sigma$ (left panel), $\alpha-\Sigma$ (middle panel), and $\dot{M}-\alpha$ (right panel) at $r=15$ $r_{g}$ in the simulation.  There is a clear positive, linear trend in the $\dot{M}$ dependence on both $\alpha$ and $\Sigma$.  Surprisingly, there is a negative trend in the $\Sigma$-$\alpha$ relationship which we believe is a caused by magnetic buoyancy of more strongly magnetized gas in the disk.
  	} \label{fig-ams}
\end{figure*}

Figure \ref{fig-ams} shows the $\dot{M}-\Sigma$, $\alpha-\Sigma$, and $\dot{M}-\alpha$ correlations..  We find that, indeed, the mass accretion rate scales with both $\alpha$ and $\Sigma$, validating the \emph{a priori} assumptions that went into the phenomenological motivation of the propagating fluctuations model.  The Pearson correlation coefficient between $\Sigma$ and $\dot{M}$ is $0.87$ and between $\alpha$ and $\dot{M}$ it is $0.70$, indicating these are both statistically strong correlations.  Additionally, the correlation coefficient of $\alpha$ with $\Sigma$ is $r=-0.66$, which is considered a moderately strong anti-correlation.  While the strong positive correlations of $\dot{M}$ with $\alpha$ and $\Sigma$ confirm the underlying assumptions of the propagating fluctuations model, the negative $\alpha$-$\Sigma$ correlation is a slightly different than expected.  We believe the anti-correlation between $\alpha$ and $\Sigma$ can be attributed to the magnetic buoyancy in the disk.  In the higher density regions, pressure balance in the disk displaces magnetically dominated (lower $\beta$) gas upwards, decreasing the local field strength, thereby decreasing $\alpha$ and causing the negative trend.

\subsection{Radial Coherence}

The strongest evidence for propagating fluctuations in our simulation comes from radial coherence and frequency dependent phase shifts of the $\dot{M}$ variability.  At the heart of the propagating fluctuations model is the predication that modulations in the accretion rate at larger radii will be seen at the inner radii with a time-lag set by the viscous inflow time.  The coherence function provides the cleanest way to assess the causal connection in the disk.  

Adopting the convention of \citep{1999ApJ...510..874N}, we will consider $\dot{M}$ at two radii, $s_1(t)$ and $s_2(t)$, with Fourier transforms $S_1(f)$ and $S_2(f)$, respectively.  The coherence of these two signals, given by 
\begin{equation}\label{eqn-coherence}
\gamma^{2}=\frac{|\langle S_1^{*}(f)\rangle \langle S_2(f)\rangle|^{2}}{\langle |S_1(f)|^{2}\rangle \langle |S_2(f)|^{2}\rangle},
\end{equation}
is a real-valued positive function that has a maximum of unity if and only if $s_2(t)$ is related to $s_1(t)$ via simple a linear transfer function, 
\begin{equation}\label{eqn-transfer}
s_2(t) = \int_{- \infty}^{\infty} \mathcal{T}_{r}(t-\tau) s_1(\tau) d\tau
\end{equation}
where $\mathcal{T}_{r}$ is some transfer function.  In the other extreme, $\gamma^2=0$ implies no linear relationship between $s_1(t)$ and $s_2(t)$.  In general, $\gamma^2$ can be considered a measure of the fraction of variability in $s_2(t)$ that is coherently related to $s_1(t)$ and has values between $0$ (completely incoherent) and $1$ (perfectly coherent).   For non-zero coherence, we can compute the cross-spectrum $S_1(f)^{*}S_2(f)$. The complex phase $\Phi(f)$ of the cross spectrum gives the frequency-dependent phase shift of the coherent parts of $s_1(t)$ and $s_2(t)$.

\begin{figure}[t]
  \centering
  \subfigure[Coherence]{\includegraphics[width=0.5\textwidth]{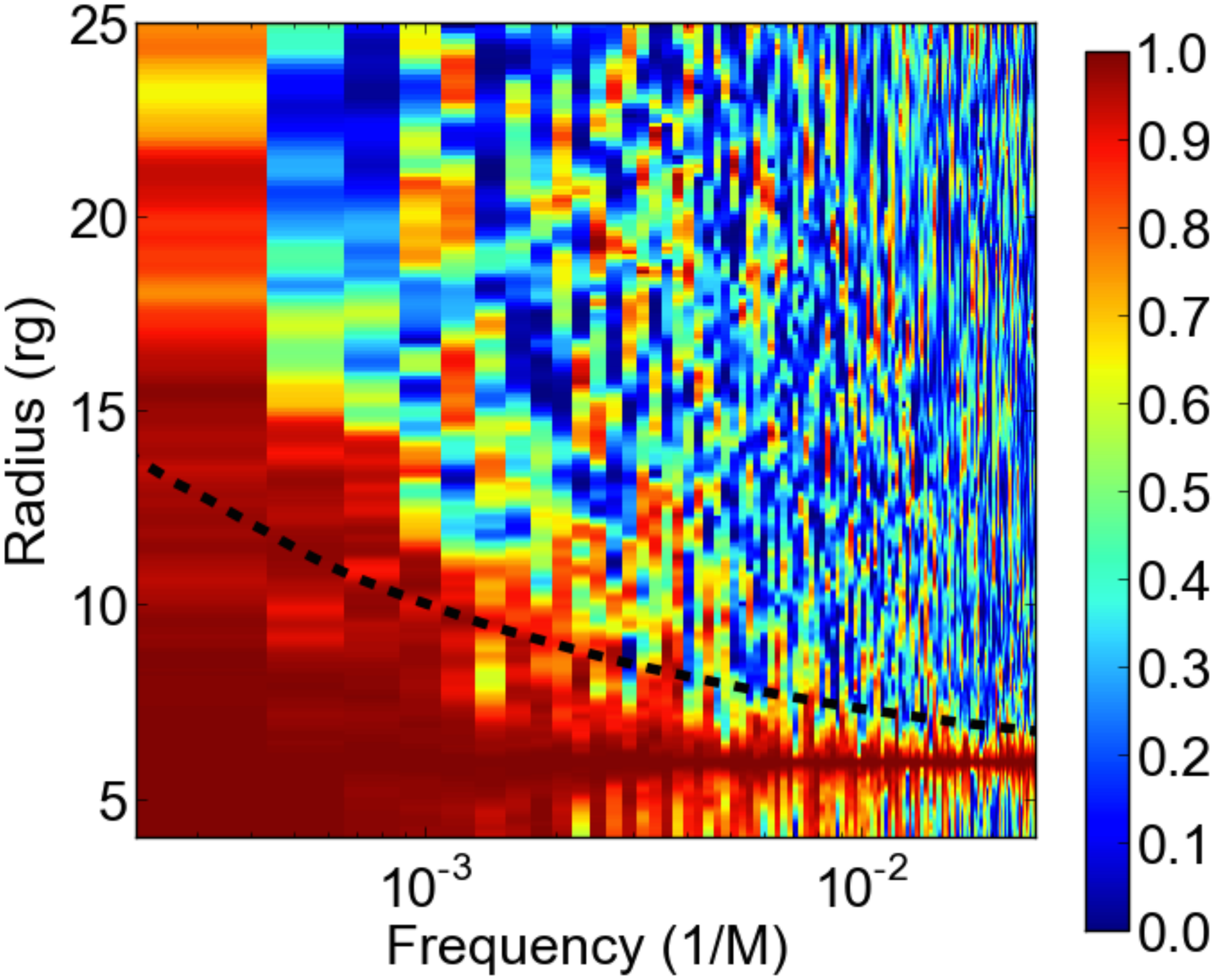}}
  \subfigure[Phase]{\includegraphics[width=0.5\textwidth]{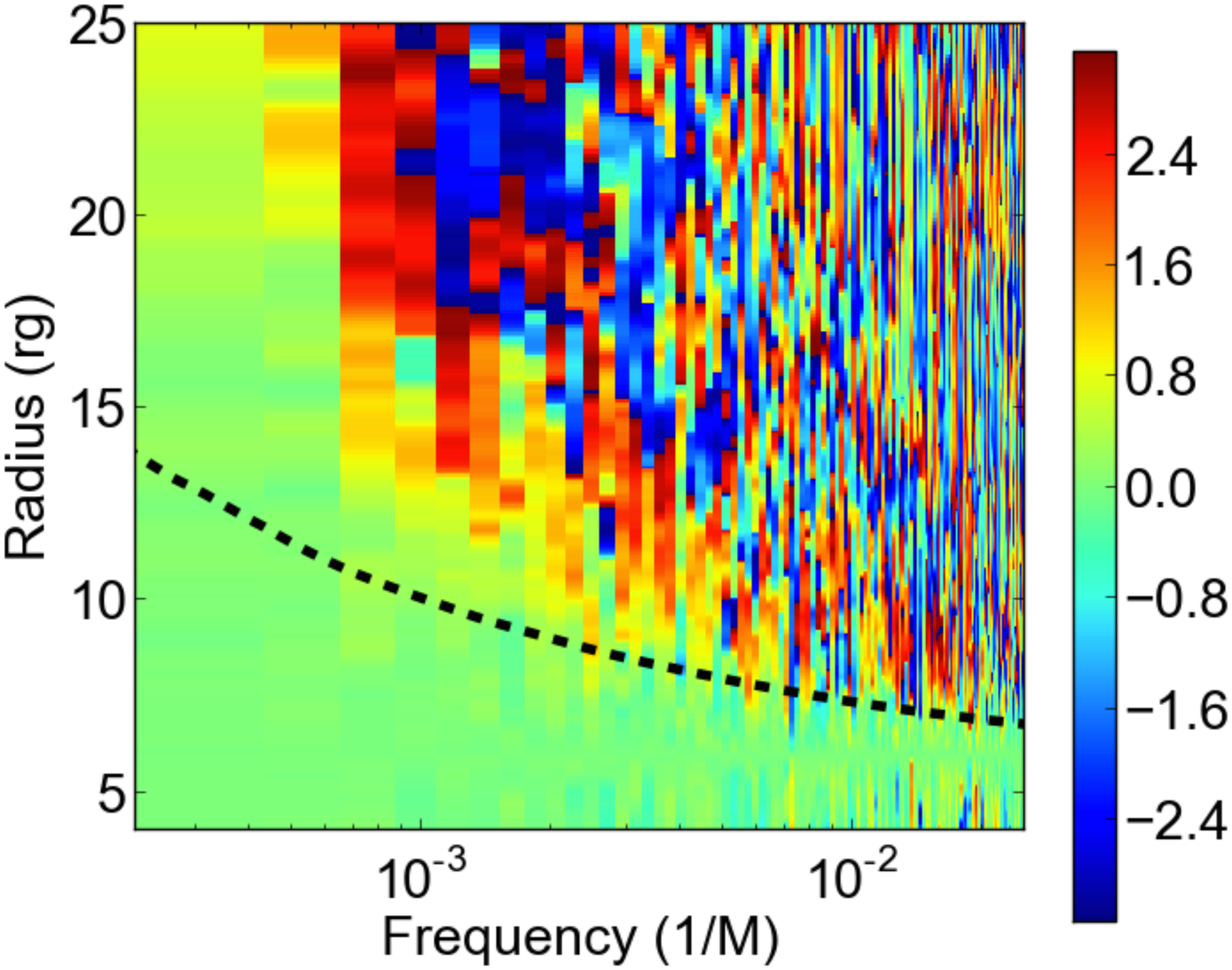}}
  \caption{Coherence (left panel) and phase (right panel) maps.  The black dotted line indicates the local viscous frequency calculated from the average radial velocity of the gas.  The region of high coherence and net phase shift is enveloped by the viscous frequency indicating fluctuations above the viscous frequency at a given radii are quickly damped out while fluctuations below the viscous frequency are preserved.
  } \label{fig-coh_phase}
\end{figure}

The coherence function and phase shifts are shown in Figure \ref{fig-coh_phase}.  The coherence function was calculated with the ISCO taken to be the inner, reference radii.  The $\dot{M}$ ``signal" was broken into three segments for averaging and the coherence function was calculated with each radial bin.  Similar to \citet{2014ApJ...791..126C}, we find the coherent regions are found at frequencies below that of the local viscous frequency.  The viscous time can hence be interpreted as a minimum timescale on which inward communication of flow fluctuations can occur.

\begin{figure}[t]
  \centering
  \subfigure[Phase Difference]{\includegraphics[width=0.42\textwidth]{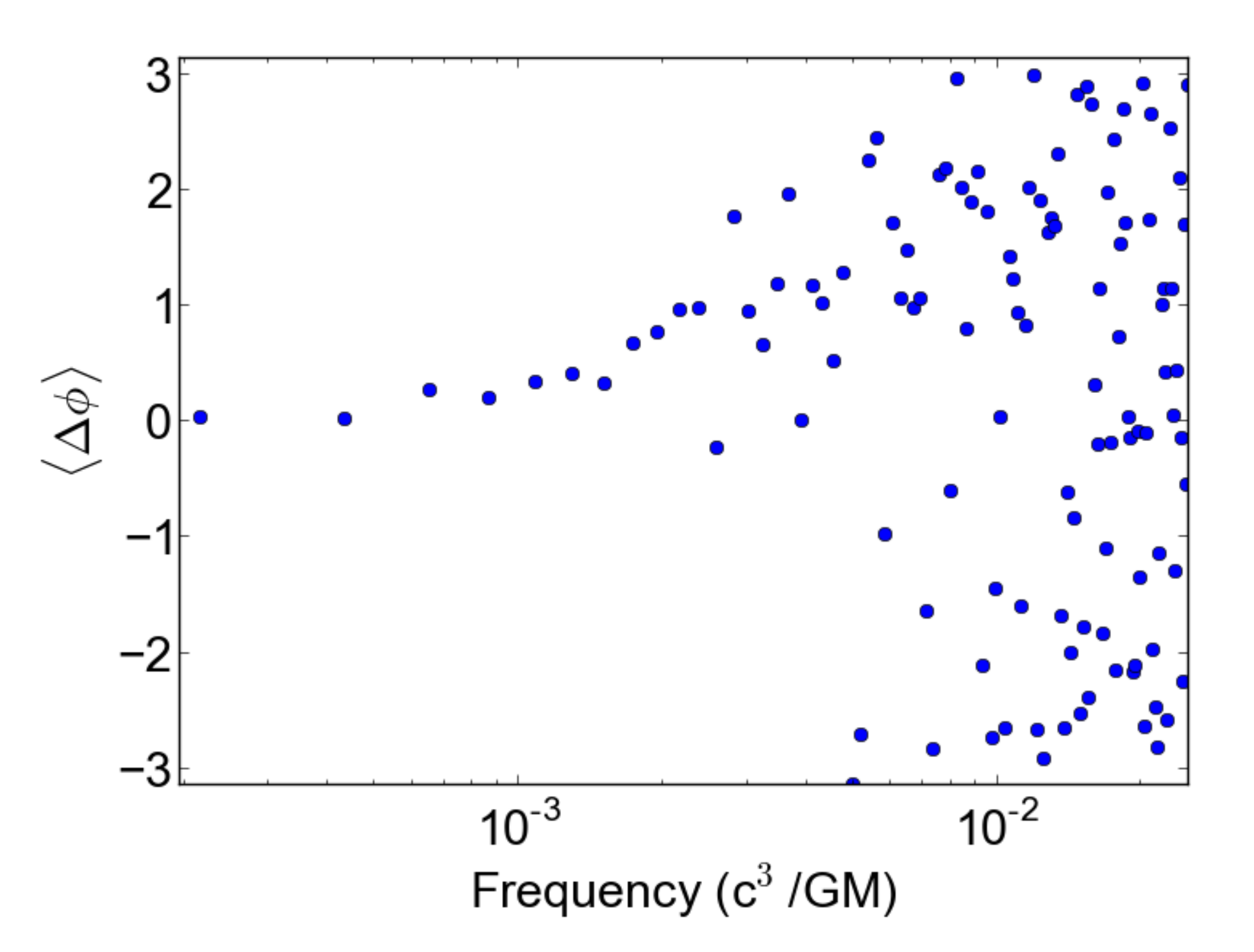}}
  \subfigure[Time Lags]{\includegraphics[width=0.42\textwidth]{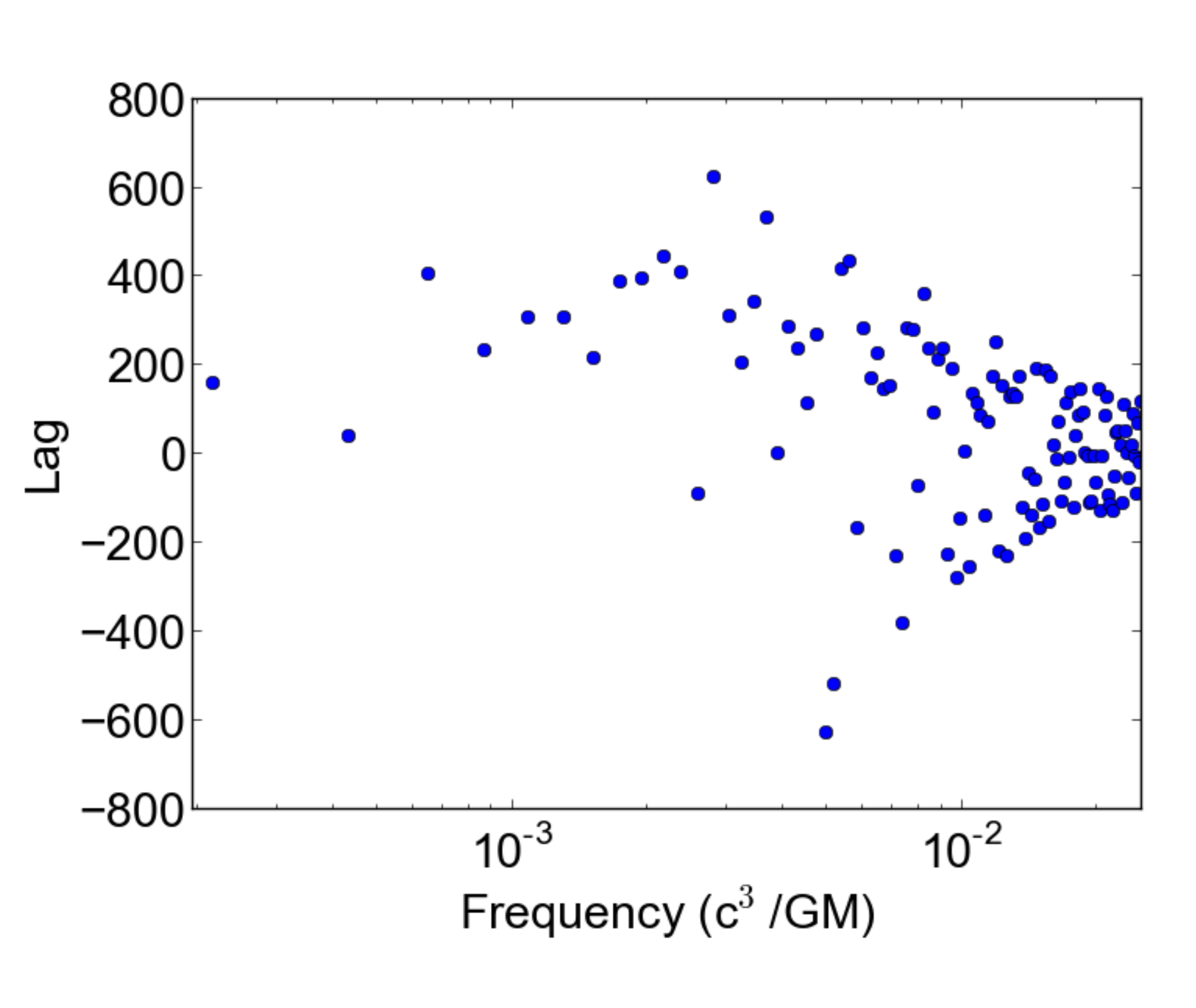}}
  \caption{Frequency dependent phase differences and time lags between $\dot{M}$ at $6 \: r_{g}$ and $10 \: r{g}$.  At frequencies greater than $\omega \approx 3\times10^{-3}$ the coherence breaks down, as can be seen by the large scatter in the two plots.
  } \label{fig-phase_lag_slice}
\end{figure}

From the average phases, time-lags can be determined.  The frequency-dependent time lag at a given radii is $\tau(f)=\Phi(f)/2 \pi f$.  Figure \ref{fig-phase_lag_slice} shows the coherence function calculated between $\dot{M}$ at the ISCO and 10 $r_{g}$.  Given the length of our simulation, only the inner regions have been run for a viscous time, so we only show the structure of the phases and time lags calculated at these two radii.  From the two plots we can see fluctuations in $\dot{M}$ have phase shifts that increase with frequency, but a constant time lag up until the fluctuations become incoherent at $\omega \approx 3\times10^{-3}$.  The constant time lags of the fluctuation in $\dot{M}$ at different radii was seen in the semi-analytic model of \citet{2014ApJ...791..126C}.  This implies that the behavior of the propagating fluctuations in our MHD simulation is consistent with those of the one-dimensional viscous disk.

\subsection{RMS-Flux Relationship}

We finish our analysis of $\dot{M}$ in by looking at the RMS-flux relationship.  While the RMS-flux relationship is a staple of broadband variability \citep{2012MNRAS.422.2620H}, its physical origins remain uncertain.  This relationship is of great interest because it appears to be an intrinsic property of the accretion process and, possibly, even more fundamental than the PSD of the variability.  \citet{2001MNRAS.323L..26U} originally hypothesized the RMS-flux relationship is evidence of \citet{1997MNRAS.292..679L} type propagating fluctuations in $\dot{M}$ since the RMS-flux relationship is a ubiquitous property from all black holes, regardless of mass.

\begin{figure}
  \includegraphics[width=0.5\textwidth]{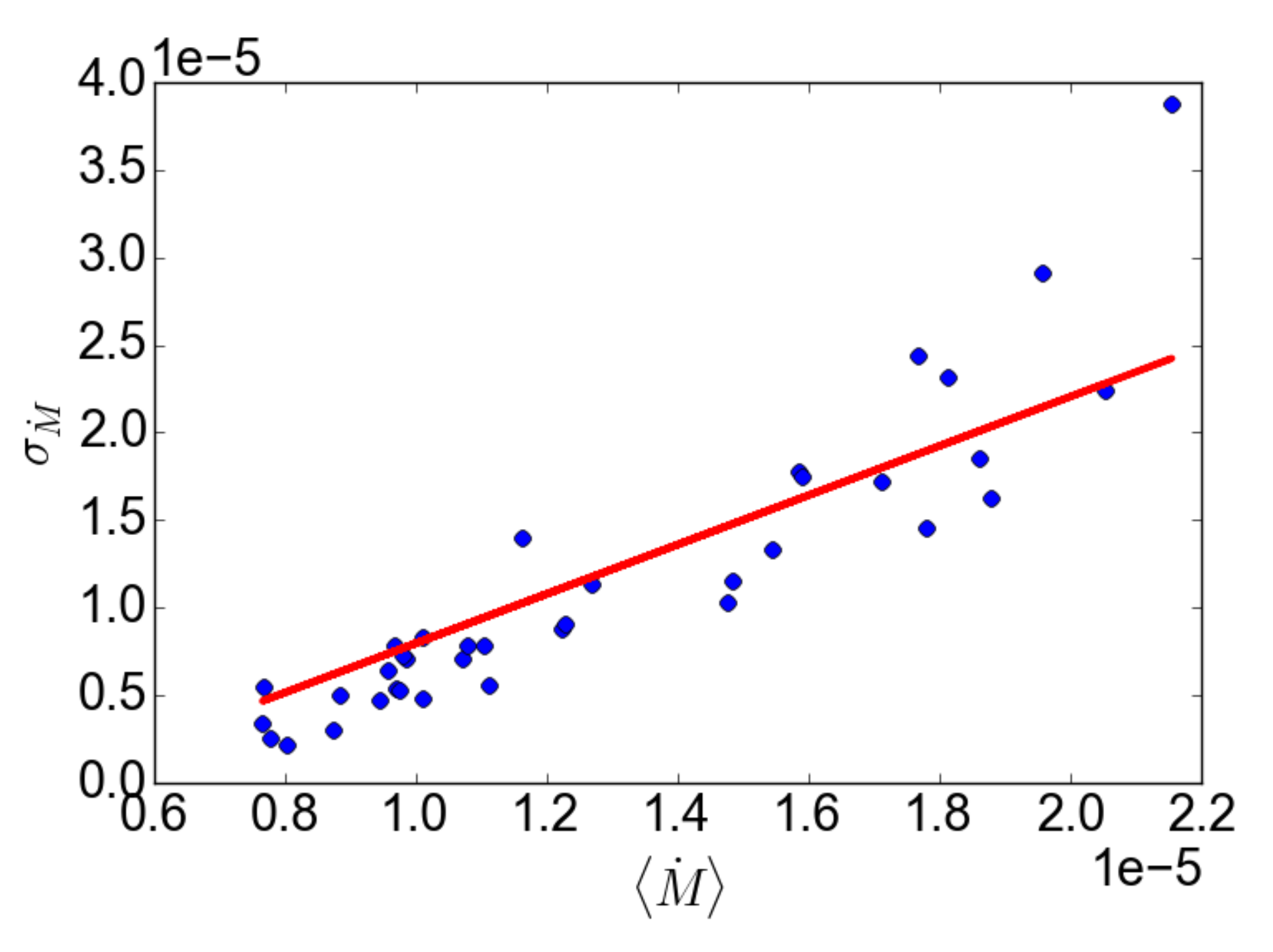}
\caption{Relationship between the RMS of $\dot{M}$ vs its average value at the ISCO calculated over 40 ISCO orbit periods of the simulation.  A strong linear dependence is present in the variability, consistent with the RMS-flux relationship observed from astrophysical black hole systems.
\label{fig-RMS_flux}}
\end{figure}

If the RMS-flux relationship in the photometric variability is truly due to the propagating fluctuations in the accretion disks of these sources, it should be seen in the $\dot{M}$ of our simulation.  To test this we calculated the mean,
\begin{equation}
\overline{\dot{M}}=\frac{1}{N}\sum_{i=1}^N\dot{M}_i,
\end{equation}
and RMS
\begin{equation}
\sigma_{\dot{M}} = \sqrt{\frac{1}{N} \sum_{i=1}^N \Big(\dot{M}_i - \overline{\dot{M}}\Big)^2},
\end{equation}
of the mass accretion rate at the ISCO over a sliding window of 40 ISCO orbits (N=20 data dumps).  Shown in Figure \ref{fig-RMS_flux} is a plot of the raw $\langle \dot{M} \rangle$ vs $\sigma_{\dot{M}}$ scatter plot.  The relationship was fit using a Bayesian, MCMC method with a linear function of the form
\begin{equation}
\hat{\sigma} = k(\langle \dot{M} \rangle + C)
\end{equation}
with $k$ = $1.4 \pm 0.4$ and $C$ = $(-6\pm 6) \times 10^{-6}$.

The (almost) one-to-one correspondence of the RMS to the average mass accretion rate is expected, as well as the intercept passing through the origin, i.e. when there is no flux there should be no variability.  However, the observed RMS-flux relationships are typically offset and best described by a model with a constant baseline flux, $F_{obs}(t)=F_{const}+F_{var}(t)$ \citep{2004A&A...414.1091G}, that varies with black hole spectral state.  Our results suggest the variable component is related to the underlying MHD turbulence of the disk and the flux offset can be attributed to disk processes not captured by our simulation.

\section{Proxy Light Curve Analysis}
\label{sec-proxy_emission}

We now discuss the global behavior of our emission proxy and how the propagating fluctuations in $\dot{M}$ might appear as photometric variability.  The full details of the radiation produced through the interaction of the accretion flow with the various emission mechanisms can only be captured by including more realistic radiation physics, which our simulation lacks.  Nevertheless, we can use our emission proxy to explore the emission variability in a broad sense.  In particular, we want to determine if the structure of the $\dot{M}$ variability can be easily observed when the emission is integrated over the entire disk.  The flux we observe from GBHBs and AGNs is, in effect, the integrated emission from a range of radii that evolve on timescales proportional to their dynamical times. Consequently, there is the possibility some of the variability could be smoothed out as lower-frequency variability from larger radii can wash-out higher-frequency variability.  Given how clear the nonlinear signal is in astrophysical sources, we want to check it is reproduced by the variability of the emission proxy from our simulation.

\begin{figure*}[t]
  \subfigure[Synthetic Light Curve]{\includegraphics[width=0.5\textwidth]{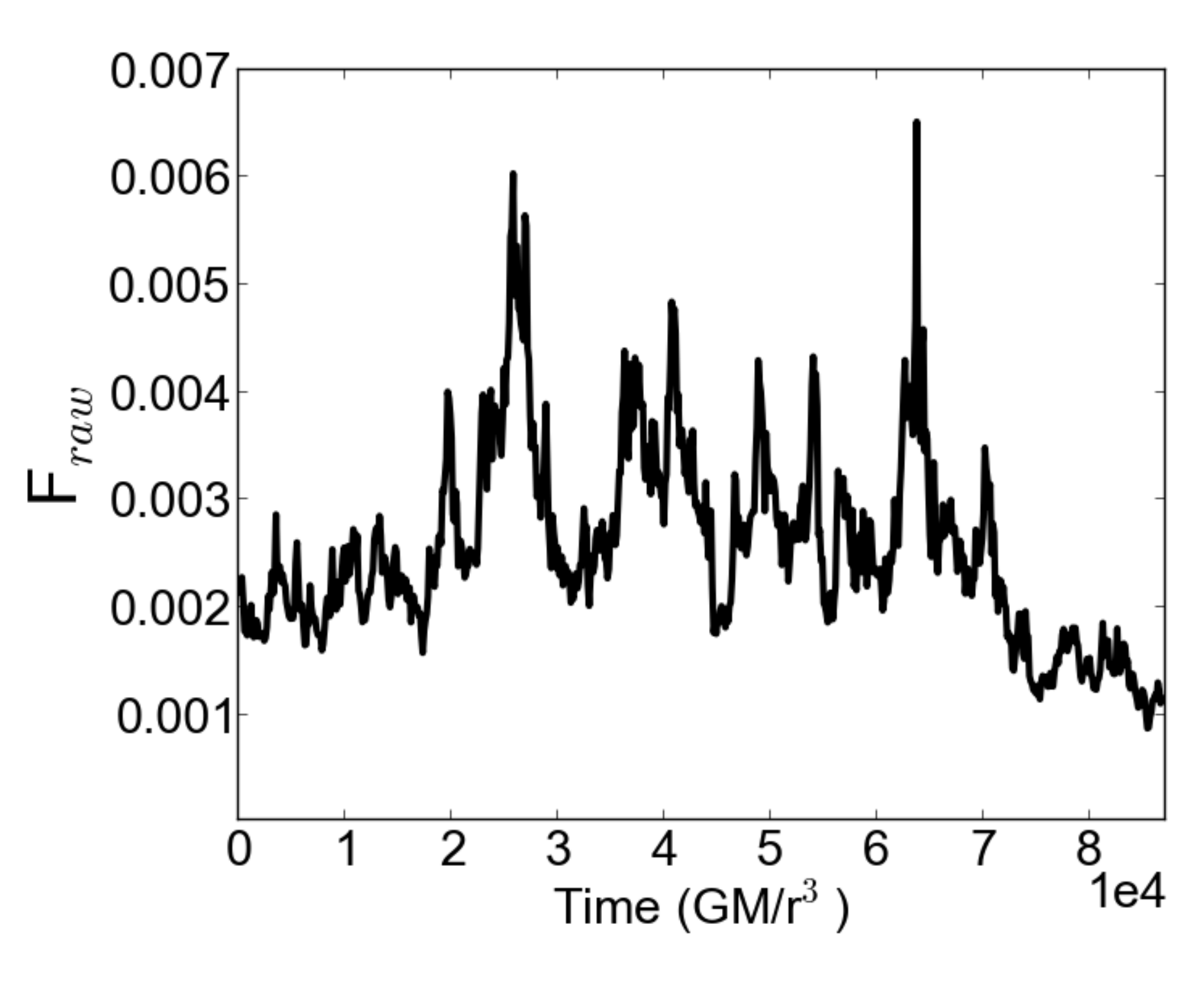}}
  \hspace{0.3in}
  \subfigure[Average Disk Emission Profile]{\includegraphics[width=0.5\textwidth]{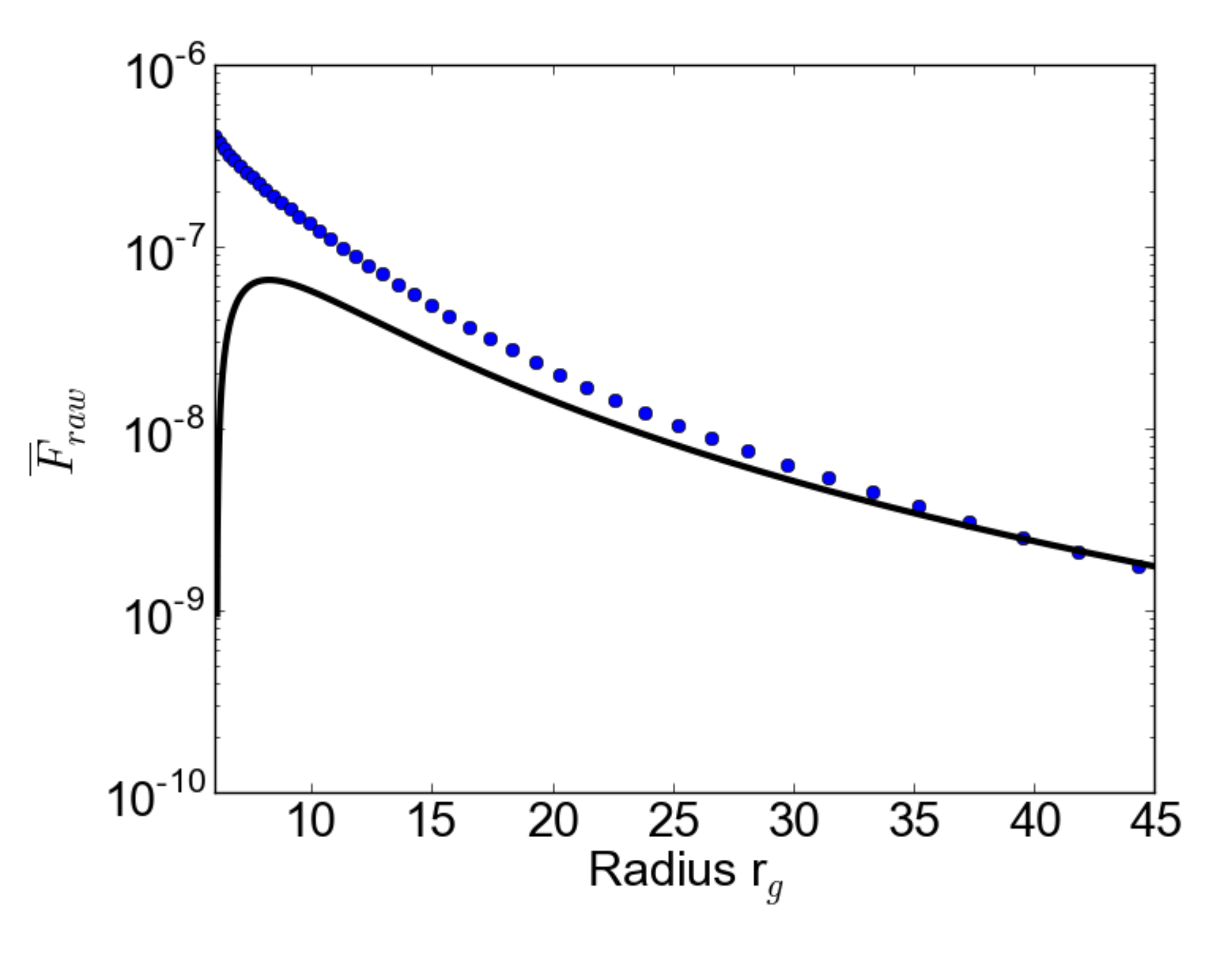}}
\caption{Synthetic light curve (left panel) and disk emission profile (right panel, blue dots).  The integrated synthetic emission from the disk displays large, rapid fluctuations and behaves similarly to $\dot{M}$ across the ISCO.  Qualitatively, the variability is quite similar to the variability of real astrophysical black hole systems.  The rapid variability indicates the inner regions of the disk have the largest contribution to the variability, which is confirmed by the radial profile of the synthetic emission proxy.
\label{fig-lc_plots}}
\end{figure*}

Figure \ref{fig-lc_plots} shows the synthetic light curve and radial profile of the average emission for our disk.  We created the light curve by integrating the dissipation from the emission proxy over the well-resolved region of the simulation ($4-45 \: r_{g}$).  Given our radial range, the synthetic light curve is composed of signals from radii with dynamical times that vary by a factor of 37.5, imitating observation with a broad-band filter.  Several aspects of the disk emission are apparent.  First, the light curve is similar to the time trace of $\dot{M}$ at the ISCO.  The signal is aperiodic and flares in the light curve correspond to large accretion events.  Second, any flares in the light curve quickly decay down to a well-defined, stationary average. The property of mean-reversion is of statistical interest as a way to model its behavior.  Finally, and worth the most discussion, is the dominate role emission from the inner region plays in driving variability.  This is more clearly seen in the radial profile of the emission per unit area of the disk, but can also be inferred by the high-frequency flares.  In Figure \ref{fig-lc_plots} the disk profile of \cite{1973A&A....24..337S} is shown for comparison.  The two profiles trace each other well at larger radii, but diverge in the innermost regions.  In the canonical $\alpha$-disk the energy flux goes to zero at the ISCO because it was believed at the time that no torques could be felt from the gas in the inner plunging region.  MHD simulations have shown that, in fact, torques can be felt across the ISCO and radiation can originate from closer to the black hole than the ISCO \citep{2001ApJ...548..348H, 2002ApJ...566..164H, 2002ApJ...573..754K}.  In addition to the physical effect of torques spanning the ISCO region, there may also be a nonphysical contribution to our profile because we assume any energy from stresses is immediately converted into radiated energy.  In reality, there is a time delay between the injection and the dissipation as the energy is transported through the turbulent cascade.  In the inner regions near the black hole the material could be rapidly swept into the hole before it has had time to completely radiate the injected energy.  This would lead to an overestimation of the emission from  the inner regions by our emission proxy due to this advection of energy.

\subsection{Signatures of Propagating Fluctuations}

In our analysis of the mass accretion rate in the simulation we found the variability of $\dot{M}$ had near-identical structure to the universally observed flux variability from black hole systems.  We will now focus on assessing how well $\dot{M}$ fluctuations translate into fluctuations in disk luminosity using the diagnostics used in Section \ref{sec-prop_flucs}.  

\begin{figure}
  \subfigure[Histogram of Light Curve]{\includegraphics[width=0.5\textwidth]{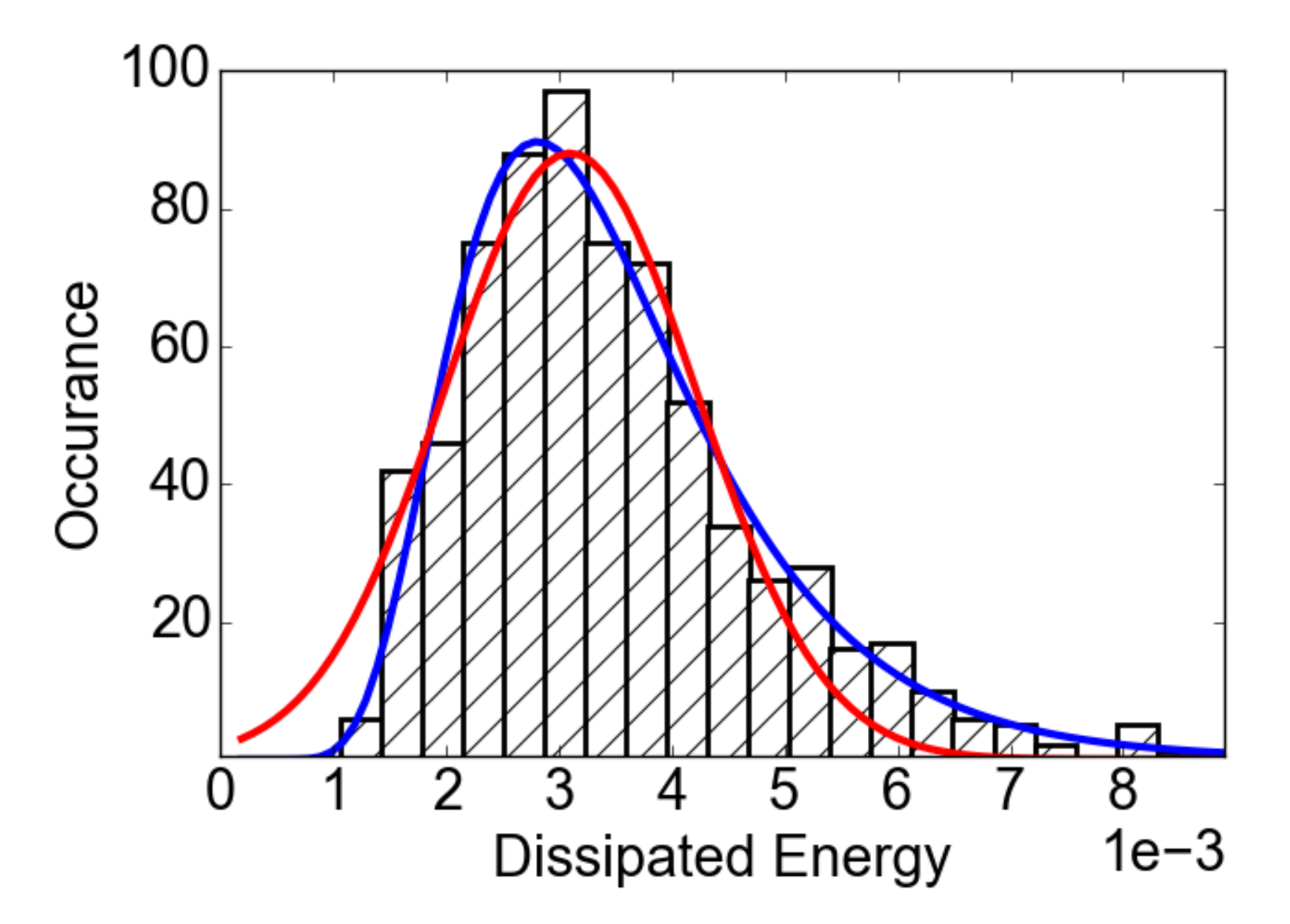}}
  \subfigure[Light Curve Coherence]{\includegraphics[width=0.5\textwidth]{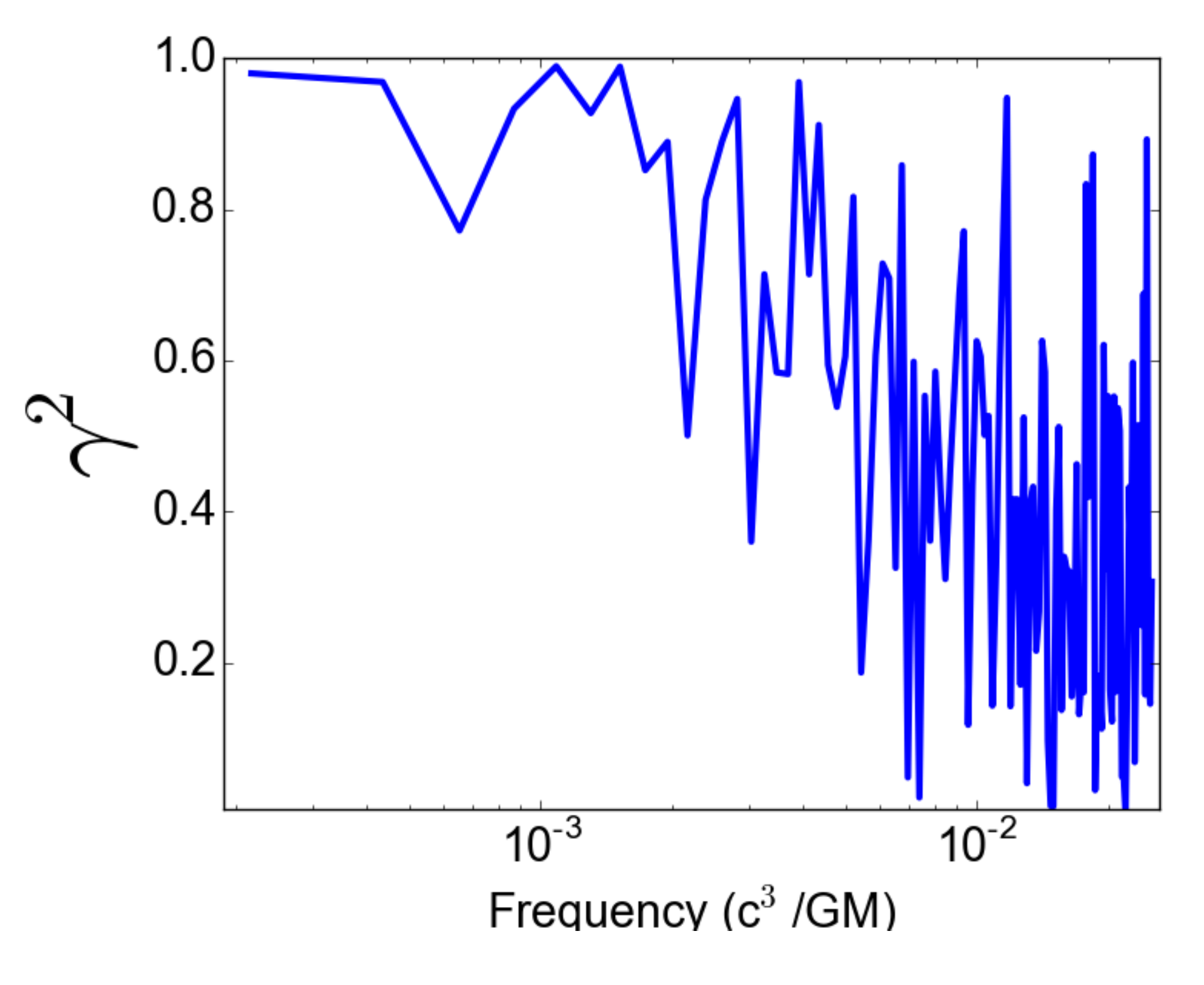}}
  \subfigure[RMS-Flux Relationship]{\includegraphics[width=0.5\textwidth]{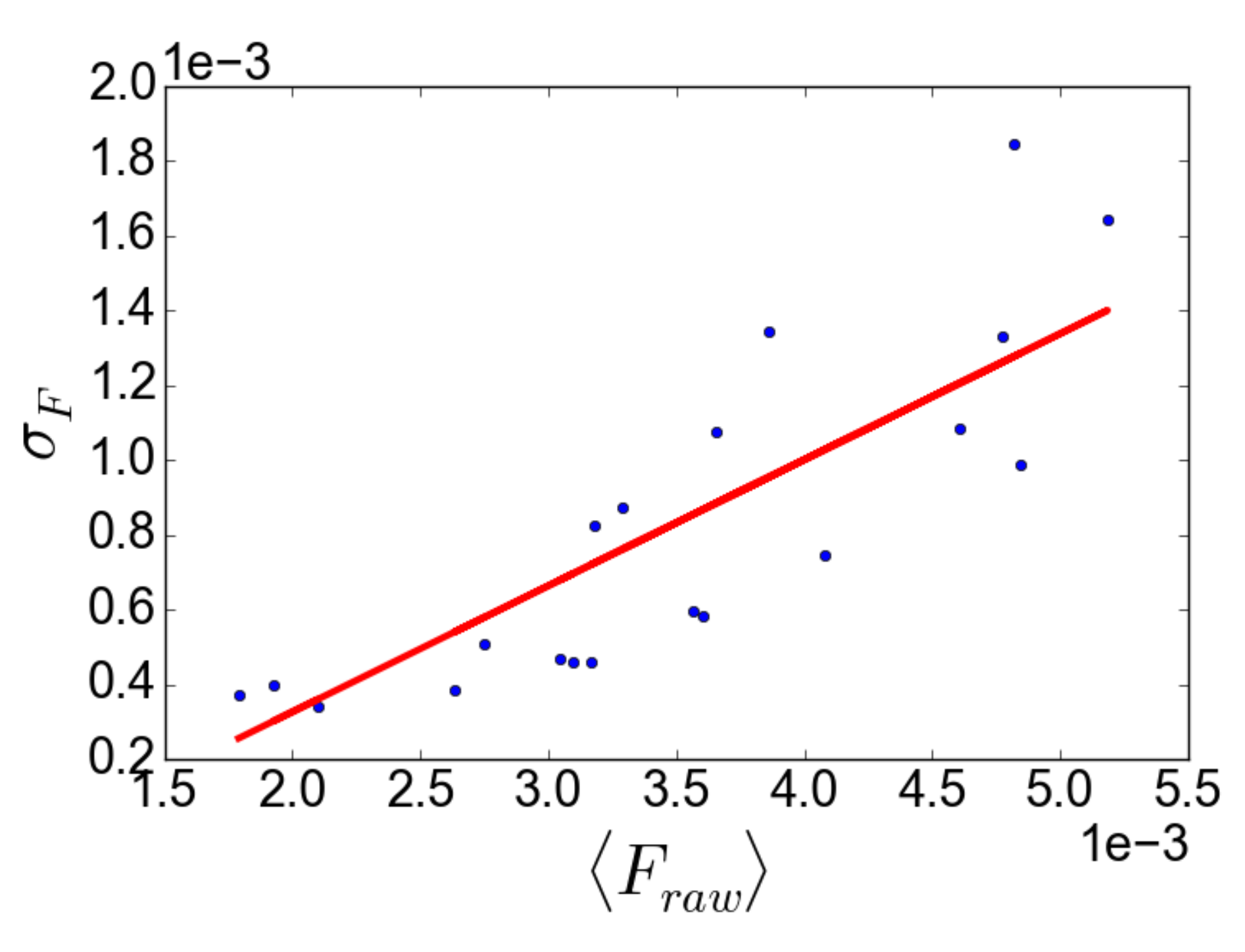}}
\caption{Histogram of light curve (top), light curve coherence (middle), and RMS-flux relationship (bottom).  The structure of the synthetic light curve variability is similar to that of $\dot{M}$ at the ISCO, as is seen in Figures \ref{fig-dist}, \ref{fig-coh_phase}, and \ref{fig-RMS_flux}.
\label{fig-lc_diagnostics}}
\end{figure}

Figure~\ref{fig-lc_diagnostics} (upper and middle panels) shows the flux histogram and RMS-flux relationship for the total (``broad-band'') proxy light curve shown in Fig.~\ref{fig-lc_plots}.  The preferred fit to the flux histogram is a log-normal distribution with parameters $\mu=-5.7$ and $\sigma=-0.39$.  The $\chi^{2}/D.O.F$ of this fit is $18.1/24$ which is statistically well-fit and much better than that of a normal distribution with $\chi^{2}/D.O.F$ of this fit is $98.9/24$.  The skewness of this distribution is $\gamma_{1}=1.28$, indicating the histogram of the light curve is more highly skewed than the histogram of the $\dot{M}$ and the log-normal distribution is more pronounced.  Naively, the increased skewness is surprising given that the light curve is integrated over the disk and presumably additively combining many random processes.  According to the central limit theorem, this should drive the distribution to be more Gaussian.  However, since $\dot{M}$
is radially coherent, the fluctuations of the emission proxy are not independent, as is assumed in a Gaussian process.  

The full synthetic light curve also has a linear relationship between the RMS of the variability and the average flux value.  As we did with $\dot{M}$, the light curve was broken into 20 sections and the standard deviation and mean values were calculated for each.  The distribution is well fit by a linear function with $k$ = $0.34 \pm 0.07$ and $C$ = $(-3\pm 3) \times 10^{-4}$.  The RMS-flux relationship of the synthetic emission is flatter than the linear relationship between the RMS of $\dot{M}$ and the average value.  In fact, the slope of our linear fit is remarkably similar to those seen in GBHBs, specifically Cyg X-1 and SAX J1808.4-3658 \citep{2001MNRAS.323L..26U}.  While it is tempting to make a direct comparison, the statistics on our measurement prevent drawing any deeper conclusions.  However, given how much the difference in the slopes of the RMS-flux relationship of $\dot{M}$ compared to the RMS-flux relationship of the light curve, it does appear that integrating over the disk plays a significant role.  By integrating over the entire disk, the low-frequency emission adds a slowly varying component to the flux, thereby decreasing the percentage of the flux from the fast varying inner regions and decreasing the slope.

An examination of coherence requires us to create two light curves to imitate observation with two bands (for instance hard and soft X-rays).  We formed these light curves by integrating the emission proxy in the regions spanning $4-10 r_{g}$ and $10-15 r_{g}$.  Then, as we did with $\dot{M}$, the coherence function was calculated by dividing the light curves into three sections, resulting in the coherence function shown in Fig.~\ref{fig-lc_diagnostics} (bottom panel).  The coherence shows the same behavior as found for the mass accretion rate, with highly-coherent signals at low-frequencies, and a falling coherence at higher-frequencies.  De-coherence is seen at $\omega \approx 300 \: c^3/GM$ which, while a relatively high-frequency, agrees with the coherence and phase plots of $\dot{M}$ fluctuations with the ISCO in Figure \ref{fig-coh_phase}.

\subsection{Timing Properties}

Having established the signatures of propagating fluctuations in $\dot{M}$ are present in the variability of our emission proxy, we will now look at the PSD and the statistical behavior of the variability by fitting a simple Ornstein-Uhlenbeck (OU) process.

\begin{figure}
  \includegraphics[width=0.5\textwidth]{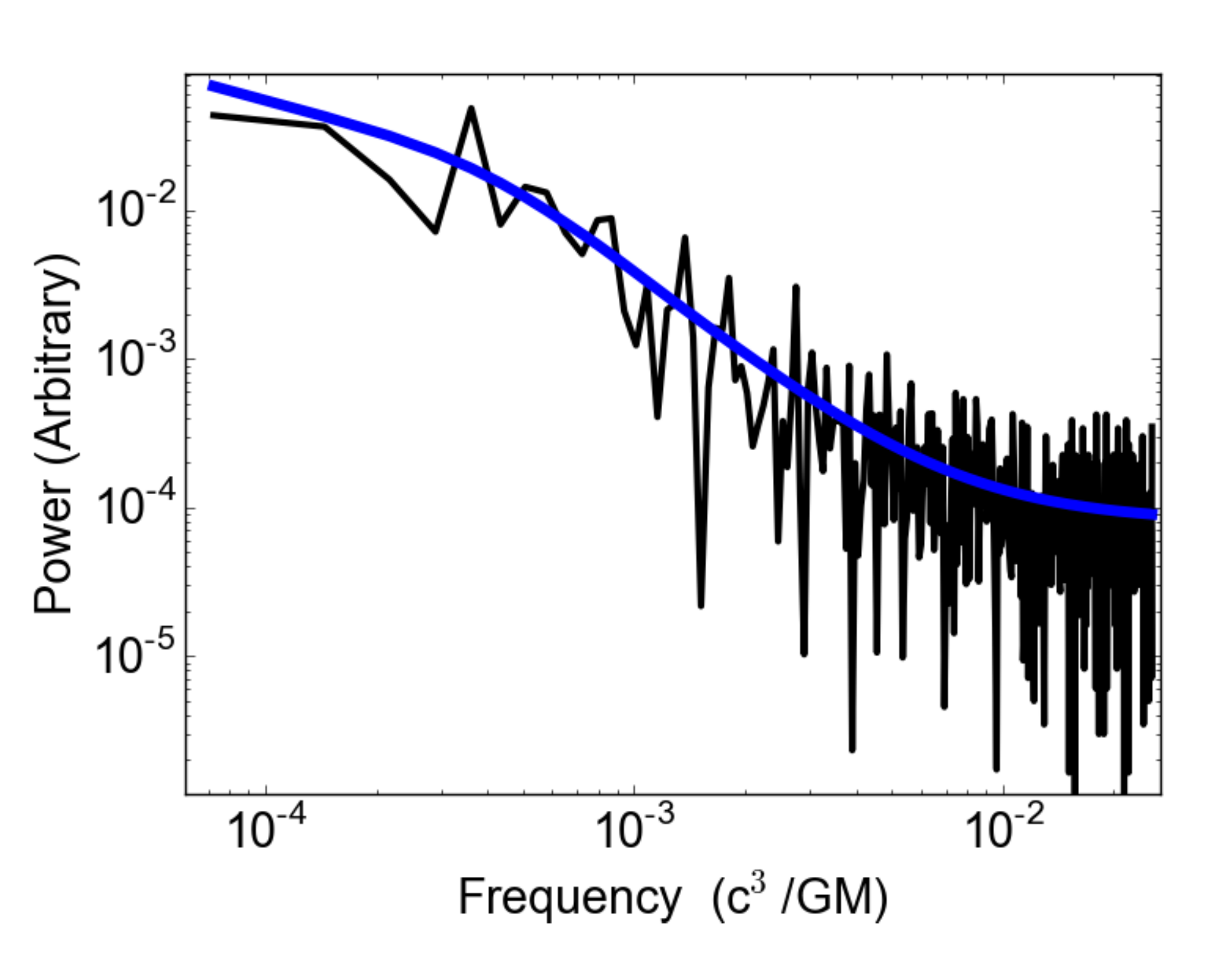}
\caption{Synthetic light curve PSD with broken power law fit (blue line).
\label{fig-plaw_fit}}
\end{figure}

Shown in Figure \ref{fig-plaw_fit} is the PSD of the synthetic light curve fit with a broken power law.  The last 512 data points of our light curve were used to calculate the PSD which ensures no phase dependencies were present.  Following the method described in \citet{2009ApJ...692..869R}, the PSD was fit with a power law plus white-noise component and broken power law plus white-noise component of the form 
\begin{equation}
A(\omega) = \begin{array}{ll}
K\omega^{-\Gamma_{1}} & \mbox{if $\omega < \omega_{brk}$} \\
K\omega_{brk}^{\Gamma_{2}-\Gamma_{1}}\omega^{\Gamma_{2}} & \mbox{if $\omega > \omega_{brk}$} \end{array}
\end{equation} 
where K is a normalization constant, $\Gamma_{1}$ is the low frequency slope, $\Gamma_{2}$ is the high frequency slope, and $\omega_{brk}$ is the break frequency, to determine which model provides the best description of the PSD.  The power density was assumed to have an exponential probability distribution with the probability of measuring a given power between $p$ and $p+dp$ expressed by, \begin{equation}
P(p)dp = \frac{1}{p_{0}} e^{-p/p_{0}}dp,
\end{equation} where p$_{0}$ is the mean power.  For an individual bin, the likelihood of a measuring power $p_{obs,i}$ is\begin{equation}
\mathcal{L}_{i} = (1/p_{mod,i}) \textrm{exp}(p_{obs,i} / p_{mod,i}).
\end{equation}  For the entire PSD, the likelihood is the product of the individual likelihoods.  Since this can often be very large or very small, the standard statistic used is the log likelihood, \begin{equation}
\textrm{ln} \: \mathcal{L} = \sum_i [-\textrm{ln} p_{mod,i}-p_{obs,i}/p_{mod,i}].
\end{equation}  A Markov Chain Monte Carlo (MCMC) algorithm was used to conduct the fitting and find the parameters that maximize the log likelihood of each PSD model.

The PSD is well fit by both a broken power law with $\Gamma_{1} = 0.5 \pm 0.3$, $\Gamma_{2}=1.8 \pm 0.2$ , and $\omega_{brk}=(2.9 \pm 1.4) \times 10^{-4}$ and a single power law with $\Gamma=1.38 \pm 0.18$.  However, the broken power law provides a much better statistical fit with $\Delta \textrm{ln} \: \mathcal{L} = 5.08$ at the expense of two degrees of freedom.  This translates into a $\Delta \chi^{2} = 10.16$.

Hence, while the underlying PSD of $\dot{M}$ at the ISCO in our simulation has $\Gamma=1$, the integrated emission proxy has a steeper PSD.  This is simply due to the integrated contributions from across the accretion disk, with additional low frequency power from large radii being added to the PSD.   It is not immediately clear what physics sets the break frequency $\omega_{brk}$, corresponding to a timescale of $2000-6500 \: r_g/c$, although it approximately corresponds to the cooling/thermal timescale of the inner regions of the disk.  In order to ensure that the low-frequency break is not imposed by the truncation of our emission proxy at $r=45r_{g}$, we generated light curves and the corresponding PDSs truncating at $r=20,30, 40r_{g}$.   In all cases, the break persisted and the frequency was unchanged.

In addition to the PSD, the statistical properties of the variability of the synthetic light curve are of interest and provide an additional method of characterizing the synthetic light curve.  Recent work has shown a damped random-walk, and more specifically an Ornstein-Uhlenbeck (OU) process, provides a very good statistical model of GBHB and AGN variability \citep{2009ApJ...698..895K, 2010ApJ...708..927K, 2010ApJ...721.1014M, 2011ApJ...730...52K}.   The OU process is part of a class of statistical models called \emph{continuous time, first order autoregressive}, CAR(1), processes.  CAR(1) processes can take many forms and behave very different depending on their model parameters.  The OU process describes a noisy relaxation process where a system is stochastically perturbed from a stationary mean and decays back to the mean with a relaxation time $\tau$ \citep{1996AmJPh..64..225G}.  The stochastic differential equation describing such a process is: \begin{equation}
dX(t) = \lambda(\mu-X(t))dt+\sigma dW_{t}
\end{equation} where $\lambda=1/\tau$, $\mu$ is the mean and $\sigma$ is a measure of the average magnitude of volatility per $\sqrt{dt}$.  This equation has the following direct solution: \begin{equation}
X_{i+1}=X_{i} e^{-\lambda t} + \mu (1-e^{-\lambda \delta}) + \sigma \sqrt{\frac{1-e^{-2 \lambda \delta}}{2 \lambda}}N_{0,1}
\end{equation} where $\delta$ is the time step and $N_{0,1}$ is a zero-centered, Gaussian distribution with a variance of 1.

The even sampling of our light curve provides some simplicity in fitting the OU process.  $X_{i+1}$ is linearly dependent on $X_{i}$ and takes the form: \begin{equation}
X_{i+1} = A X_{i}+b + \epsilon,
\end{equation} where $\epsilon$ is a normal-random term that is independent and identically distributed.  From a simple linear fit, we can solve for $\lambda$, $\mu$, and $\sigma$ where \begin{equation}
\lambda = \frac{- \textrm{ln } a}{\delta}, \end{equation} \begin{equation}
\mu = \frac{b}{1-a}, \& \end{equation} \begin{equation}
\sigma = sd(\epsilon) \sqrt{\frac{-2 \textrm{ ln } a}{\delta (1-a^2)}}.
\end{equation}  

Since the OU process is inherently a Gaussian process we must therefore perform the fit to the synthetic light curve in log-space since our ``flux" distribution is log-normally distributed.  Thinking in terms of real observations, this is equivalent to fitting in magnitude space.  For our light curve, the best fit parameters are $\lambda = (3.5\pm1.2)\times10^{-3}$, $\mu=(3.3\pm0.6)\times10^{-3}$ and $\sigma=(6\pm3)\times10^{-5}$.  The $\lambda$ parameter corresponds to a decay time of $t_{decay}=286 \:GM/c^{3}$.

\section{Discussion}
\label{sec-discussion}

In our simulation we find that \cite{1997MNRAS.292..679L} type ``propagating fluctuations" quickly develop and are sustained throughout the entire simulation.  The multiplicative manner by which fluctuations in $\dot{M}$ combine and grow defines the behavior and subsequent evolution of the disk.  The most significant consequence from the multiplicative combination of $\dot{M}$ fluctuations is that the accretion history of a given parcel of gas is retained, allowing for the apparent propagation of fluctuations on a viscous time.  The $\dot{M}$ variability at the ISCO in our simulation bears striking resemblance to variability observed from astrophysical black holes and shares the same phenomenological properties.  In particular, it reproduces the log-normal flux distributions, RMS-flux relationship, and radial coherence ubiquitously observed from GBHBs and AGNs.  These nonlinear features have typically been difficult for variability models to account for, but are a consequence of a scenario where the mass accretion rate is proportional to the multiplicative product of a stochastically varying viscosity and the local surface density.  In our simulated disk they develop naturally from the MRI driven turbulence. 

Additionally, we use an emission proxy to generate a disk-integrated synthetic light curve to verify that the nonlinear characteristics of the $\dot{M}$ variability will actually translate into observable features in the disk emission.  While rudimentary, our emission proxy follows the turbulent energy injected to the gas, which we assume is quickly radiated away since we are considering a radiatively efficient disk.  We find that, indeed, the nonlinear features of $\dot{M}$ are present when the entire disk integrated emission is considered and are not lost with the addition of low-frequency variability originating from gas at large radii. 

Our work was able to address several issues raised by \citet{2014ApJ...791..126C} regarding the timescale on which the effective $\alpha$ parameter of the disk fluctuates and how well an MHD disk can really be described by a strictly viscous disk on long time scales.  The previous efforts to model propagating fluctuations in mass accretion rate of \cite{1997MNRAS.292..679L} using a linearized model and \citet{2014ApJ...791..126C} using a one dimensional viscous disk model relied on the \emph{a priori} assumption that the low-frequency fluctuations in effective $\alpha$ occurred on a viscous time.  However, this is very long for the magnetic field to vary.  When \citet{2014ApJ...791..126C} allowed $\alpha$ to vary on a dynamical time, the typically assumed evolutionary timescale for the MRI, they found that the fluctuations were too rapid and propagating fluctuations did not develop.  We find the disk dynamo can provide the necessary intermediate timescale modulation of the effective $\alpha$ to drive propagating fluctuations in mass accretion rate.  Dynamo action is universally observed in modern accretion disk simulations and seems to be intimately tied to the MHD physics of a differentially rotating plasma.  Thus, the dynamo can provide a natural mechanism to introduce variability in the effective $\alpha$.  While the ubiquity of the dynamo has been established, its presence is typically neglected when considering accretion disk evolution.  Our results highlight the significant impact the dynamo can have on the disk and that it is a fundamental ingredient in the driving of propagating fluctuations.

The properties of the $\dot{M}$ and light curve variability from our global, \emph{ab initio} MHD simulation are very similar to those from the simple viscous disk with stochastic viscosity parameter of \citet{2014ApJ...791..126C}.  From this similarity, we can conclude that on long timescales our MHD disk acts in a viscous manner.  This is not surprising, but there is not a trivial connection between the internal MHD stresses that transport angular momentum and the expected viscous behavior of the disk, even though it has largely been intuited from empirical results.  

Our MHD simulation captures gross aspects of real disk variability, but it is not without limitations.  In particular, we use simple prescriptions to approximate the affects of general relativity and radiative physics.  As a first effort at modeling propagating fluctuations in mass accretion rate in an MHD disk we are mainly concerned with capturing the dynamics of the disk and the interplay of the accretion flow with the stochastically fluctuating effective $\alpha$.  To do this, a long, high-resolution simulation is required.  Therefore,  the priority of where computational resources are used shifts away from more complex physics and into the duration and resolution of the simulation.

While the focus of our analysis has been on understanding the signatures of propagating fluctuations around black holes, we do not expect the effects of general relativity and radiation physics to have a significant impact on the characteristics of the disk evolution are primarily concerned with.  In fact, the phenomenology we seek to understand is also readily observed in cataclysmic variable (CV) systems \citep{2012MNRAS.427.3396S, 2012MNRAS.421.2854S, 2014MNRAS.445.1031S}.  These systems are non-relativistic and not hot enough to be largely affected by radiation pressure in the disk.  The similarity of the broad-band variability between  CVs, GBHBs, and AGNs implies that propagating fluctuations are a generic feature of highly ionized accretion disks.  The fluctuations resulting from the underlying accretion processes behave consistently across many orders of magnitude of mass, temperature and evolutionary time implying that the growth of propagating fluctuations is not unique to an specific class of system, but rather intrinsically tied to the angular momentum transport resulting from MHD turbulence.  Therefore, we believe that the results from this simulation are robust enough to provide a qualitative description the accretion mechanism for highly ionized accretion disks. 

\section{Conclusions}
\label{sec-summary}

We have performed the first examination of the propagating fluctuation picture using \emph{ab initio} MHD turbulent models of disk accretion.  The propagating fluctuations model has served to explain the universally observed nonlinear features of photometric variability from accreting black holes of all masses.  In our simulation, propagating fluctuations in $\dot{M}$ form and behave in a manner consistent with those of the viscous disk model of \citet{2014ApJ...791..126C}.  Our main results are summarized here: 

\begin{enumerate}

\item The effective $\alpha$-parameter in our simulated disk is highly variable.  The variability has two components:  a high-frequency component due to local fluctuations from the MRI and a lower frequency component due to modulation of the stress by the disk dynamo.

\item The lower frequency variability in the effective $\alpha$-parameter from the disk dynamo drives fluctuations in $\dot{M}$ by modulating the internal disk stress on an intermediate timescale.  As was shown in \citet{2014ApJ...791..126C}, if fluctuations in $\alpha$ are too rapid, fluctuations will be damped out and propagating fluctuations will not grow.  The dynamo frequency is low enough that propagating fluctuations can easily grow.  Previous efforts directed at understanding the role of propagating fluctuations in $\dot{M}$ relied on \emph{ad hoc} prescriptions for the variability timescale.  We show, for the first time, that the dynamo fills this role and provides a natural driver for fluctuations in $\dot{M}$.

\item The $\dot{M}$ in our simulation is positively correlated with $\Sigma$ and the effective $\alpha$.  In the analytic disk equations the mass accretion rate is proportional to the product of the local surface density and the local viscosity.  These correlations occur in our \emph{ab initio} simulation, suggesting that they naturally result from the interaction of the MRI with accretion flow.

\item The $\dot{M}$ variability at the ISCO of our simulation resembles the photometric variability observed from GBHBs and AGNs.  The $\dot{M}$ variability is log-normally distributed, has a linear RMS-flux relationship, and displays radial coherence.  These features are empirically observed and have been attributed to propagating fluctuations in $\dot{M}$.  We find that they are, in fact, properties of accretion flow when propagating fluctuations in $\dot{M}$ are present.

\item Using an emission proxy to generate a disk integrated, synthetic light curve, we find that the qualitative properties of the accretion flow are reflected in the observable emission.  While the nonlinear signatures are preserved in the light curve, integrating over the disk adds additional low-frequency power from more distant radii.  However, since inner region of the disk is brightest, the high frequency variability is preferentially weighted and the nonlinear properties of $\dot{M}$ are preserved.

\end{enumerate}

\acknowledgements

We would like to thank Mark Avara, M. Coleman Miller, Phil Uttley, and Christian Knigge for the many insightful discussions throughout the course of this work.  JDH is grateful for the financial support from the UMD-Goddard Joint Space Science Institute (JSI) through his Graduate Fellowship.  This research was also supported by NASA through grants NNX15AC40G and NNX14AJ04G.

\end{document}